\documentclass[a4paper,11pt]{article}

\renewcommand{\thetable}{\Roman{table}}

\pdfoutput=1 

\usepackage{jheppub} 
\usepackage[T1]{fontenc} 
\usepackage{lineno,hyperref}
\usepackage{caption}
\usepackage{subcaption}
\usepackage{soul}
\title{\boldmath Methods and restrictions to increase the volume of resonant rectangular-section
haloscopes for detecting dark matter axions}

\author[*,a]{J.M.~Garc\'ia-Barcel\'o,}
\author[a]{A.~\'Alvarez~Melc\'on,}
\author[a]{A.~D\'iaz-Morcillo,}
\author[b]{B.~Gimeno,}
\author[a]{A.J.~Lozano-Guerrero,}
\author[a]{J.~Monzo-Cabrera,}
\author[a]{J.R.~Navarro-Madrid,}
\author[a]{P.~Navarro}

\affiliation[a]{Department of Information Technologies and Communications, Technical University of Cartagena, 30203 - Cartagena, Spain}
\affiliation[b]{Instituto de F\'isica Corpuscular (IFIC), CSIC-University of Valencia, 46980 - Valencia, Spain}

\affiliation[*]{Corresponding author}

\emailAdd{josemaria.garcia@upct.es}

\abstract
{Haloscopes are resonant cavities that serve as detectors of dark matter axions when they are immersed in a strong static magnetic field. In order to increase the volume and improve its introduction within dipole or solenoid magnets for axion searches, various haloscope design techniques for rectangular geometries are discussed in this study. The volume limits of two types of haloscopes are explored: based on single cavities and based on multicavities. For both cases, possibilities for increasing the volume in long and/or tall structures are presented. For multicavities, 1D geometries are explored to optimize the space in the magnets. Also, 2D and 3D geometries are introduced as a first step for laying the foundations for the development of these kind of topologies. The results prove the usefulness of the developed methods, evidencing the ample room of improvement in rectangular haloscope designs nowadays. A factor of three orders of magnitude improvement in volume compared with a single cavity based on WR-90 standard waveguide is obtained with the design of a long and tall single cavity. Similar procedures have been applied for long and tall multicavities. Experimental measurements are shown for prototypes based on tall multicavities and 2D structures, demonstrating the feasibility of using these types of geometries to increase the volume in real haloscopes.}

\begin{document}
\maketitle
\flushbottom

\section{Introduction}
\label{Sec:Introduction}
Axions and other particles consistent with the Standard Model that might be part of dark matter have sparked a lot of attention in the recent decades. The strong Charge Conjugation-Parity issue \cite{Peccei:1977Jun,Peccei:1977Sep} might be solved by axions, the particles predicted by Weinberg \cite{Weinberg:1978} and Wilczek \cite{Wilczek:1978}. Five years later, it was predicted that axions could possibly be a dark matter candidate using the misalignment idea \cite{Preskill:1983,Abbott:1983,Dine:1983}.\\

Over the last thirty years, numerous research groups have developed experimental systems to search for dark matter axions \cite{Irastorza:2018dyq}. These experiments make use of the inverse Primakoff effect \cite{Primakoff:1951}. In turn, depending on the origin of the axion source, these detection techniques can be divided into three types: Light Shining through Walls (LSW), helioscopes and haloscopes. The first one generates axion particles by itself (artificially), while helioscopes and haloscopes are based on external natural sources (the sun and the galactic halo, respectively). All of them use the axion-photon conversion determined by a strong external static magnetic field. In addition, by making use of high quality factor resonators (like microwave cavities), this transformation can be boosted in the case of haloscopes \cite{Sikivie:1983ip}.\\

Several components make up the entire axion detection system. To begin, because of the extremely low axion-photon coupling, a cryogenic environment with temperatures in the Kelvin range is required to reduce the thermal noise. Second, the received radio frequency (RF) power of the haloscope is amplified, filtered, down-converted and detected by a coupled receiver, adding very low noise levels. Finally, the receiver performs an Analog-Digital conversion and a Fast Fourier Transform for data post-processing \cite{RADES_paper3}.\\

The major goals of an axion detection system are to enhance the axion-photon detection RF power and to improve the axion-photon conversion sensitivity of the haloscope. This power is determined by the axion characteristics as well as the haloscope (cavity in this case) parameters, as shown in the following equation \cite{RADESreviewUniverse}:
\begin{equation}
\label{eq:Pd}
    P_d \, = \, g^2_{a\gamma}\, \frac{\rho_a}{m_a} \, B_e^2 \, C \, V \, \frac{\beta}{\left(1+\beta\right)^2} \, Q_0
\end{equation}
where $\beta$ is the extraction coupling factor (with $\beta=1$ for critical coupling operation regime to achieve the maximum power transfer), $g_{a\gamma}$ is the unknown axion-photon coupling coefficient, $\rho_a$ is the dark matter density, $m_a$ is the axion mass (proportional to the working frequency of the experiment), $B_e$ is the magnitude of the static external magnetic field $\vec{B}_e$, $C$ is the form or geometric factor, $V$ is the haloscope or cavity volume and $Q_0$ its unloaded quality factor. Here the $Q_0$ is assumed to be much lower than the axion quality factor ($Q_a\approx10^6$) \cite{kim_CAPP_2020}. It should be emphasized that the external static magnetic field ($\vec{B}_e$) depends on the magnet employed in the experiment (dipole or solenoid) and its spatial distribution and polarization must be considered in order to boost the axion-photon conversion. In addition, $\beta=1$ (which is known as the critically coupling condition) is achieved employing only one port. In a practical set up, there is usually a second port in the haloscope, but it is very undercoupled during the data taking operation. In fact, this second port is only employed for testing and for the electromagnetic characterization of the cavity resonance.\\

The form factor provides the coupling between $\vec{B}_e$ and the radio frequency electric field ($\vec{E}$) induced into the cavity by the axion-photon interaction. It can be written as:
\begin{equation}
\label{eq:C}
    C \, = \, \frac{|\int _V \, \vec{E} \cdot \vec{B}_e \, dV|^2}{\int_V \, |\vec{B}_e|^2 \, dV \int_V \, \varepsilon_r \, |\vec{E}|^2 \, dV}
\end{equation}
where $\varepsilon_r$ is the relative electric permittivity filling the cavity medium (generally air or vacuum). On the other hand, from equation~\ref{eq:Pd} the axion-photon conversion sensitivity of the haloscope for a given signal-to-noise ratio ($\frac{S}{N}$) can be obtained as \cite{RADESreviewUniverse}
\begin{equation}
\label{eq:ga}
    g_{a\gamma} \, = \,  \left(\frac{\frac{S}{N} \, k_B \, T_{sys} \, \left(1+\beta\right)^2}{\rho_a \, C \, V \, \beta \, Q_0}\right)^{\frac{1}{2}}\frac{1}{B_e}\left(\frac{m_a^3}{Q_a \, \Delta t}\right)^{\frac{1}{4}}
\end{equation}
where $k_B$ is the Boltzmann constant, $T_{sys}$ is the noise temperature of the system and $\Delta t$ is the time window used in the data taking. Then, the factors that can be adjusted and optimized in the design of a haloscope are $\beta$, $C$, $V$ and $Q_0$.\\

The main objective of this work is to analyze the possibilities of increasing the volume of a haloscope based on rectangular cavities to effectively improve the axion detection sensitivity. In addition, other important concepts are discussed such as the improvement in mode clustering or mode separation, which is a key feature of a microwave-cavity haloscope to avoid the degradation of the form factor and the quality factor \cite{RADES_paper2} as the following sections show. The maximum volume allowed in a haloscope design depends mainly on four factors: the cavity shape (rectangular or cylindrical), the operation electromagnetic mode and frequency, whether the multi-cavity concept is used or not, and the geometry and type of magnet (and hence the direction of the magnetic field) where the axion measurement campaign will be carried out.\\

As commented before, this work is focused on rectangular geometries. The resonant frequency of this kind of cavities for $TE_{mnp}$ and $TM_{mnp}$ modes is expressed as
\begin{equation}
\label{eq:fR}
    f_{mnp} = \frac{c}{2\sqrt{\mu_r\varepsilon_r}}\sqrt{\left(\frac{m}{a}\right)^2+\left(\frac{n}{b}\right)^2+\left(\frac{p}{d}\right)^2}
\end{equation}
where $c$ is the speed of light in vacuum, $\mu_r$ is the relative magnetic permeability of the medium inside the cavity ($\mu_r=\varepsilon_r=1$ is assumed in this work), $m$, $n$, and $p$ are integers that denote the number of maxima of the electric field along the $x$, $y$, and $z$ axis, respectively, and $a$, $b$ and $d$ are the width, height and length of the cavity, respectively. For $TE_{mnp}$ modes the allowed indexes are: $m= 0, \, 1, \, 2, \, ...$; $n= 0, \, 1, \, 2, \, ...$; $p= 1, \, 2, \, 3, \, ...$, although $m$ and $n$ can not be zero simultaneously. For $TM_{mnp}$ modes: $m= 1, \, 2, \, 3, \, ...$; $n= 1, \, 2, \, 3, \, ...$ and $p= 0, \, 1, \, 2, \, ...$. As indicated by this equation, resonant frequencies are dependent on the three cavity dimensions. This relationship actually suggests difficulties to increase at the same time volume and frequency, and without increasing mode clustering.\\

\begin{figure}[h]
\centering
\begin{subfigure}[b]{0.28\textwidth}
         \centering
         \includegraphics[width=1\textwidth]{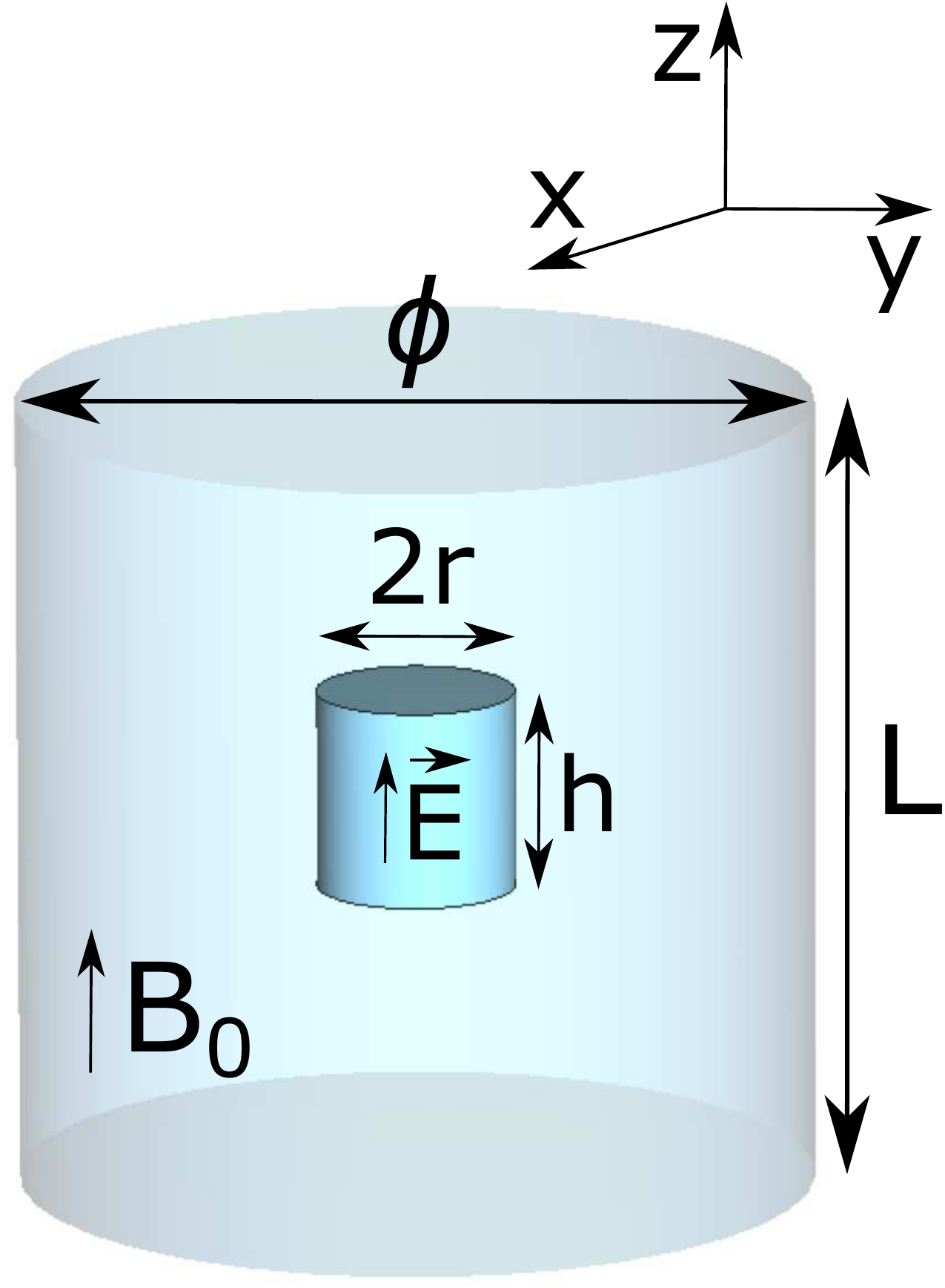}
         \caption{}
         \label{fig:Solenoid}
\end{subfigure}
\hfill
\begin{subfigure}[b]{0.6\textwidth}
         \centering
         \includegraphics[width=1\textwidth]{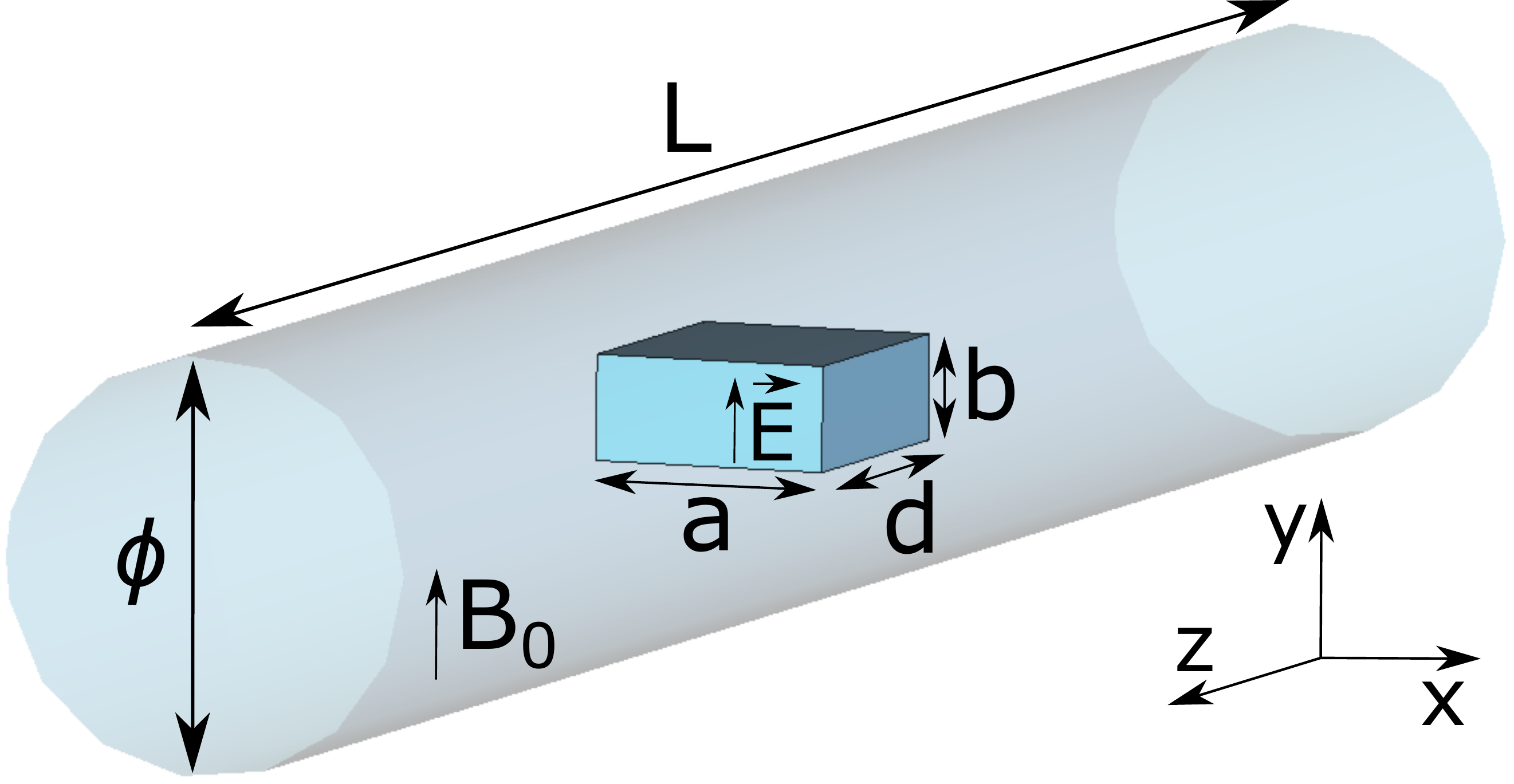}
         \caption{}
         \label{fig:Dipole}
\end{subfigure}
\caption{Examples of (a) a solenoid magnet bore with a cylindrical cavity of radius $r$ and height $h$ operating with the $TM_{010}$ cylindrical mode and (b) a dipole magnet bore with a rectangular cavity of width $a$, height $b$ and length $d$ working with the $TE_{101}$ rectangular mode. The direction of the magnetic field is mostly in the $z-$axis for the solenoid magnet and in the $y-$axis for the dipole magnet. Light blue objects represent solenoid and dipole magnets, while darker blue objects represent microwave cavities.}
\label{fig:Dipole&Solenoid}
\end{figure}

The most used magnets in dark matter axion detection experiments are solenoids, as shown in Figure~\ref{fig:Solenoid}.
Particularly, they are used by ADMX \cite{Braine:2019fqb} and HAYSTACK \cite{PhDThesis-Brubaker}. These magnets create an axial magnetic field (along $z-$axis) and the cavity is usually cylindrically shaped, aligning the direction of the electric field of the $TM_{010}$ cylindrical mode with the external magnetic field of the magnet and thus providing a good form factor. This paper will, however, discuss how rectangular cavities can also be optimized for this type of magnet. Meanwhile, powerful accelerator dipole magnets (see Figure~\ref{fig:Dipole}), such as the CAST magnet, produces a transverse magnetic field with an intensity of around $\sim9$~T, and it was available in the early stages of the RADES project. In this case, the selected cavity type was rectangular, where the electric field of the $TE_{101}$ rectangular mode is vertically polarized and, therefore, mostly parallel to the dipole static magnetic field \cite{RADESreviewUniverse}. Other example is BabyIAXO, a superconducting toroidal magnet whose magnetic field pattern can be consulted in \cite{BabyIAXO} and it can be considered in this work as a dipole magnet for simplicity. Figure~\ref{fig:Dipole&Solenoid} shows the optimum orientation of a rectangular and cylindrical cavities for dipole and solenoid magnets, respectively. In Table~\ref{tab:magnets} a description of the most common magnets used or to be used by several research groups is shown.
\begin{table}[h]
\begin{tabular}{|c|c|c|c|c|c|c|}
\hline
Magnet & Type & B (T) & T (K) & $\phi$ (mm) & L (m) & References \\ \hline\hline
CAST & Dipole & 9 & 1.8 & 42.5 & 9.25 & \cite{CAST:1999} \\ \hline
BabyIAXO & Quasi-dipole & $\sim2.5$ & 4.2 & 600 & 10 & \cite{BabyIAXO} \\ \hline
SM18 & Dipole & 11 & 4.2 & 54 & 2 & \cite{RADES-HTS} \\ \hline
Canfranc & Solenoid & 10 & 0.01 & 130 & 0.15 & \cite{Aja_2022} \\ \hline
MRI (ADMX-EFR) & Solenoid & $\sim9$ & 0.1 & 650 & 0.8 & \cite{ADMX-EFR} \\ \hline
HAYSTAC & Solenoid & 9 & 0.127 & 140 & 0.56 & \cite{PhDThesis-Brubaker} \\ \hline
CAPP-8TB & Solenoid & 8 & 0.05 & 165.4 & 0.476 & \cite{Choi:2021} \\ \hline
\end{tabular} 
\centering
\caption{\label{tab:magnets} Characteristics of different magnets for axion data taking. $\phi$ and $L$ are the diameter and the length, respectively, of the magnets.}
\end{table}
For this work, a constant magnetic field $\vec{B}_e=B_e \, \hat{y}$ in the dipole and quasi-dipole \cite{BabyIAXO} magnets has been selected as approximation for the calculation of the form factor.\\

In general, the length of the dipole magnet bore is much longer than its diameter. This is the case of the CAST and BabyIAXO magnets with diameters of $42.5$ and $600$~mm and lengths of $9.25$ and $10$~m, respectively \cite{CAST:1999,BabyIAXO}. On the other hand, the length and diameter of the solenoid magnet bores are quite similar. This is the case of the Canfranc and MRI (ADMX-EFR) magnets, with diameters of $130$ and $650$~mm, and lengths of $150$ and $800$~mm, respectively \cite{Aja_2022,ADMX-EFR}. Then, the first idea to take advantage of the bore in dipole magnets is to increase the total length of the haloscope. With this purpose, the length of the sub-cavities can be increased employing the multicavity concept \cite{RADES_paper1,RADES_paper2}. However, it will be shown that the use of new topologies including tall structures on both dipole and solenoid magnets are also very interesting concepts.\\

In section~\ref{Sec:Ind} the limits in volume in a haloscope based on a single cavity (in terms of mode separation or mode clustering) are demonstrated. In section~\ref{Sec:Multicav} some important concepts are introduced to design a haloscope based on the multicavity concept in order to increase its volume as much as possible. In section~\ref{Sec:2D3D} a first step for laying the foundations for the development of the previous structures to exploit even more the space available in the bore of the magnets is further elaborated. Finally, in section~\ref{Conclusions} the conclusions and future prospects of this work are examined.

\section{Single cavities}
\label{Sec:Ind}
For rectangular cavities working in dipole magnets the $TE_{101}$ mode is selected since it is the one that maximizes the form factor described in equation~\ref{eq:C}. For this mode, the height of the cavity $b$ does not affect the resonant frequency $f_{TE_{101}}$ since $n=0$, so it can be increased as desired in order to increase the cavity volume. However, there is a limit where the cavity height cannot be increased due to the proximity of the higher modes with $n\neq0$ (closeness with the $TE_{111}$ in this case), which may hinder the correct identification of the mode and can even reduce in some cases the form factor.\\

Also, studying equation~\ref{eq:fR}, it is observed that the best option to increase the length of the haloscope without decreasing the resonant frequency is by reducing slightly the width for the $TE_{101}$ mode. This reduction is small compared to the gained length, so the total volume will increase. Additionally, when the length of the cavity is much larger than its width the resonant frequency becomes almost independent on the cavity length $f_{TE_{101}} \approx \frac{c}{2a}$. Here, again, length limit is imposed by the proximity of the next resonant mode (mode clustering with the $TE_{102}$ in this case). In the following sections, several strategies for increasing the volume of haloscopes without lowering the resonant frequency of the $TE_{101}$ mode will be discussed.\\

\subsection{Long cavities}
\label{SubSec:IndLong}
\begin{figure}[h]
\centering
\begin{subfigure}[b]{0.49\textwidth}
         \centering
         \includegraphics[width=1\textwidth]{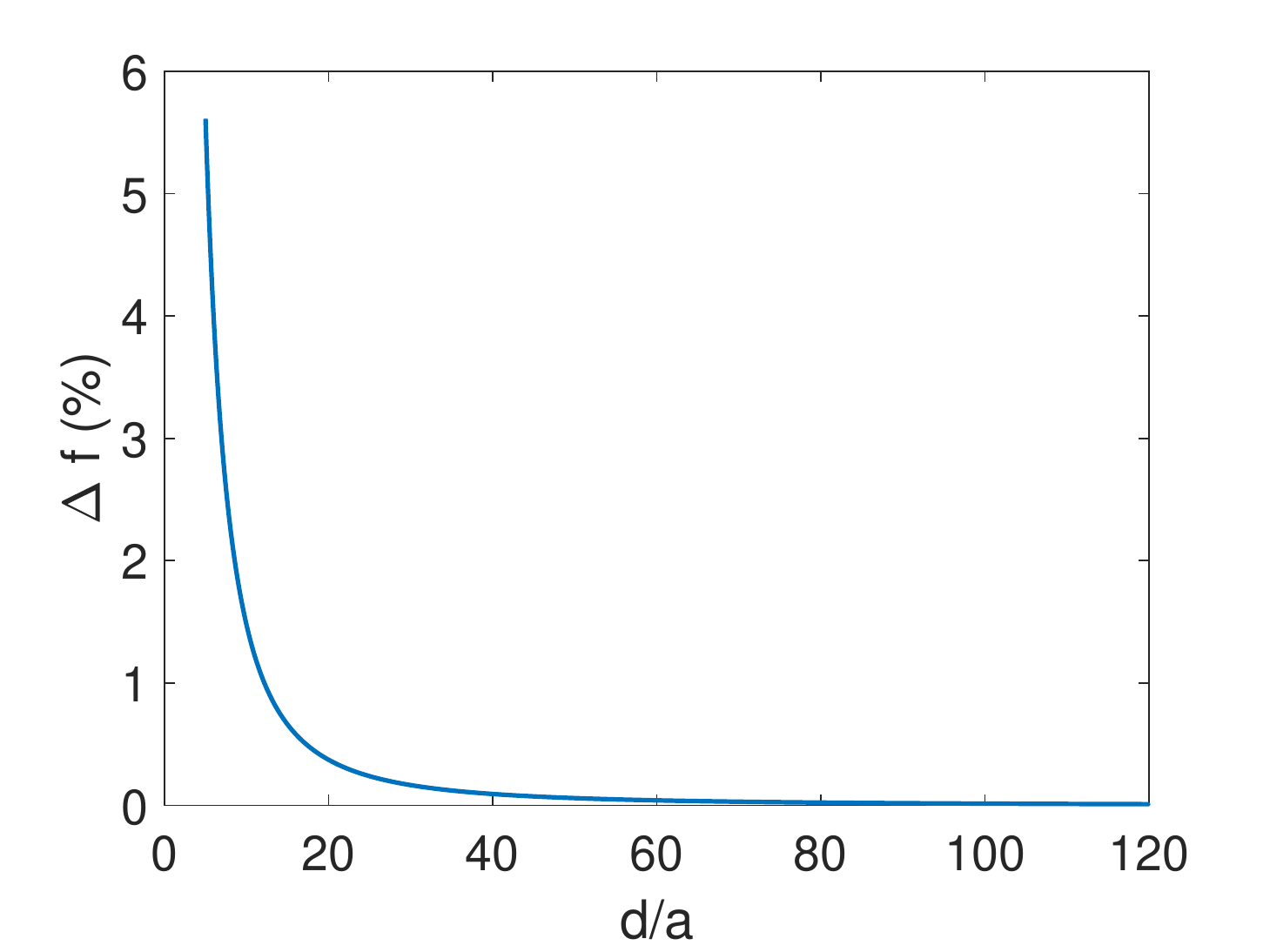}
         \caption{}
         \label{fig:ModeMixing_vs_d-a}
\end{subfigure}
\hfill
\begin{subfigure}[b]{0.49\textwidth}
         \centering
         \includegraphics[width=1\textwidth]{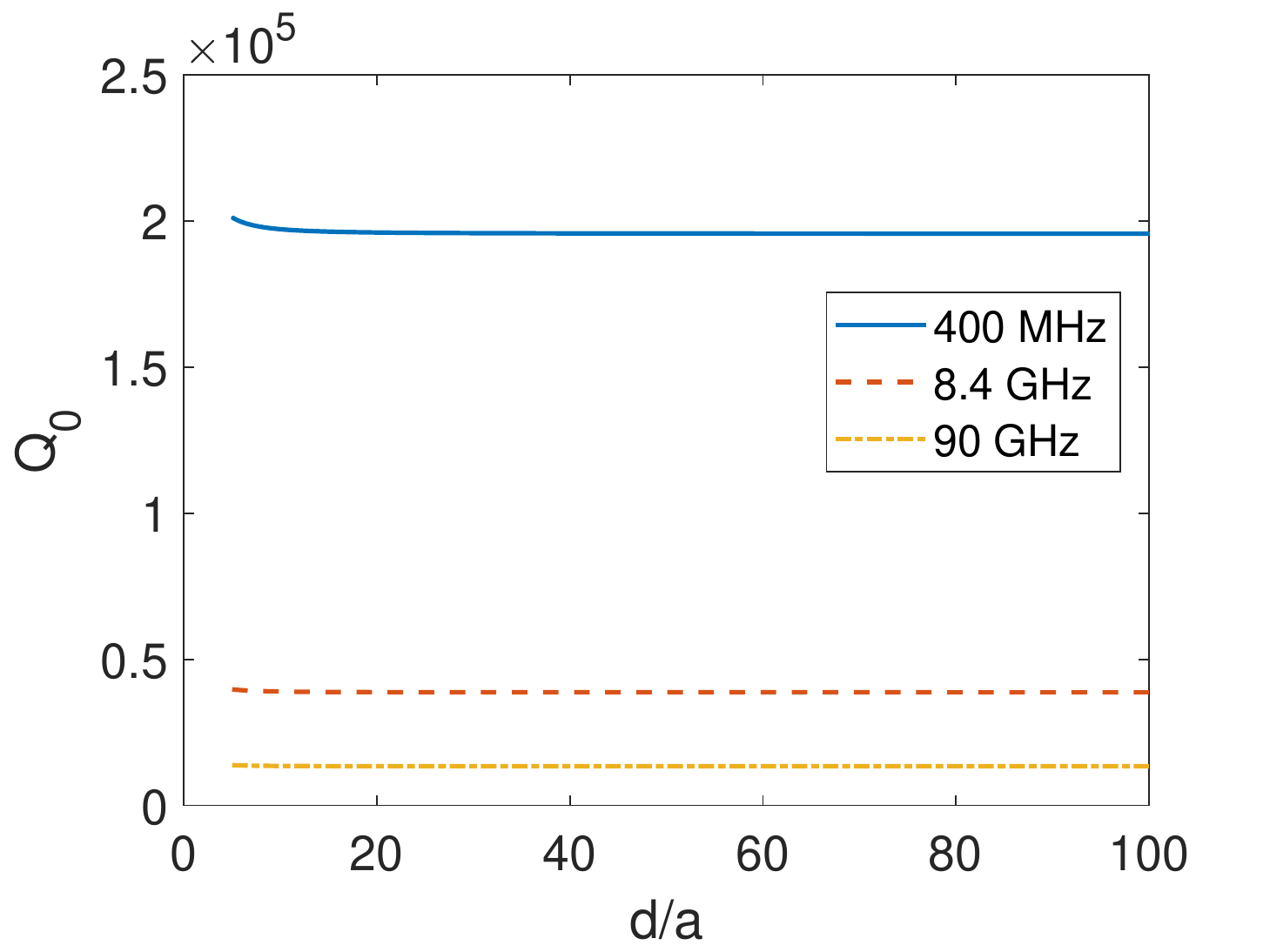}
         \caption{}
         \label{fig:Q0_vs_d-a_Bands}
\end{subfigure}
\caption{(a) Relative mode separation between modes $TE_{101}$ and $TE_{102}$ of a rectangular cavity as a function of $d/a$ for any frequency and (b) $Q_0$ of the $TE_{101}$ mode as a function of $d/a$ for three frequencies ($0.4$, $8.4$ and $90$~GHz).}
\label{fig:ModeMixing_and_Q0_vs_d-a}
\end{figure}

As previously stated, the limitation to increase the length, $d$, of a rectangular single cavity is based on the mode separation between the modes $TE_{101}$ and $TE_{102}$. Figure~\ref{fig:ModeMixing_vs_d-a} plots the relative mode separation ($\Delta f=\frac{|f_{axion}-f_{neighbour}|}{|f_{axion}|}\times 100$~$\%$, where $f_{axion}$ is the resonant frequency of the mode induced by the axion-photon conversion and $f_{neighbour}$ is the resonant frequency of the closest mode) for a rectangular cavity as a function of $d/a$, which is valid for any resonant frequency of $TE_{101}$ and height $b$.\\

The results show a rapid decrease of the mode separation when $d/a$ increases. If the next mode is far enough away in frequency, the form factor will be the theoretical maximum for any cavity size: $C_{TE_{101}}=64/\pi^4 = 0.657$, obtained from equation~\ref{eq:C}.\\

The unloaded quality factor of a $TE_{10p}$ mode in a rectangular waveguide cavity resonator without dielectric losses can be expressed as \cite{Balanis:1989}
\begin{equation}
\label{eq:Q0}
    Q_{0} = \frac{1}{2} \sqrt{\frac{\pi \ \sigma}{\varepsilon_0 \ f_{10p}}}\frac{b\left(a^2+d^2\right)^{3/2}}{ad\left(a^2+d^2\right)+2b\left(a^3+d^3\right)}
\end{equation}
where $\sigma$ is the electrical conductivity of the cavity walls ($\sigma=2\times 10^9$~S/m is assumed, which corresponds with copper at cryogenic temperatures), $\varepsilon_0\approx8.854\times 10^{-12}$~F/m is the vacuum electric permittivity, $f_{10p}$ is the resonant frequency of a $TE_{10p}$ mode, and $p$ is the number of the sinusoidal variations along the longitudinal $z-$axis ($p=1$ for the working mode). In addition, equation~\ref{eq:Q0} shows that the unloaded quality factor decreases with higher frequencies, which is equivalent to reduce the cavity width. From Figure~\ref{fig:Q0_vs_d-a_Bands} it can be concluded that the $Q_0$ parameter is also length independent for high $d$ values.\\

\begin{figure}[h]
\centering
\begin{subfigure}[b]{0.49\textwidth}
         \centering
         \includegraphics[width=1\textwidth]{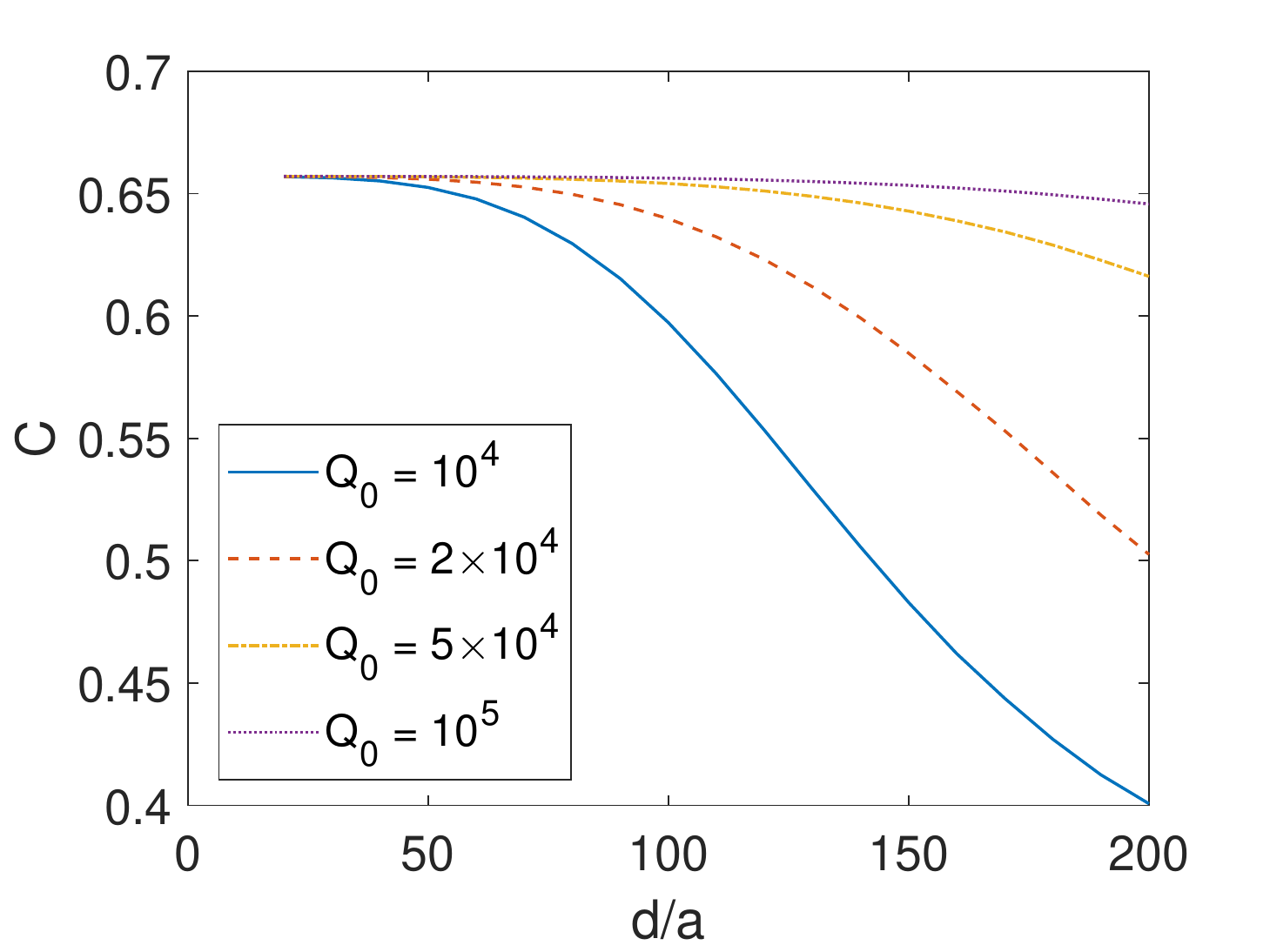}
         \caption{}
         \label{fig:C_vs_d-a_vs_Q0}
\end{subfigure}
\hfill
\begin{subfigure}[b]{0.49\textwidth}
         \centering
         \includegraphics[width=1\textwidth]{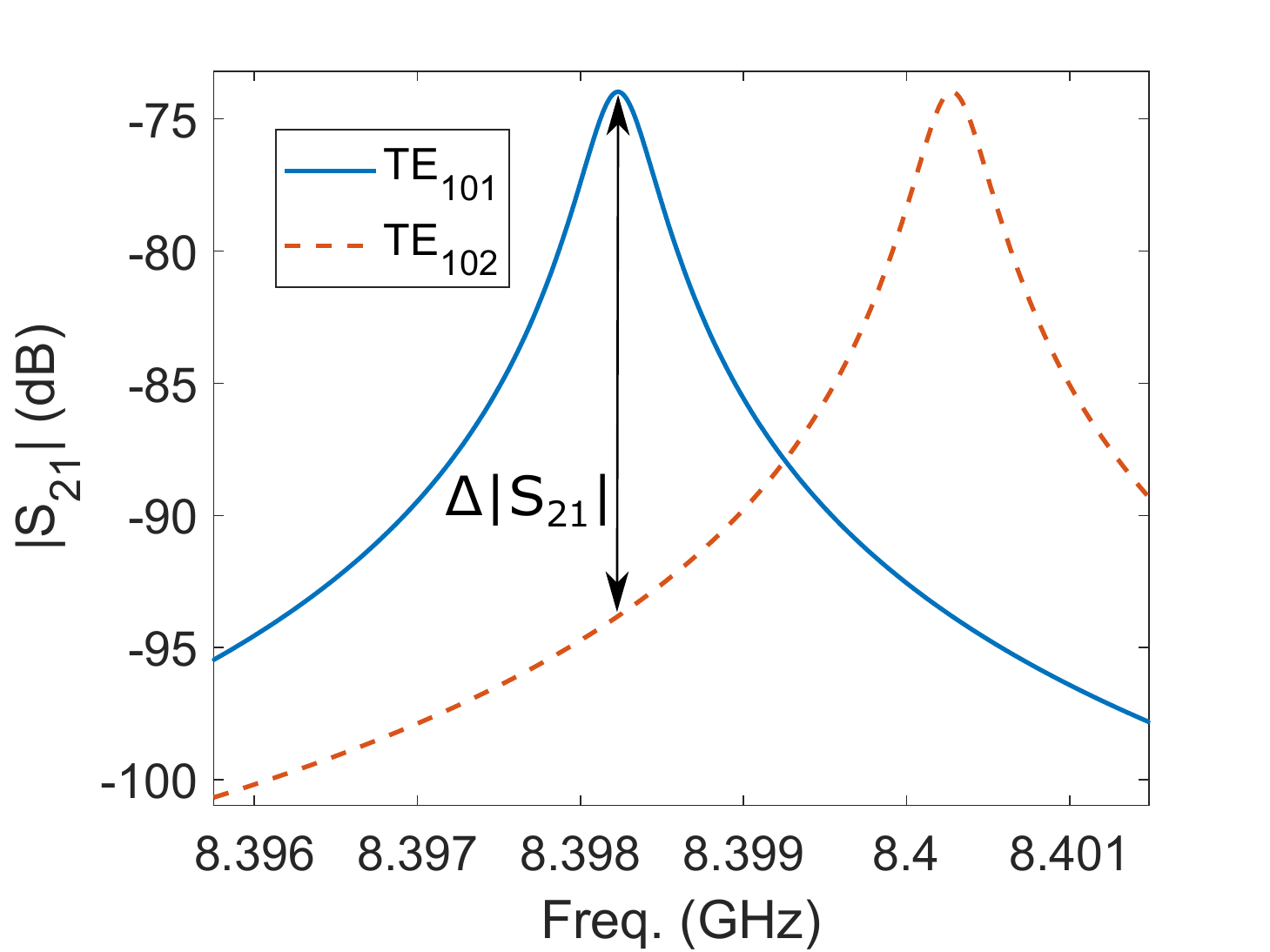}
         \caption{}
         \label{fig:DeltaS21dB}
\end{subfigure}
\caption{(a) Form factor versus $d/a$ for several $Q_0$ values and (b) example of two close resonances with amplitude difference of $\Delta |S_{21}|=20$~dB at the resonant frequency of the mode $TE_{101}$.}
\label{fig:C_vs_d-a_vs_Q0_and_ExampleDeltaS21dB}
\end{figure}

The minimum accepted mode separation (mode clustering) depends on the measured quality factor, which in turn, depends on the cavity material and the quality of manufacturing process. Larger $Q_0$s lead to sharper resonances and hence modes can get closer in frequency. In general, in a conservative approach, we can expect that the unloaded quality factor of the fabricated prototype will be half of the theoretical due to manufacturing tolerances in the fabrication process (roughness at inner walls, quality in soldering, metallic contact if screws are used). As a quantification of the mode clustering on the energy loss, in Figure~\ref{fig:C_vs_d-a_vs_Q0} the form factor versus $d/a$ for several $Q_0$ values is plotted.
This plot shows how for high $Q_0$ values, the detriment in $C$ is lower. The form factor in Figure~\ref{fig:C_vs_d-a_vs_Q0} has been computed with equation~\ref{eq:C} taking into account the perturbation of the electric field (and thus its $C$ detriment) due to the influence of the electric field of the next resonant mode ($TE_{102}$ mode in this case) when they are very close. As a consequence of this behaviour, the electric field influence of the next mode is higher if the difference in the transmission parameter $S_{21}$ of both resonances of modes $TE_{101}$ and $TE_{102}$ at $f_{rTE_{101}}$ ($\Delta|S_{21}|$) is lower. A form factor of $C=0.65$ is selected as our minimum accepted reduced value due to mode clustering. It is considered that the cavity length might be increased even if $C$ decreases, as long as the $Q_0 \times V \times C$ factor (the haloscope figure of merit) continues rising. However, to be conservative, this value is kept as a reasonable limit. This bound also ensures the right measurement of the resonant frequency $f_r$ and unloaded quality factor $Q_0$ in the experiment. Anyway, if the response of the cavity exhibits two resonances very close or even combined (due to lower than expected quality factors), there are methods to extract the original shape of each resonance and compute the relevant two parameters ($f_r$ and $Q_0$) \cite{Q0_Characterization}. In Figure~\ref{fig:DeltaS21dB} an example of two resonances with $Q_0=2\times10^4$ and $d=1400$~mm (or $d/a\approx79$) for $8.4$~GHz is shown, which provides a form factor of $C=0.65$ ($\sim99\%$ of $C_{max}=0.657$). The graphs from Figure~\ref{fig:C_vs_d-a_vs_Q0_and_ExampleDeltaS21dB} help to choose the guard frequency to avoid a high form factor detriment.\\

To illustrate the discussion, an X band example is designed with $a=17.85$~mm, $b=10.16$~mm and $d=1400$~mm. For this design, the distance between the axion mode and its first neighbour is $2.05$~MHz (or $0.024$~$\%$). Moreover, the volume is $V = 253.9$~mL, which means an improvement of a $38$ factor from a standard WR-90 cavity ($a=22.86$~mm, $b=10.16$~mm and $d=28.55$~mm, with volume $V = 6.73$~mL). In Table~\ref{tab:IndLongQVC} a summary of the obtained improvements in this comparison is shown.\\

\begin{table}[h]
\begin{tabular}{|c|c|c|c|c|c|c|}
\hline
$a$ (mm) & $b$ (mm) & $d$ (mm) & $V$ (mL) & $Q_0$ & $C$ & $Q_0 \times V \times C$ (L) \\ \hline\hline
$22.86$ & $10.16$ & $28.55$ & $6.63$ & $4.6\times10^4$ & $0.657$ & $200.37$ \\ \hline 
$17.85$ & $10.16$ & $1400$ & $253.9$ & $3.9\times10^4$ & $0.65$ & $6436.37$ \\ \hline
\end{tabular}
\centering
\caption{\label{tab:IndLongQVC} Comparison of the operational parameters of a standard rectangular resonant cavity employed for resonating at $8.4$~GHz with a very long cavity (large $d$) for the same resonant frequency.}
\end{table}

It can be observed in Table~\ref{tab:magnets} that this long cavity fits perfectly in a dipole magnet as CAST. However, for a solenoid magnet the cavity length should be reduced to fit with the bore diameter. For example, in MRI (ADMX-EFR) a maximum length $d\approx\phi_{MRI}=650$~mm is imposed. In fact, the longitudinal axis of the cavity should be oriented in the radial axis of the solenoid magnet ($x$ or $y-$axis in Figure~\ref{fig:Solenoid}) due to its magnetic field direction, as it was explained in the previous section. In this case, there is a lot of unused space along the longitudinal axis of the solenoid magnet ($z-$axis in Figure~\ref{fig:Solenoid}). Anyway, the increased volume from a standard cavity is still high.\\

For the limit case shown in Table~\ref{tab:IndLongQVC}, the haloscope sensitivity achieved is very good at one frequency, but special care must be taken if a certain frequency range is to be swept because there are likely to be many mode crossings. Therefore, when designing one of these structures, its dimensions will be limited by a tradeoff between the volume achieved and the number of mode crossings tolerated, again taking into account the dimensions of the magnet chosen.

\subsection{Tall cavities}
\label{SubSec:IndTall}
\begin{figure}[h]
\centering
\begin{subfigure}[b]{0.49\textwidth}
         \centering
         \includegraphics[width=1\textwidth]{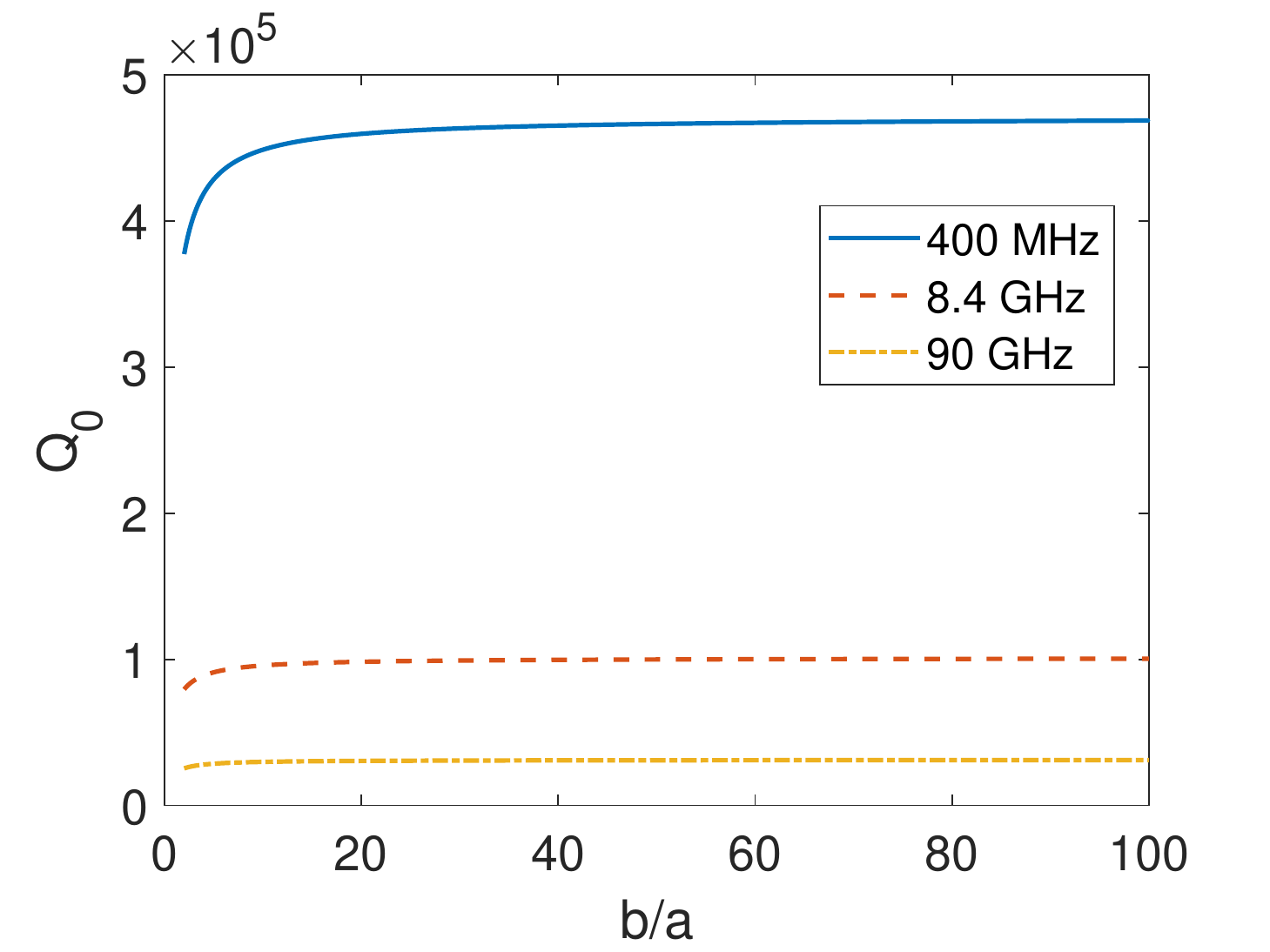}
         \caption{}
         \label{fig:Q0_vs_b-a_Bands}
\end{subfigure}
\hfill
\begin{subfigure}[b]{0.49\textwidth}
         \centering
         \includegraphics[width=1\textwidth]{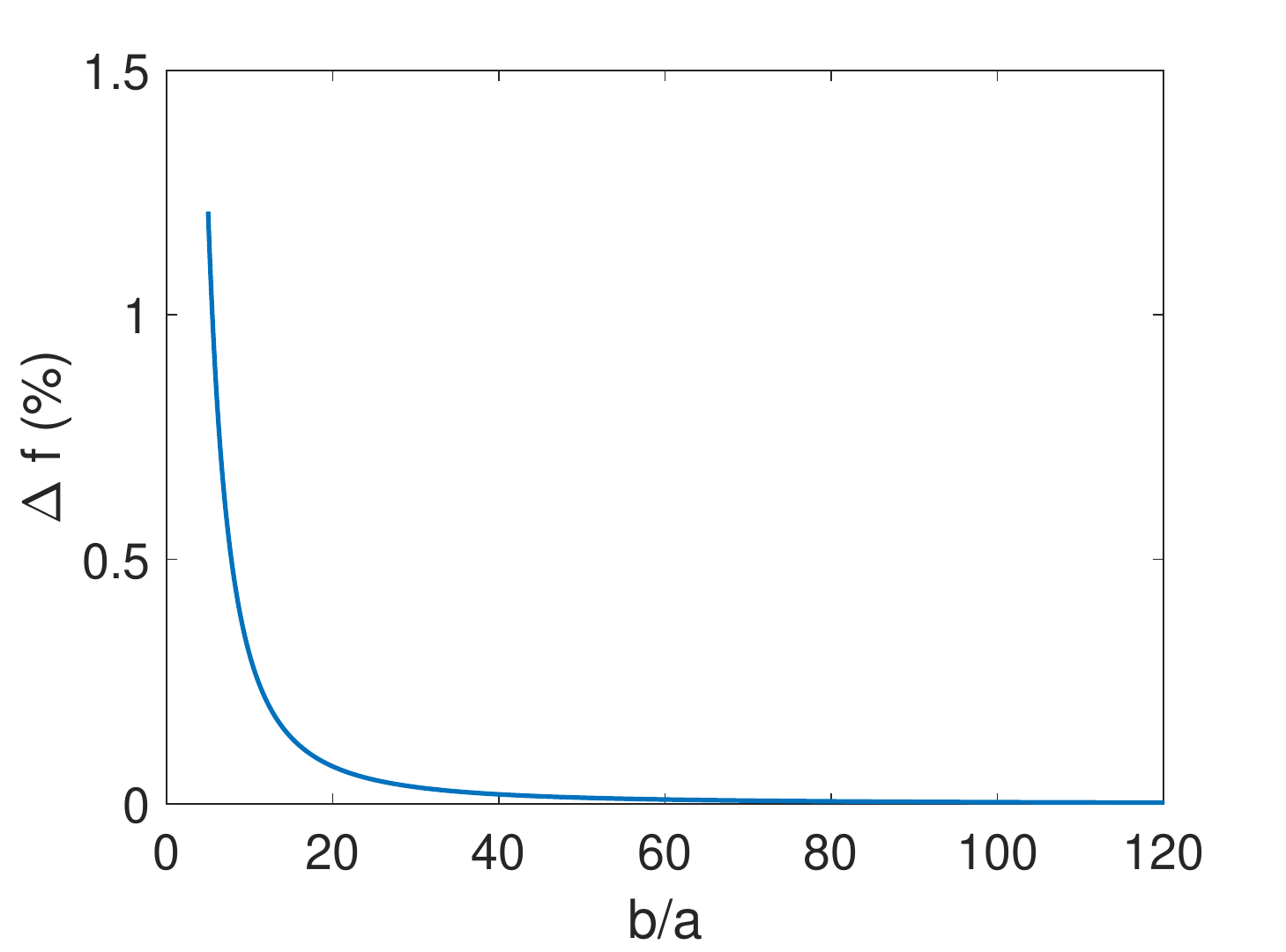}
         \caption{}
         \label{fig:ModeMixing_vs_b-a}
\end{subfigure}
\caption{(a) $Q_0$ of the $TE_{101}$ mode as a function of $b/a$ for three frequencies ($0.4$, $8.4$ and $90$~GHz), and (b) relative mode separation between the modes $TE_{101}$ and $TE_{111}$/$TM_{111}$ of a single cavity as a function of $b/a$ for $d=28.55$~mm (X band).}
\label{fig:Q0_vs_b_and_ModeMixing_vs_b-a}
\end{figure}

Similarly to the longitudinal dimension of a single resonant cavity, the vertical dimension $b$ can be increased up to a limit imposed by the proximity of the next modes (mode clustering between the $TE_{101}$ with the $TE_{111}/TM_{111}$). In the case of the tall cavities, the width $a$ is not reduced. With the increasing of the waveguide height ($b$), the $Q_0$ value is increased up to half of the limit, as shown in Figure~\ref{fig:Q0_vs_b-a_Bands}.
For example, at $8.4$~GHz (X band) a $Q_0=10^5$ is obtained for heights $b$ between $500-2000$~mm. For $\sim400$~MHz (UHF band) and $\sim90$~GHz (W band) the quality factor takes values around $Q_0=4.7\times10^5$ and $Q_0=3.1\times 10^4$, respectively, as it can be seen in Figure~\ref{fig:Q0_vs_b-a_Bands}. For completeness, in Figure~\ref{fig:ModeMixing_vs_b-a} the frequency proximity with the nearest mode for X band frequencies ($d=28.55$~mm) is also represented. Similarly to the plot in Figure~\ref{fig:ModeMixing_vs_d-a}, the results show a behaviour with a rapid increase of the mode separation for low values of $b/a$, while for high values $\Delta f$ starts to stabilise at values close to zero.\\

\begin{figure}[h]
\centering
\includegraphics[width=0.6\textwidth]{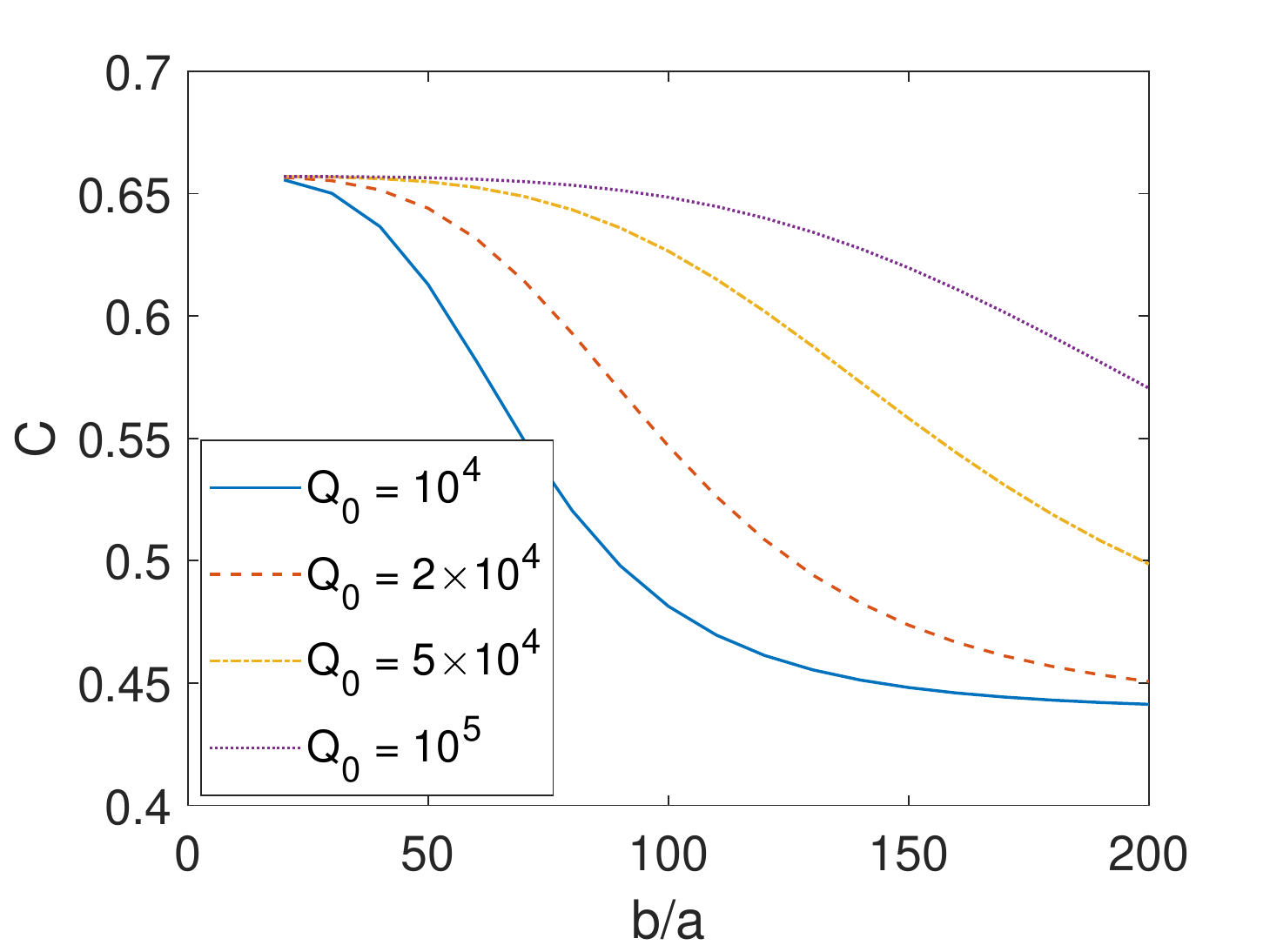}
\caption{Form factor versus $b/a$ for several $Q_0$ values.}
\label{fig:C_vs_b-a_vs_Q0}
\end{figure}

To find the minimum accepted mode separation, the same limit of $C_{min}=0.65$ is imposed. The form factor versus $b/a$ for several $Q_0$ values taking into account that now the electric field contribution that affects negatively $C$ is the one from the $TE_{111}$ mode is plotted in Figure~\ref{fig:C_vs_b-a_vs_Q0}.
As it was the case for long cavities this plot shows how for high $Q_0$ values the detriment in $C$ is lower.\\

If the same analysis is repeated as for the long cavities at X band to determine the minimum accepted mode separation, a limit value of $b=1500$~mm (or $b/a=66$) is obtained, taking into account an unloaded quality factor after fabrication of $Q_0^{real}\approx50000$ (half of the theoretical one).\\

In this example, with $a=22.86$~mm, $b=1500$~mm and $d=28.55$~mm the distance between the axion mode and its first neighbour is $0.59$~MHz (or $0.007$~$\%$). For this case the volume is $V = 978.98$~mL, which means an improvement of a $148$ factor from a standard WR-90 cavity. A summary of these improvements is shown in Table~\ref{tab:IndTallQVC}.\\

\begin{table}[h]
\begin{tabular}{|c|c|c|c|c|c|c|}
\hline
$a$ (mm) & $b$ (mm) & $d$ (mm) & $V$ (mL) & $Q_0$ & $C$ & $Q_0 \times V \times C$ (L) \\ \hline\hline
$22.86$ & $10.16$ & $28.55$ & $6.63$ & $4.6\times10^4$ & $0.657$ & $200.37$ \\ \hline
$22.86$ & $1500$ & $28.55$ & $978.98$ & $10^5$ & $0.65$ & $6.4\times10^4$ \\ \hline
\end{tabular}
\centering
\caption{\label{tab:IndTallQVC} Comparison of the properties of a standard rectangular resonant cavity employed for resonating at $8.4$~GHz with a very tall cavity (large $b$) for the same resonant frequency.}
\end{table}

Focusing on Table~\ref{tab:magnets}, it can be observed how the height of this tall cavity has to be decreased until it fits into the longitudinal axis of a solenoid magnet. For example, in MRI (ADMX-EFR) a maximum height $b=L_{MRI}=800$~mm is imposed. Anyway, the gained volume from a standard cavity is again very high. For a dipole magnet the only option to have a substantial benefit is BabyIAXO, whose $\phi_{BabyIAXO}=600$~mm diameter bore can be used to fit this tall structure in the radial orientation ($y-$axis in Figure~\ref{fig:Dipole}). With this scenario, there is a lot of unused space on the longitudinal axis of this magnet ($z-$axis in Figure~\ref{fig:Dipole}), that can also be exploited with the novel ideas proposed on the next sections.\\

Similarly to the previous section, for the case shown in Table~\ref{tab:IndTallQVC}, the sensitivity value obtained in the haloscope is considerably high at one frequency, but many mode crossings could appear if a tuning system is employed. Thus, at the designing step, the dimensions will be restricted by a tradeoff between the increase in the volume, the number of mode crossings, and the magnet.

\subsection{Large cavities}
\label{SubSec:IndLongAndTall}
\begin{figure}[h]
\centering
\begin{subfigure}[b]{0.49\textwidth}
         \centering
         \includegraphics[width=1\textwidth]{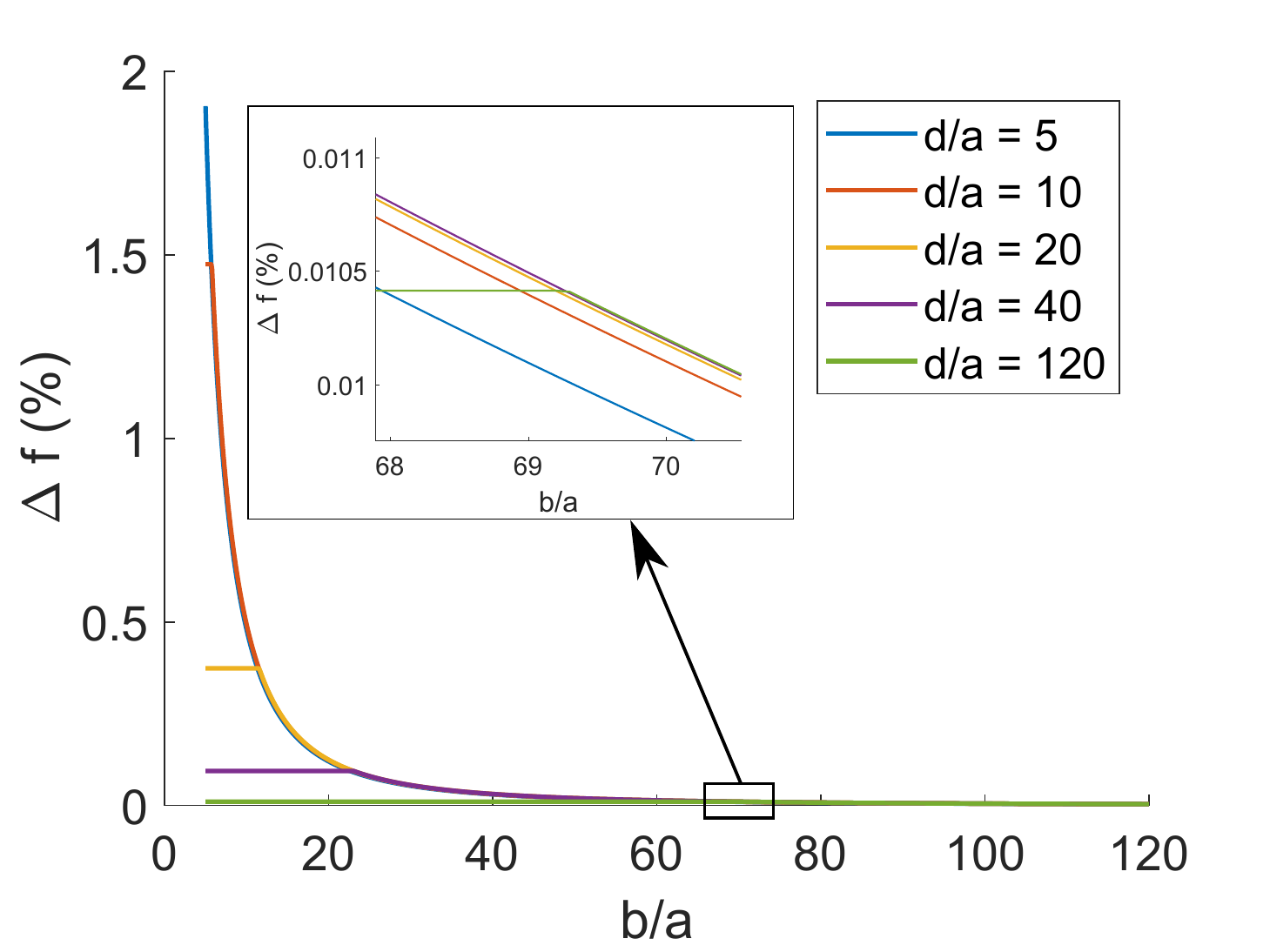}
         \caption{}
         \label{fig:ModeMixing_vs_b-a_vs_d-a}
\end{subfigure}
\hfill
\begin{subfigure}[b]{0.49\textwidth}
         \centering
         \includegraphics[width=1\textwidth]{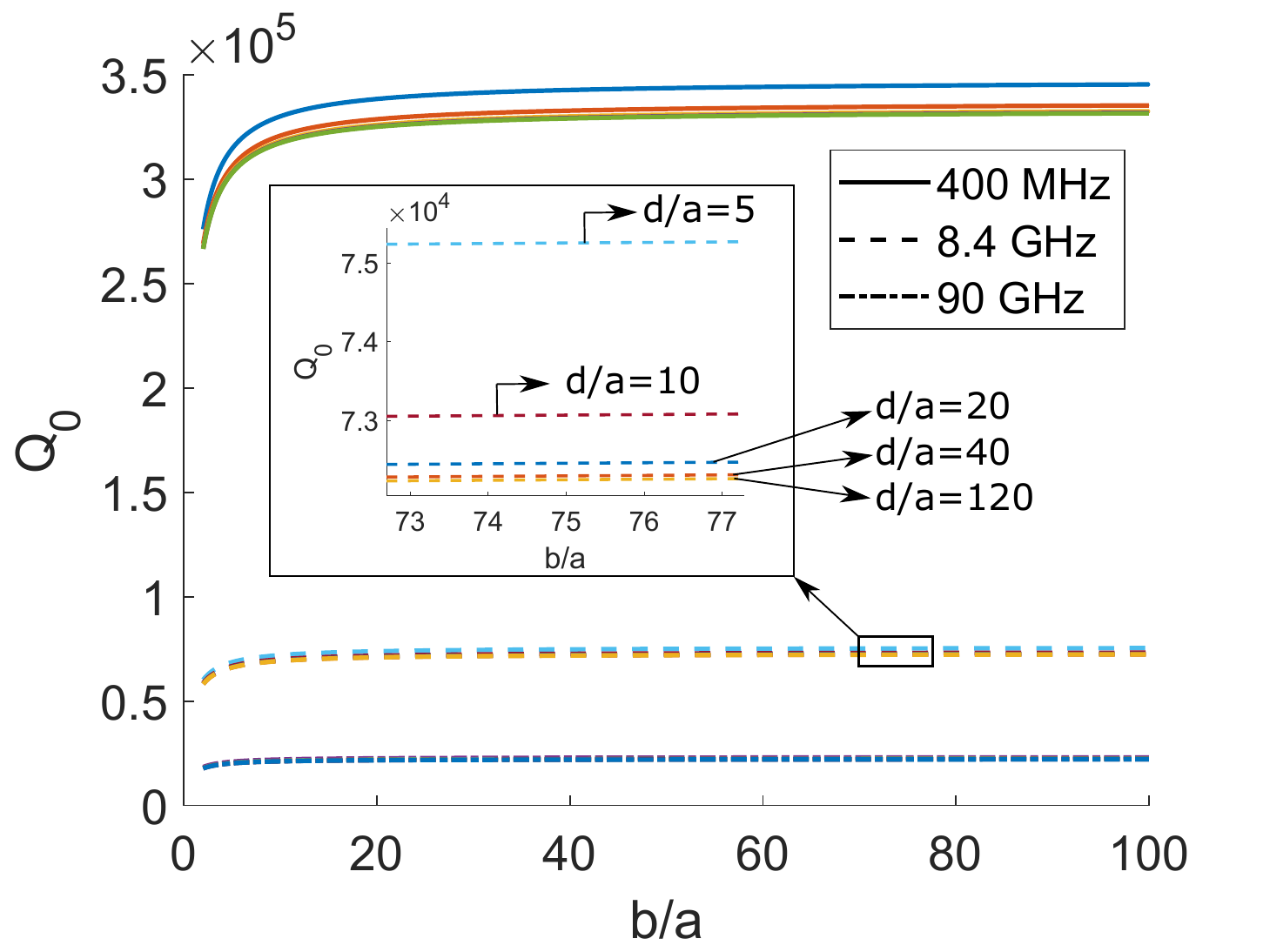}
         \caption{}
         \label{fig:Q0_vs_b-a_vs_d-a_Bands}
\end{subfigure}
\caption{(a) Relative mode separation between the modes $TE_{101}$ and $TE_{102}$ or $TE_{111}$ (the closest one, depending on the $b/a$ and $d/a$ values) for X band frequencies ($a=17.85$~mm). (b) Quality factor of the $TE_{101}$ mode as function of $b/a$ for three frequencies ($0.4$, $8.4$ and $90$~GHz) for five $d/a$ cases. In both pictures, the insets depict a zoom to differentiate all the $d/a$ cases.}
\label{fig:Q0_and_S21_LongAndTall}
\end{figure}

The last idea for increasing the volume of a single cavity is to increase both length and height dimensions at the same time. As mentioned above, to maintain the same resonant frequency in a very long cavity the width should be slightly reduced. On the other hand, the resonant frequency does not depend on the height as explained in the previous sections. Then, the mode clustering problem needs now to consider two mode approximations to our working mode: $TE_{102}$ (because of the longitudinal dimension $d$) and $TE_{111}$ (because of the vertical dimension $b$). The relative mode separation follows the behaviour from Figure~\ref{fig:ModeMixing_vs_b-a_vs_d-a}, which shows the case for X band frequencies ($d=28.55$~mm).
The results show once again a rapid decrease of the mode separation when $d/a$ and/or $b/a$ increases.\\

The behaviour of the quality factor is depicted in Figure~\ref{fig:Q0_vs_b-a_vs_d-a_Bands}. For the X band example, a width of $a=17.85$~mm is necessary for maintaining $f_r=8.4$~GHz. For $d$ and $b$ between $500-2000$~mm the cavity provides a $Q_0=7.2\times10^4$, as it is shown in the inset of Figure ~\ref{fig:Q0_vs_b-a_vs_d-a_Bands}. Note how the $Q_0$ is a bit lower as compared to the tall cavity because the width has been slightly reduced in order to compensate the increase of length.\\

Moreover, in order to fix the minimum mode separation, we accept a form factor of $C=0.65$. Now the electric field contributions that affect unfavorably $C$ are both from the $TE_{102}$ and $TE_{111}$ modes. The behaviour of the form factor with $d/a$, $b/a$ and $Q_0$ follows a similar performance compared to Figures~\ref{fig:C_vs_d-a_vs_Q0} and \ref{fig:C_vs_b-a_vs_Q0}. For $8.4$~GHz, the limit is reached with the values $b=1100$~mm and $d=1600$~mm, taking into account an unloaded quality factor after fabrication of $Q_0^{real}\approx3.6\times10^4$ (half of theoretical). In Table~\ref{tab:IndLongAndTallQVC} a summary of the achieved improvements is collected. It can be seen an impressive enhancement in the $Q_0 \times V \times C$ factor of $7336$ versus the standard rectangular resonant cavity.\\

\begin{table}[h]
\begin{tabular}{|c|c|c|c|c|c|c|}
\hline
$a$ (mm) & $b$ (mm) & $d$ (mm) & $V$ (mL) & $Q_0$ & $C$ & $Q_0 \times V \times C$ (L) \\ \hline\hline
$22.86$ & $10.16$ & $28.55$ & $6.63$ & $4.6\times10^4$ & $0.657$ & $200.37$ \\ \hline
$17.85$ & $1100$ & $1600$ & $3.14\times10^4$ & $7.2\times10^4$ & $0.65$ & $1.47\times10^6$ \\ \hline
\end{tabular}
\centering
\caption{\label{tab:IndLongAndTallQVC} Comparison of the operational parameters of a standard rectangular resonant cavity employed for resonating at $8.4$~GHz with a very long and tall cavity (large $d$ and $b$) for the same resonant frequency.}
\end{table}

With these results and analyzing the data from Table~\ref{tab:magnets}, it can be seen how in dipole magnets the best orientation for this kind of cavities is obtained by matching both longitudinal axis of the cavity and magnet bore since they have the highest dimension values, and both electric field of the cavity and magnetic field of the magnet are aligned. For example, in BabyIAXO the vertical cavity dimension can be increased until $b=600$~mm ($y-$axis in Figure~\ref{fig:Dipole}), and the length can be extended to its limit $d=1600$~mm ($z-$axis in Figure~\ref{fig:Dipole}), which with $a=17.85$~mm gives a volume of $V=1.71\times10^4$~mL representing an improvement of $2580$ in volume compared with a standard cavity. For a solenoid magnet, the height dimension must match the longitudinal axis of the bore ($z-$axis in Figure~\ref{fig:Solenoid}), and the haloscope length can occupy all the bore radial axis ($x$ or $y-$axis in Figure~\ref{fig:Solenoid}). For example, at the MRI (ADMX-EFR) solenoid magnet a haloscope of $a=17.85$~mm, $b=800$~mm and $d=650$~mm, which implies a volume of $V=9282$~mL, could be installed. This means reducing by half the volume compared to the case in the BabyIAXO dipole magnet. However, this reduction can be compensated by the lower working temperature (lower $T_{sys}$ in equation~\ref{eq:ga}) and higher magnetic field values employed in ADMX (see Table~\ref{tab:magnets}).\\

\begin{figure}[h]
\centering
\begin{subfigure}[b]{0.8\textwidth}
         \centering
         \includegraphics[width=1\textwidth]{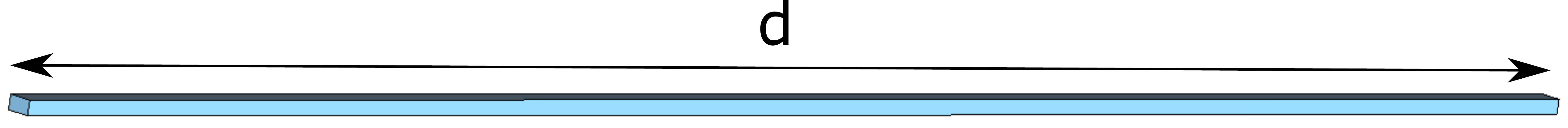}
         \caption{}
         \label{fig:LargeStructure_d1400}
\end{subfigure}
\hfill
\begin{subfigure}[b]{0.2\textwidth}
         \centering
         \includegraphics[width=0.26\textwidth]{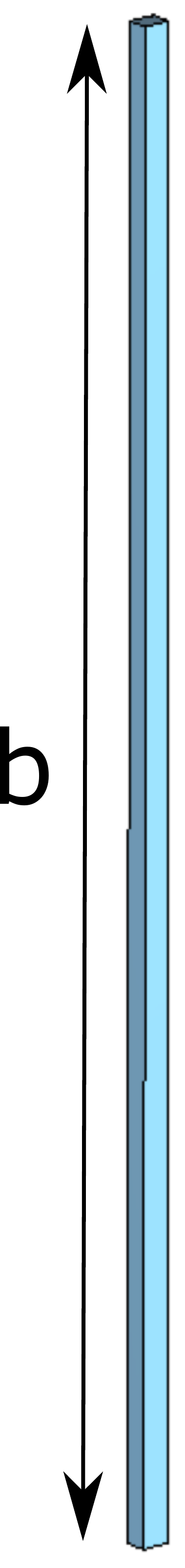}
         \caption{}
         \label{fig:TallStructure_b1500}
\end{subfigure}
\hfill
\begin{subfigure}[b]{0.76\textwidth}
         \centering
         \includegraphics[width=0.7\textwidth]{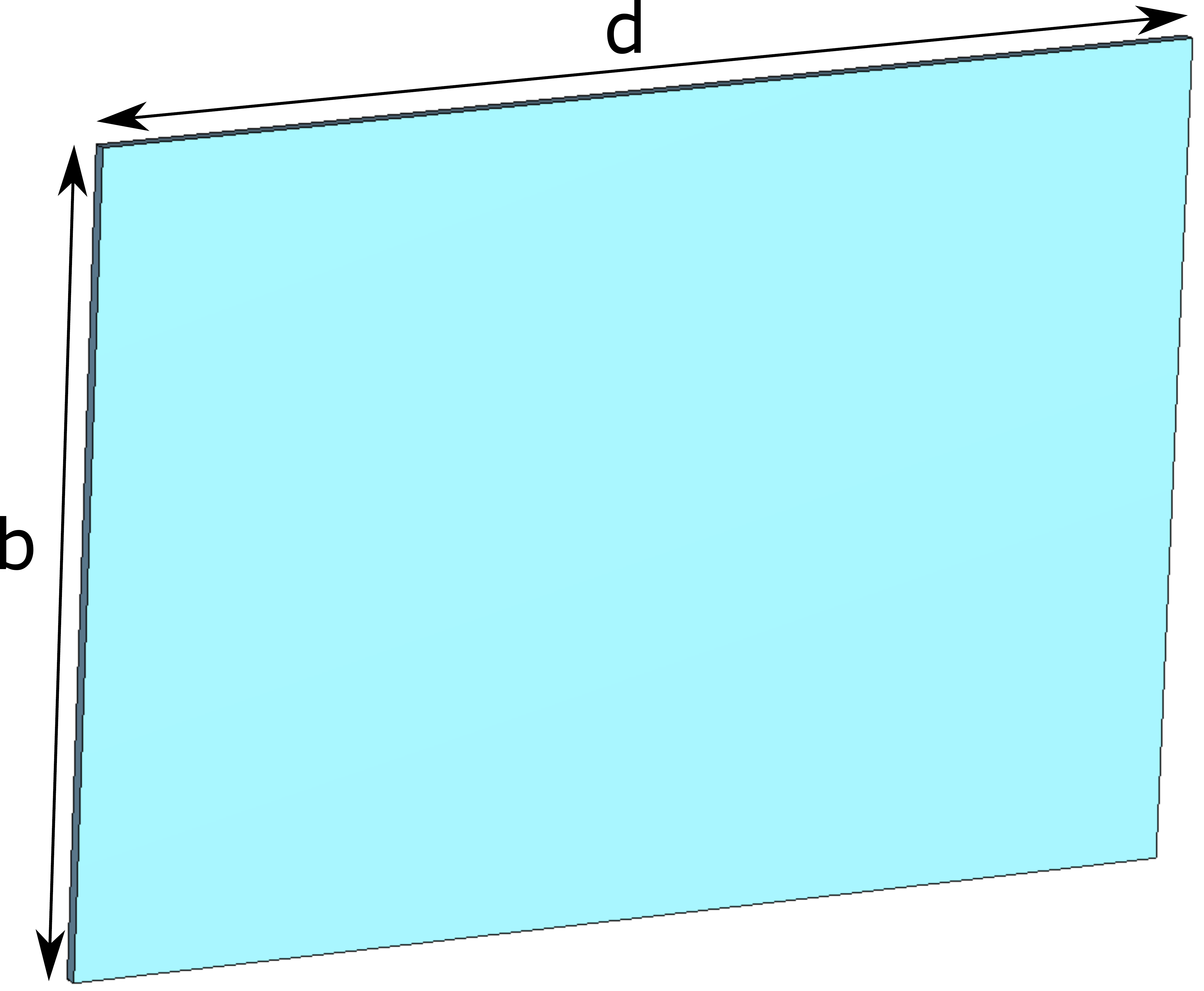}
         \caption{}
         \label{fig:LargeAndTallStructure_b1100d1600}
\end{subfigure}
\hfill
\caption{(a) Long cavity of $d=1400$~mm, (b) tall cavity of $b=1500$~mm and (c) long and tall structure of $b=1100$~mm and $d=1600$~mm.}
\label{fig:LongTallAndLongAndTall}
\end{figure}

Figure~\ref{fig:LongTallAndLongAndTall} shows for reference several drawings of each type of cavity with its limit dimensions: long, tall, and large (long and tall) haloscopes.
\\

Finally, as occurs for tall or long cavities, for the case shown in Table~\ref{tab:IndLongAndTallQVC}, the haloscope sensitivity obtained is very good at one frequency, but many mode crossings could appear with frequency tuning. Thus, the final dimensions of the designed cavity should consider the tradeoff between the volume, the number of mode crossings, and the magnet bore size.

\section{1D multicavities}\label{Sec:Multicav}
The RADES team has been employing the multi-cavity concept over the last six years in order to increase the volume of haloscopes in the longitudinal axis without decreasing the frequency \cite{RADESreviewUniverse}. In contrast to the long cavity concept, the $z-$axis multicavity designs can make use of wider rectangular waveguides (for example WR-90 for X band).\\

\begin{figure}[h]
\centering
\includegraphics[width=0.8\textwidth]{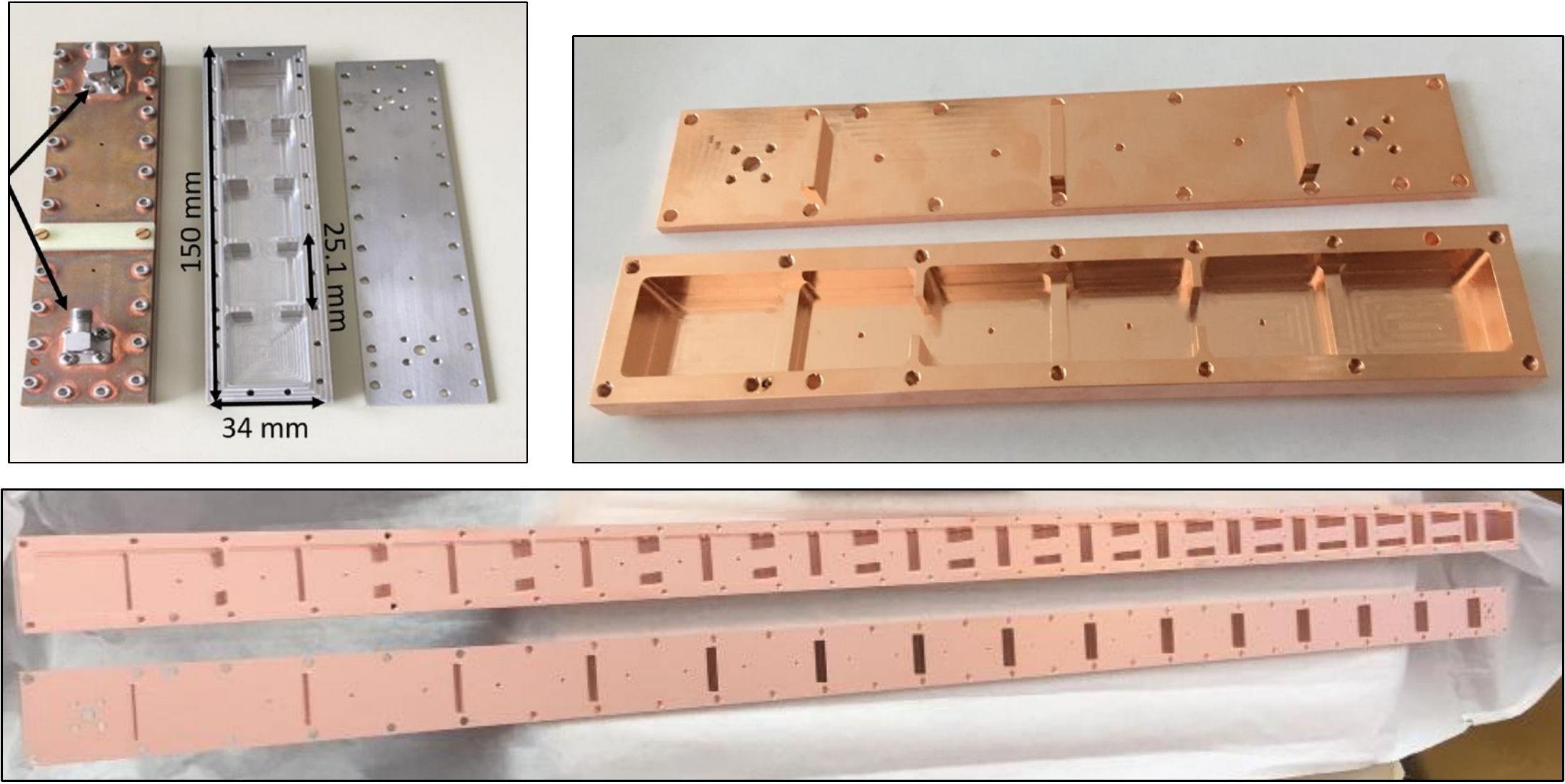}
\caption{Manufactured RADES cavities: all-inductive irises with five subcavities (top-left), alternating inductive/capacitive irises with six subcavities (top-right) and alternating inductive/capacitive irises with thirty subcavities (bottom).}
\label{fig:RADEScavities}
\end{figure}

Several small haloscope prototypes have been designed and manufactured by this experimental group. Among them, an all inductive structure based on five subcavities and two alternating structures with two different number of subcavities ($N=6$ and $N=30$, where $N$ is the number of subcavities) are shown in Figure~\ref{fig:RADEScavities}.
Results of the first two structures are presented in \cite{RADES_paper1,RADES_paper2,RADES_paper3,RADESreviewUniverse,3CavityRADES:2019}.\\

For the design of the haloscope multicavity structures, the coupling matrix has been employed as a supporting tool. The theoretical concepts of this method can be found in \cite{Cameron,RADES_paper1}. In the case on 1D multicavities, the following matrix has been employed for the development of the geometrical parameters in the studied structures of this work:
\begin{gather}\label{eq:CouplingMatrix_1D}
 \bf{M} =
  \begin{pmatrix}
   \Omega_1 & M_{1,2} & 0 & 0 & \cdots & 0 & 0 & 0 \\
   M_{1,2} & \Omega_2 & M_{2,3} & 0 & \cdots & 0 & 0 & 0 \\
   0 & M_{2,3} & \Omega_3 & M_{3,4} & \cdots & 0 & 0 & 0 \\
   0 & 0 & M_{3,4} & \Omega_4 & \cdots & 0 & 0 & 0 \\
   \vdots & \vdots & \vdots & \vdots & \ddots & \vdots & \vdots & \vdots \\
   0 & 0 & 0 & 0 & \cdots & \Omega_{N-2} & M_{N-2,N-1} & 0 \\
   0 & 0 & 0 & 0 & \cdots & M_{N-2,N-1} & \Omega_{N-1} & M_{N-1,N} \\
   0 & 0 & 0 & 0 & \cdots & 0 & M_{N-1,N} & \Omega_N
   \end{pmatrix}
\end{gather}
where $M_{i,j}$ are the impedance inverters values in the normalised low-pass prototype network and $\Omega_q$ is the difference of the resonant frequency in the $q$-th subcavity with respect to the axion frequency \cite{Cameron}. $M_{i,j}$ is related to the physical interresonator coupling $k$ selected in the design. In order to extract its value a low-pass to band-pass transformation ($\Omega=\left(\frac{f}{f_{axion}}-\frac{f_{axion}}{f}\right)\frac{1}{f_B}$, where $f_B=\frac{BW}{f_{axion}}$ is the fractional bandwidth and $BW$ the bandwidth) is usually carried out \cite{Cameron}. In this paper a bandwidth $BW=100$~MHz is employed for all the multicavity designs. With these considerations, the relation with the coupling value is given by \cite{Cameron}:
\begin{equation}\label{eq:Mij_kij_fB}
    M_{i,j}=\frac{k_{i,j}}{f_B},
\end{equation}
where $k_{i,j}$ is the physical coupling value between the resonators $i$ and $j$. More details about these parameters can be found in \cite{Cameron}. The $\Omega_q$ values can be extracted with the condition $\bf{M}\times\bf{1}_N^T=\bf{0}_N^T$, where $\bf{1}_N$ is a $1$-vector of size $N$ and $\bf{0}_N$ is a $0$-vector of size $N$ \cite{Cameron,RADES_paper1}. The matrix dimension ($N\times N$) depends on the number of subcavities. In addition, as it can be seen, the values of the elements outside the three main diagonals of the matrix are zero. This translates into the fact that resonators that are not contiguous have no physical coupling.\\

At first glance, it might be thought that there would be no problem of resonant mode clustering since the $TE_{102}$ mode is far away as it has a small subcavity length. However, the multicavity structure introduces additional resonant modes that are associated with the eigenmodes of the coupled cavity system, the so-called {\em configuration modes} \cite{RADES_paper1}. These configuration modes exist for any $TE_{mnp}$ resonant mode. Their resonant frequencies become closer to the axion eigenmode as the number of subcavities increases and as the interresonator coupling value $k$ decreases. The theory and extraction methods of the physical coupling $k$ can be found in \cite{RADES_paper2}. The number of configuration modes of the coupled cavity system for each $TE_{mnp}$ mode is equal to the number of subcavities.\\

\begin{figure}[h]
\centering
\includegraphics[width=0.8\textwidth]{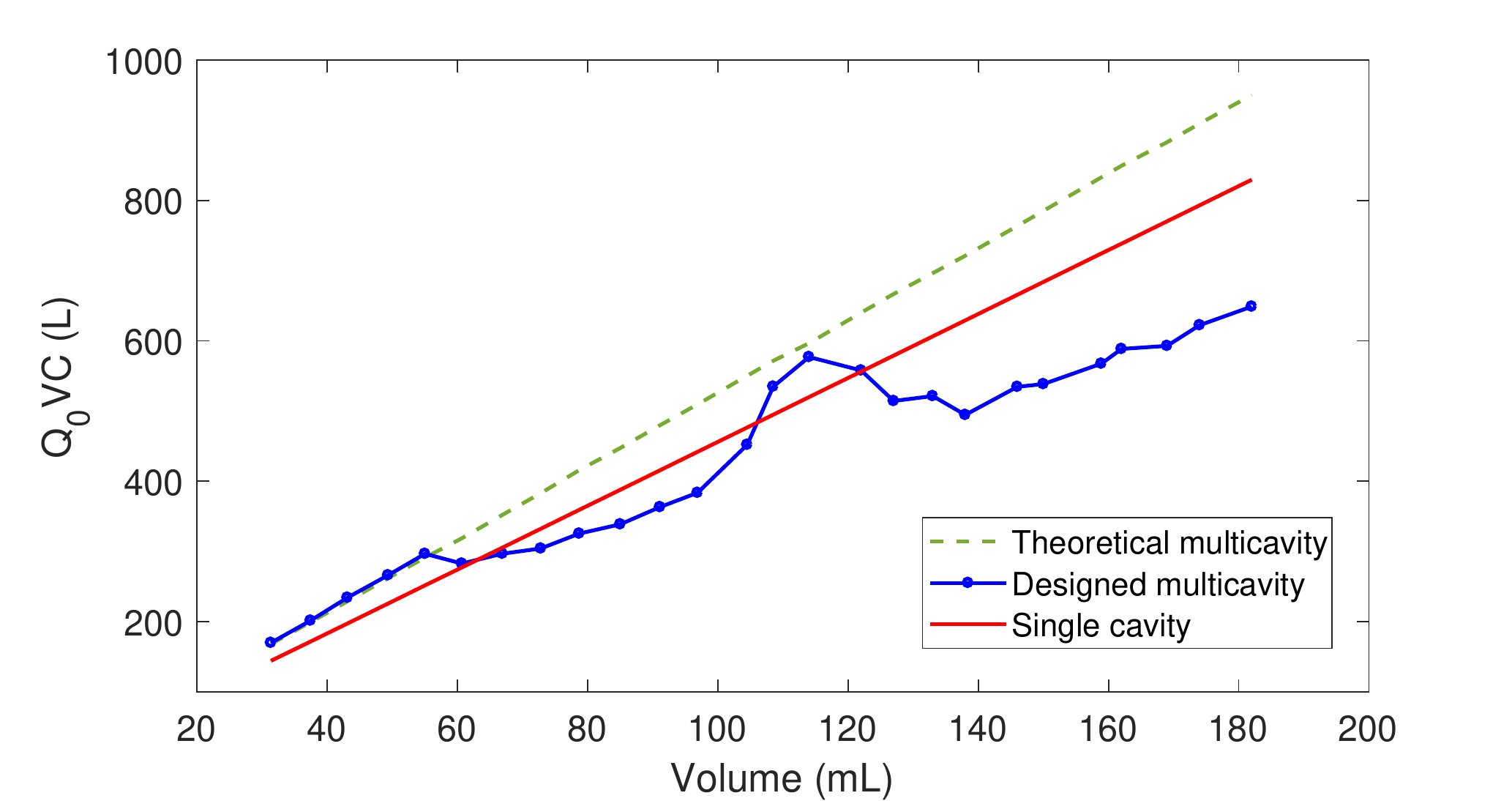}
\caption{Figure of merit $Q_0 \times V \times C$ of a large single cavity versus both theoretical and designed multicavity structures. Each dot at the designed multicavity line (blue line) corresponds with a different number of subcavities, from left to right: $N=5$ to 30 subcavities. The first case ($N=5$) correspond with the first RADES haloscope (see Figure~\ref{fig:RADEScavities} (top-left)), whose behaviour is detailed in \cite{RADES_paper1}.}
\label{fig:QVC_multicavity_vs_indi}
\end{figure}

Due to the loading effect of a coupling window \cite{Cameron}, higher interresonator couplings lead to shorter subcavity lengths in order to keep the same resonant frequency. This effect is small for the frequency of our examples ($8.4$~GHz) where the lengths could vary around $1$ or $2$~mm. However, for very high $k$ values the iris windows need to be opened significantly leading to a substantial loading effect. For other frequency bands like UHF, this effect will have to be taken into account even for relatively low values of $k$. Once again, there is a trade-off between volume and mode separation. In Figure~\ref{fig:QVC_multicavity_vs_indi} an $8.4$~GHz example can be observed to compare the figure of merit ($Q_0 \times V \times C$) of both single and multicavity designs (with $|k|=0.0377$, a similar value to the one usually employed in RADES \cite{RADES_paper2}) as a function of the total volume (increasing the length $d$ for the single cavity case and increasing the number of cavities $N$ for the multicavity case) \footnote{A similar study could be done comparing both single and multicavity structures but increasing the height $b$ and the number of cavities $N$ in the vertical direction, respectively.}.
All the simulation results in this work are obtained from the Computer Simulation Technology (CST) Studio Suite software \cite{CST} in the Frequency Domain.\\

The multicavity design procedure is based on the following steps: first, the working frequency or axion search frequency (for the $TE_{101}$ mode in our case) and a physically realisable interresonator coupling $k$ are chosen \cite{RADES_paper1}. Secondly, the coupling matrix method is applied as described in \cite{RADES_paper1}, which gives the natural frequencies of each subcavity of the array. Finally, an iterative optimization is carried out in which the subcavities are tuned to resonate at the correct frequency and the irises are adjusted to provide the chosen physical coupling.\\

Instabilities in the design results are observed due to high sensitivity in the form factor at the optimization process which becomes more complex with an increase in the number of cavities (as depicted in Figure~\ref{fig:QVC_multicavity_vs_indi}). Overcoming these difficulties would result in an improvement similar to the theoretical multicavity curve, which is better than the improvement that can be obtained with a single cavity and, therefore, the multicavity concept seems the best option for increasing the sensitivity of the axion detection system. Regarding the quality factor, it has been extracted from the study of Figure~\ref{fig:QVC_multicavity_vs_indi} that it is independent of the number of subcavities. For this comparison, the multicavity has a value slightly higher because a standard width $a=22.86$~mm is being used (for the single long cavity design it has to be reduced to $a=17.85$~mm) and the $Q_0$ depends strongly on this dimension, as it is explained in previous sections. Under these considerations (see Table~\ref{tab:IndLongQVC}), for $a=22.86$~mm the quality factor takes values around $4.6\times10^4$ and for $a=17.85$~mm it is $Q_0\approx3.9\times10^4$. Therefore, the difference in the slope of the $Q_0 \times V \times C$ behaviour between single cavities and multicavities is given by $Q_0$.\\

\begin{figure}[h]
\centering
\begin{subfigure}[b]{0.49\textwidth}
         \centering
         \includegraphics[width=1\textwidth]{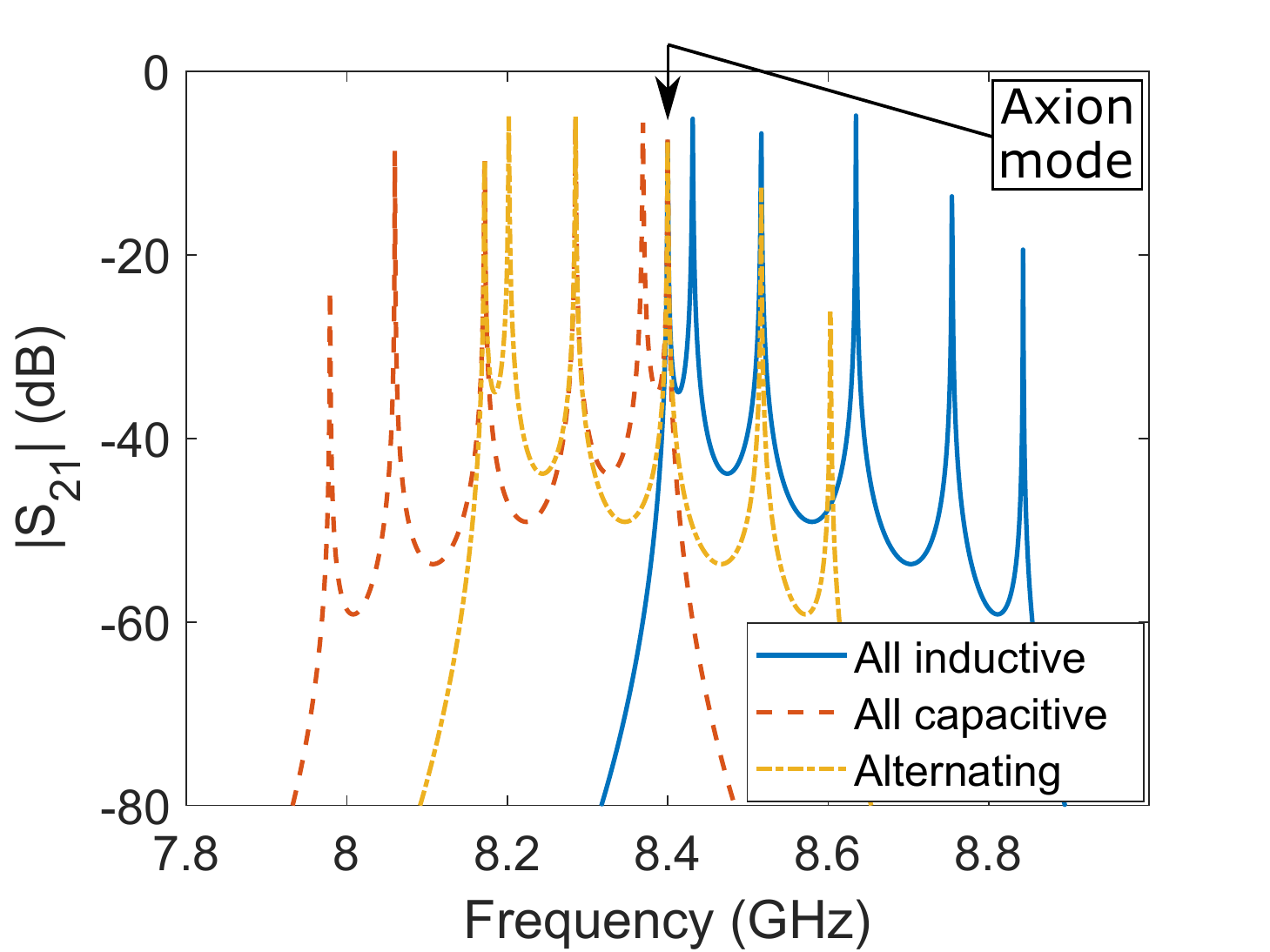}
         \caption{}
         \label{fig:Allind_vs_Allcap_vs_AltSp}
\end{subfigure}
\hfill
\begin{subfigure}[b]{0.49\textwidth}
         \centering
         \includegraphics[width=1\textwidth]{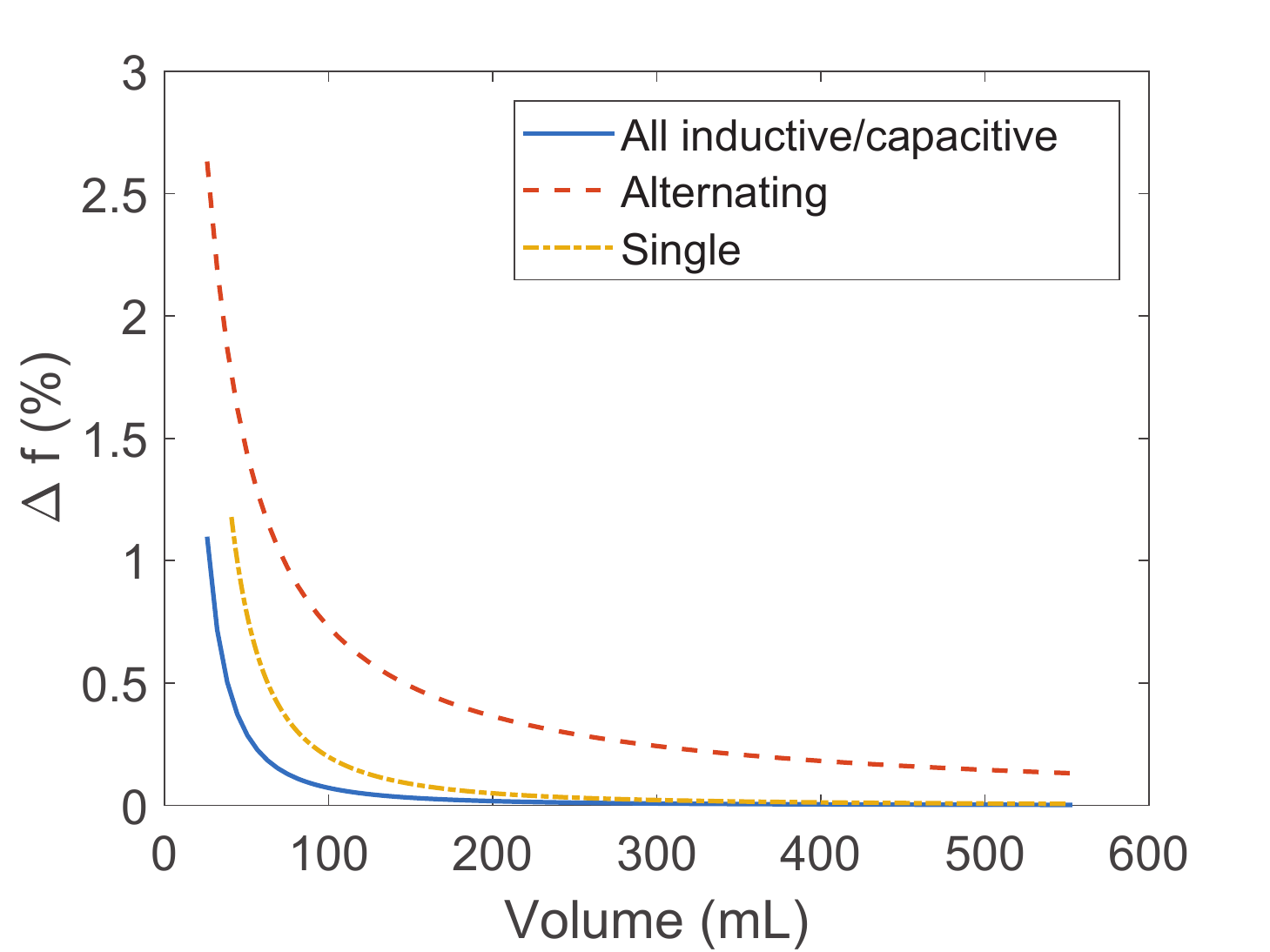}
         \caption{}
         \label{fig:Allind_vs_Allcap_vs_Alt_vs_Indi_ModeMixing}
\end{subfigure}
\hfill
\begin{subfigure}[b]{0.49\textwidth}
         \centering
         \includegraphics[width=1\textwidth]{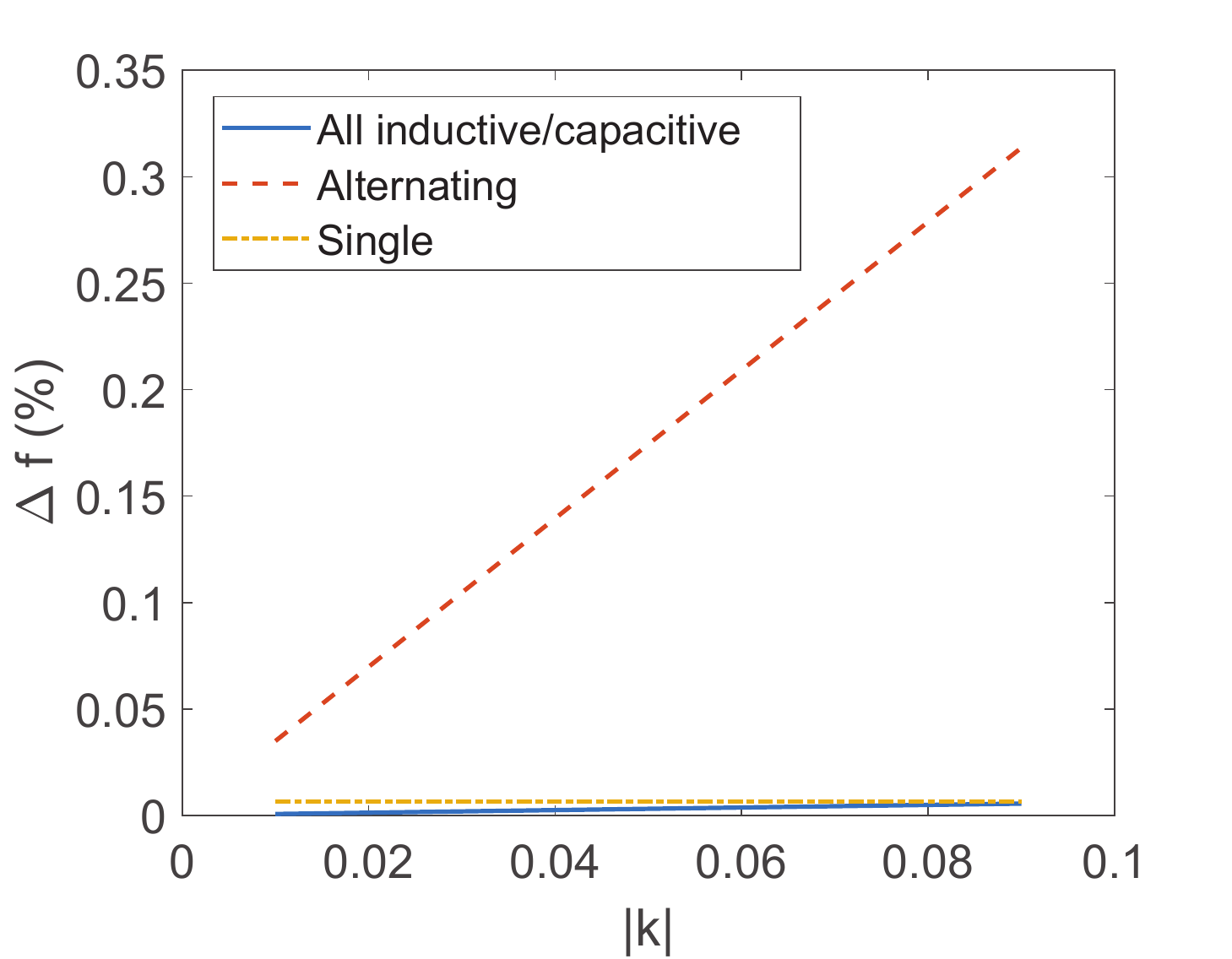}
         \caption{}
         \label{fig:Allind_vs_Allcap_vs_Alt_vs_Indi_ModeMixing_vs_k}
\end{subfigure}
\hfill
\begin{subfigure}[b]{0.49\textwidth}
         \centering
         \includegraphics[width=1\textwidth]{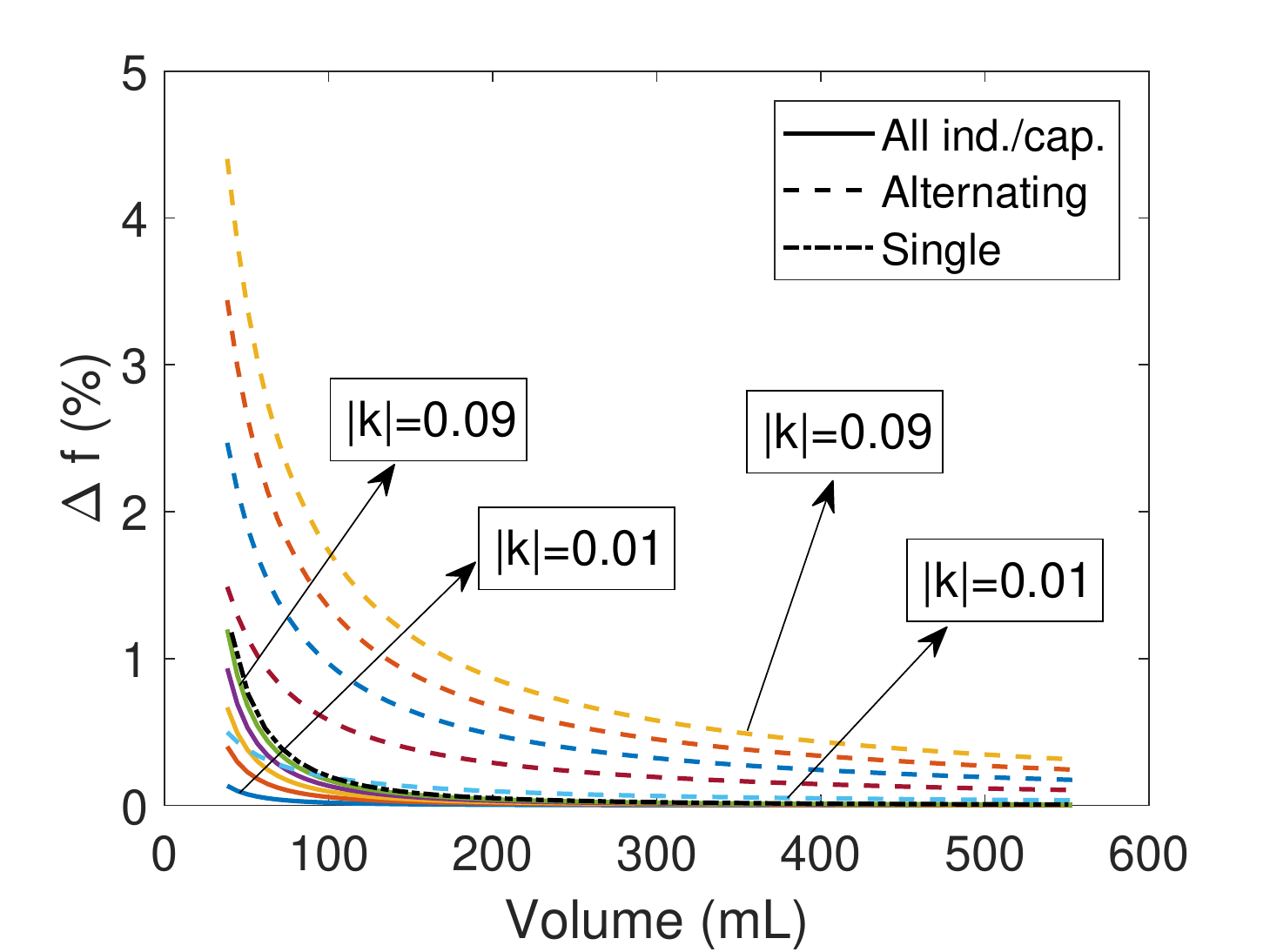}
         \caption{}
         \label{fig:Allind_vs_Allcap_vs_Alt_vs_Indi_ModeMixing_several_ks}
\end{subfigure}
\caption{Comparison of the cases all inductive irises, all capacitive irises and alternating couplings inductive/capacitive irises for a design example working at $8.4$~GHz with $|k|=0.0377$: (a) $S_{21}$ scattering parameter magnitude as a function of the frequency for a 6 subcavities multicavity of each type, (b) relative mode separation between the closest eigenmode to the axion one versus volume, (c) relative mode separation versus the absolute value of the physical coupling $k$ for $N=90$, and (d) relative mode separation between the closest mode to the axion one versus volume for several physical coupling values and types. For the all inductive/capacitive and alternating cases several $|k|$ values have been used (from bottom to top: $|k|=0.01$, $0.03$, $0.05$, $0.07$ and $0.09$). The single long cavity case has been added in (b), (c) and (d) for comparison. In (c) a single cavity length $d=2700$~mm has been employed in order to produce the same volume than the multicavity.}
\label{fig:Allind_vs_Allcap_vs_Alt}
\end{figure}

Regarding the mode clustering issue, there is a solution to shift the neighbour configuration modes of the $TE_{101}$ mode away from the axion one for the multicavity designs. This procedure is based on alternating the signs of the couplings which is practically achieved by using the two types of irises (capacitive or horizontal window, and inductive or vertical window) as discussed in \cite{RADES_paper2}. For an all-inductive design ($k<0$), the axion mode corresponds with the first configuration of the $TE_{101}$ mode, and for an all-capacitive haloscope ($k>0$) it corresponds with the last one. However, for an alternating inductive/capacitive structure the axion mode will be the central one (when there is an odd number of cavities) or the mode in the position $\frac{N}{2}+1$ (when there is an even number of cavities), where the distance between the configuration modes is higher. Figure~\ref{fig:Allind_vs_Allcap_vs_AltSp} plots an example of the $S_{21}$ scattering parameter magnitude as a function of the frequency for the three previous cases (all capacitive, all inductive or alternation of both types of irises) in a multicavity based on six subcavities with $|k|=0.0377$.
As it can be seen, the all-inductive and all-capacitive multicavities provide the axion mode at their first and last resonances, respectively, while for the alternating structure it is positioned in the position $\frac{N}{2}+1=4$. In Figure~\ref{fig:Allind_vs_Allcap_vs_Alt_vs_Indi_ModeMixing}\footnote{For this plot, a reasonable assumption has been made for the multicavity case: same subcavity volume for any $N$. In practise, the difference in length is minimum during the calculation of the final volume, which is the parameter that is represented in this plot.} the relative mode separation between the closest configuration mode to the axion one is observed for these three cases plus the single long cavity case in an X band structure. Also, in Figure~\ref{fig:Allind_vs_Allcap_vs_Alt_vs_Indi_ModeMixing_vs_k} the dependency of the relative mode separation with the physical coupling value is plotted. In addition, the behaviour of increasing the volume with different $|k|$ values and types can be observed in Figure~\ref{fig:Allind_vs_Allcap_vs_Alt_vs_Indi_ModeMixing_several_ks}. The results of the multicavity case in these plots have been generated with the formulation described in \cite{RADES_paper1} (for the all inductive/capacitive case) and \cite{RADES_paper2} (for the alternating case).\\

As it can be seen in Figure~\ref{fig:Allind_vs_Allcap_vs_Alt_vs_Indi_ModeMixing}, the alternating concept provides a great improvement in terms of mode separation. However, the manufacturing of mixed capacitive and inductive irises is complicated, which makes the construction of alternating multicavities more difficult as compared with the all inductive multicavity case. Also, although the largest frequency separations are achieved with the highest values of $|k|$, as it is depicted in Figures~\ref{fig:Allind_vs_Allcap_vs_Alt_vs_Indi_ModeMixing_vs_k} and \ref{fig:Allind_vs_Allcap_vs_Alt_vs_Indi_ModeMixing_several_ks}, in the practical design of multicavity haloscopes intermediate values of physical coupling are chosen so that the loading effect of the couplings does not reduce the subcavity lengths excessively as reported previously \cite{RADES_paper1}.\\

Another advantage of the multicavity concept compared with single cavities is that the extraction of the RF power (with a coaxial to waveguide transition, for example) in a critical coupling regime (that is $\beta=1$) is easier. This is because in a multicavity structure there is a maximum value of the electric field in each subcavity for the resonant mode, while in a single cavity there is only one maximum inside the whole structure. For multicavities, the concentration of the electric field in the center of the subcavities decreases with higher $|k|$. Thus, another trade-off between the relative mode separation (requiring high $|k|$) and the extraction of the coupling power (more efficient with low $|k|$) is found here.\\

After introducing the concept of 1D multicavity for $z-$axis connected small subcavities, in the next sections it is generalized for long, tall and large subcavities connected in different axes.

\subsection{Long subcavities}
\label{SubSec:MulticavLong}
As a novel concept for taking advantage of the volume available in the bore of a dipole or solenoid magnet, the combination of both long and multicavity concepts must be considered. This principle is based on increasing the length of the subcavities in the multicavity array, reducing slightly the width $a$ to maintain the proper operational frequency. As a small drawback, the reduction of the width in the subcavities yields to a small lowering of the quality factor, as explained previously.\\

\begin{figure}[h]
\centering
\begin{subfigure}[b]{0.69\textwidth}
         \centering
         \includegraphics[width=0.6\textwidth]{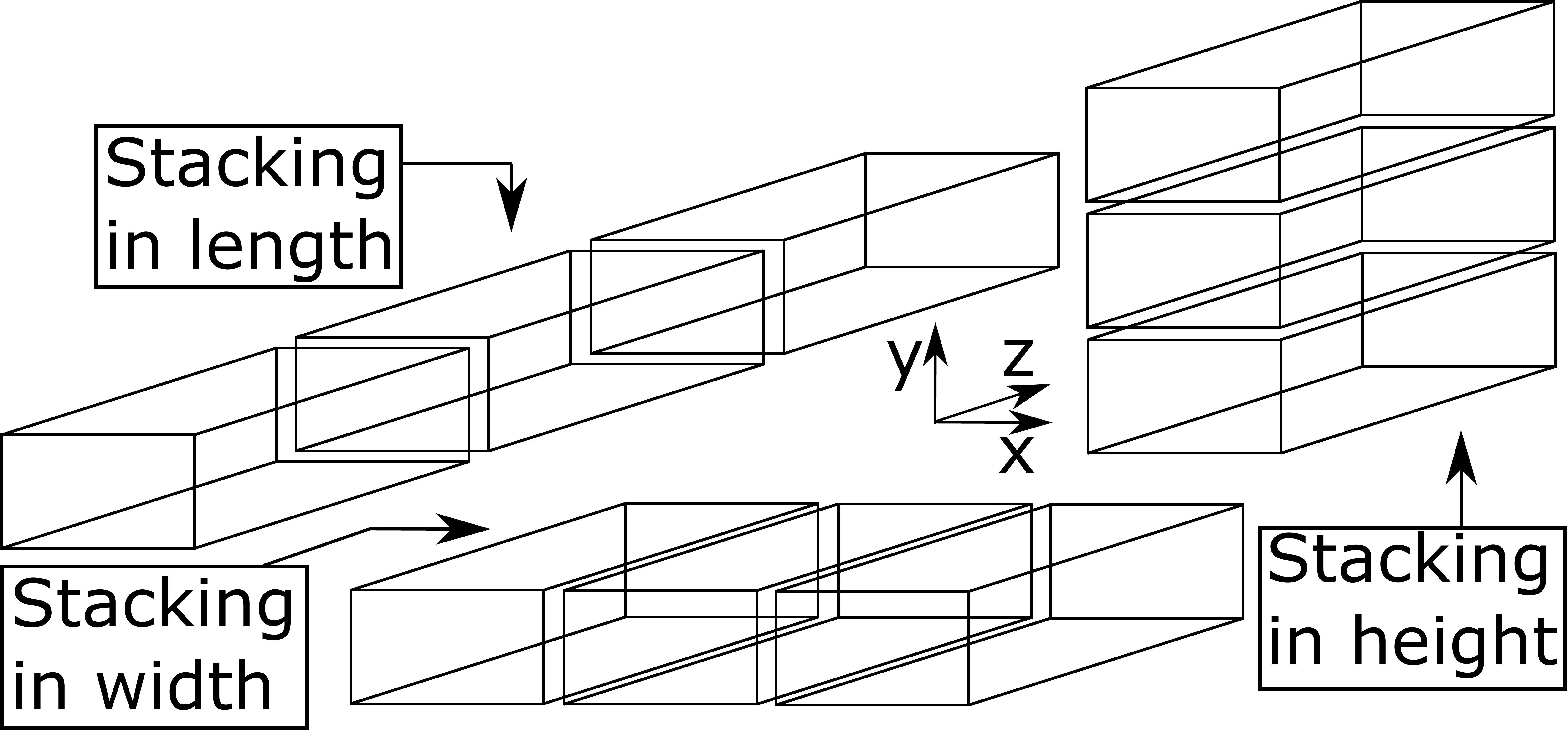}
         \caption{}
         \label{fig:MulticavitiesStackedInSeveralDirections_Long}
\end{subfigure}
\hfill
\begin{subfigure}[b]{0.32\textwidth}
         \centering
         \includegraphics[width=1\textwidth]{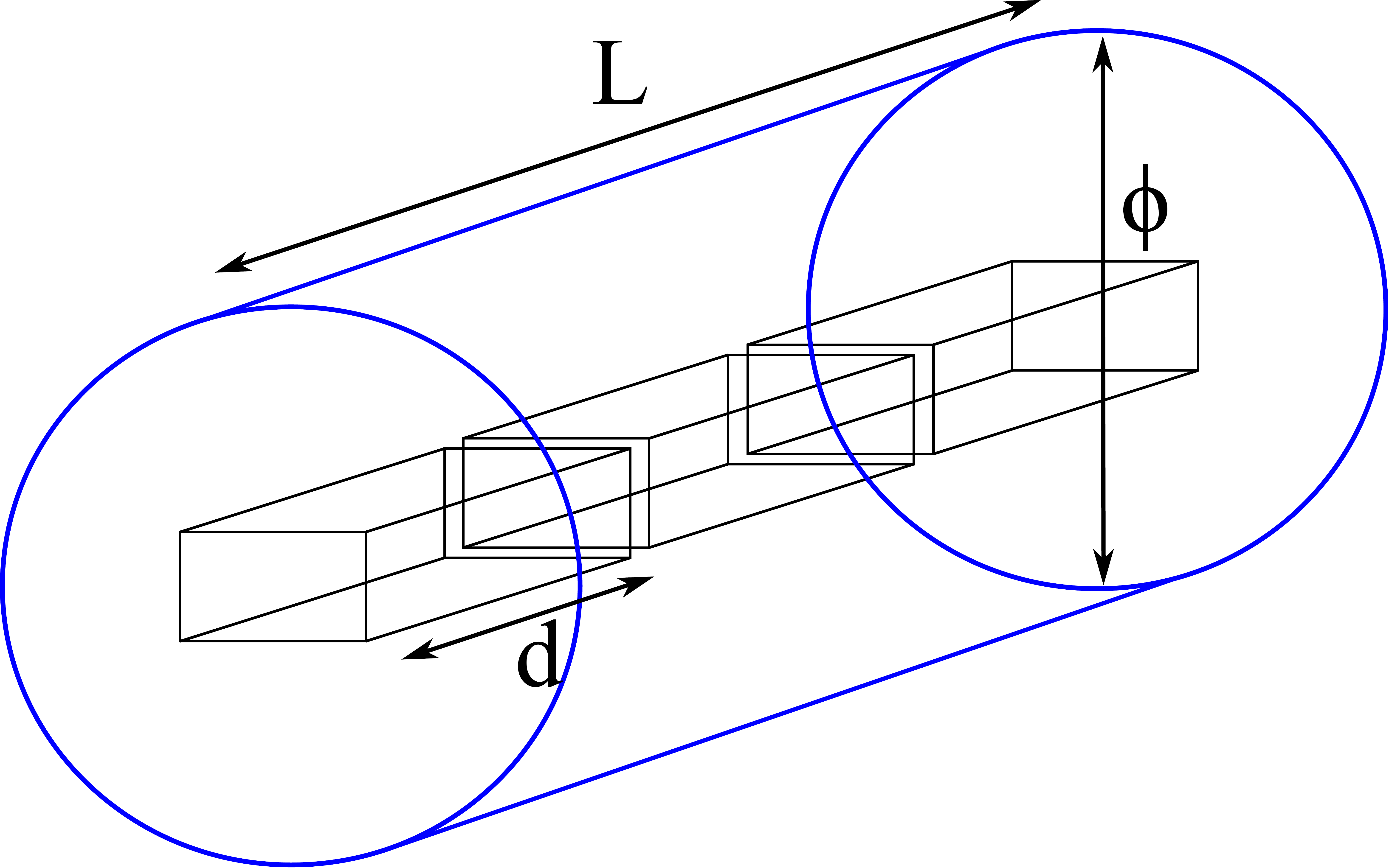}
         \caption{}
         \label{fig:MulticavitiesStackedInLong_Long_Dipole}
\end{subfigure}
\hfill
\begin{subfigure}[b]{0.32\textwidth}
         \centering
         \includegraphics[width=1\textwidth]{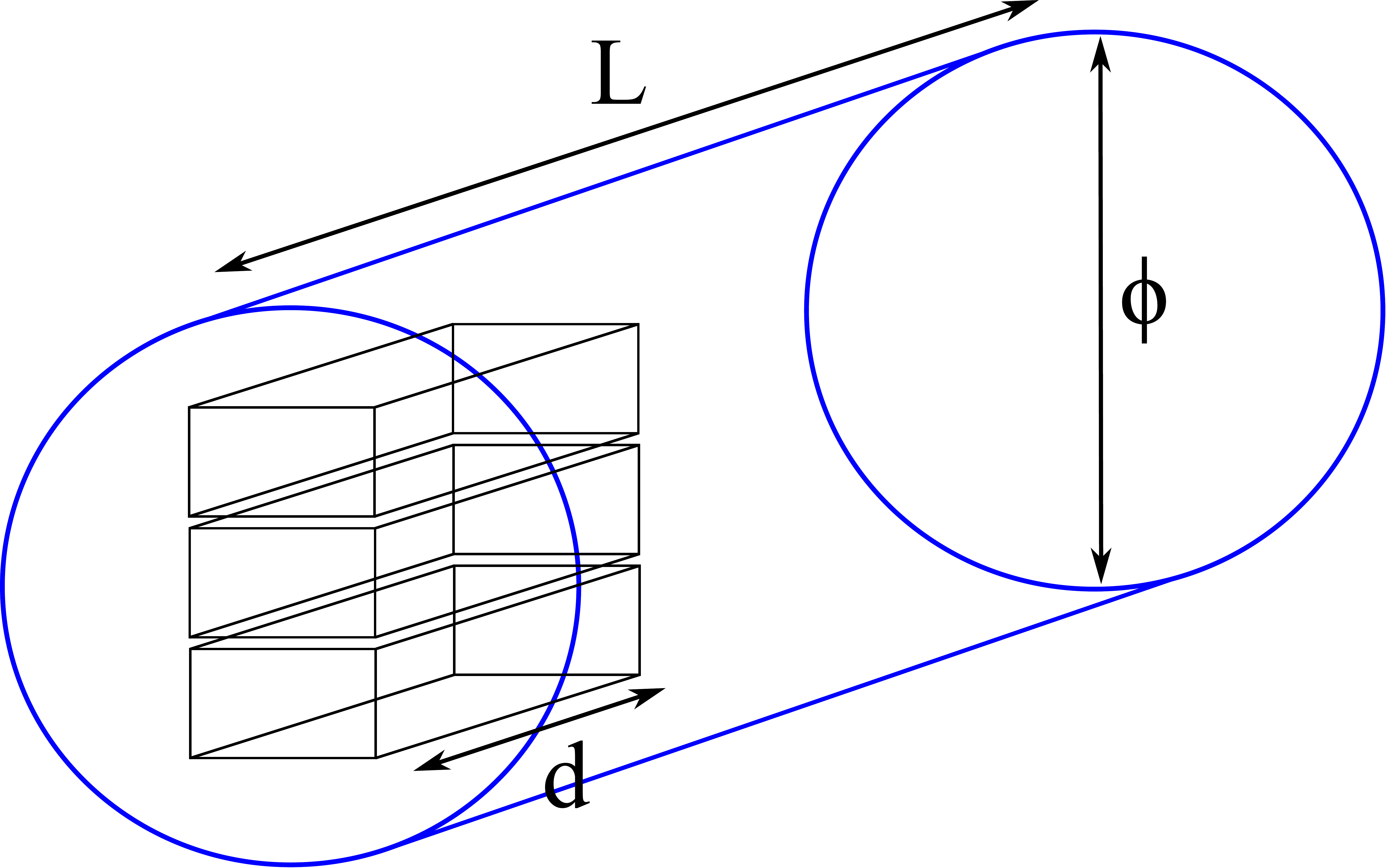}
         \caption{}
         \label{fig:MulticavitiesStackedInHeight_Long_Dipole}
\end{subfigure}
\hfill
\begin{subfigure}[b]{0.32\textwidth}
         \centering
         \includegraphics[width=1\textwidth]{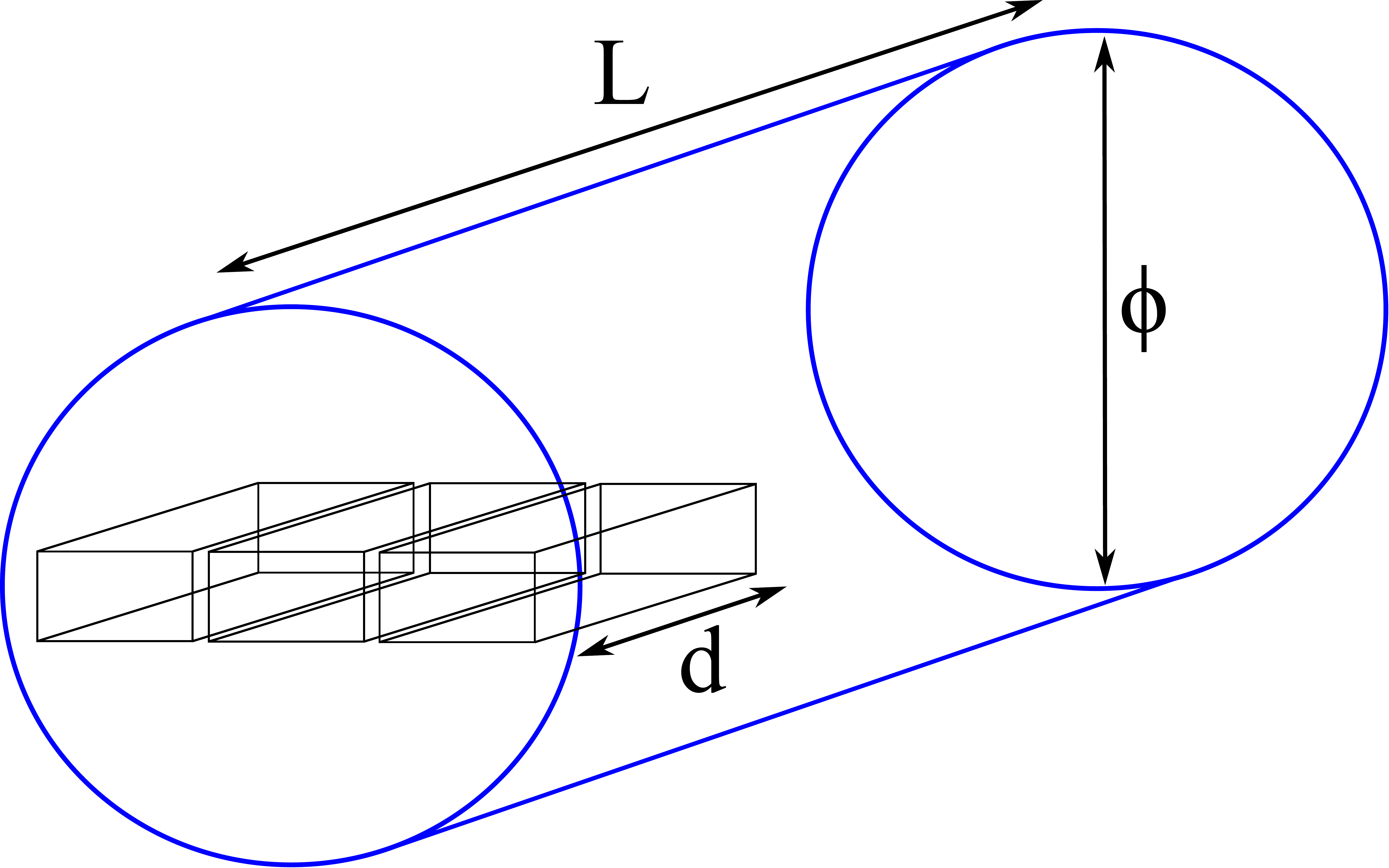}
         \caption{}
         \label{fig:MulticavitiesStackedInWidth_Long_Dipole}
\end{subfigure}
\hfill
\begin{subfigure}[b]{0.2\textwidth}
         \centering
         \includegraphics[width=1\textwidth]{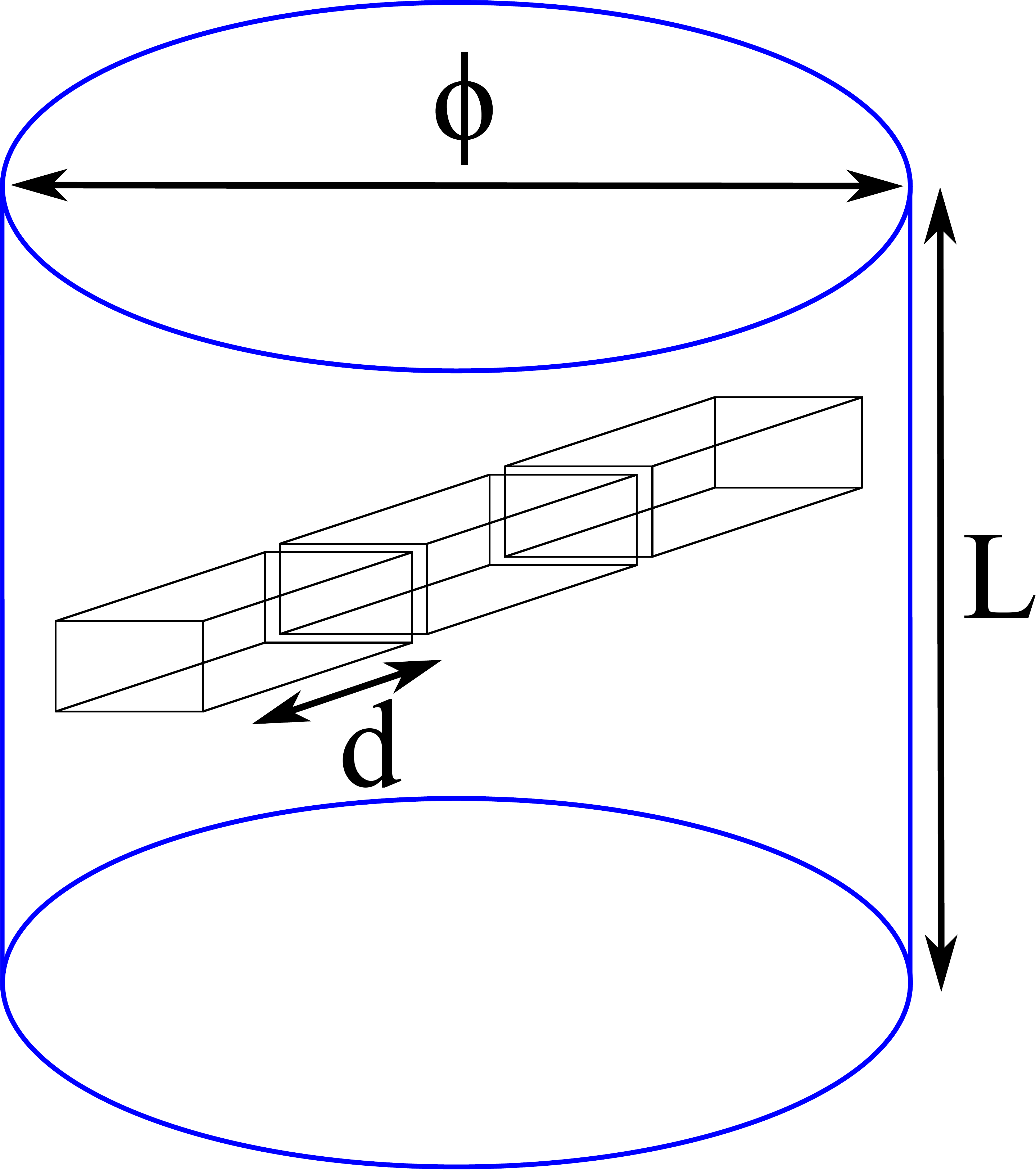}
         \caption{}
         \label{fig:MulticavitiesStackedInLong_Long_Solenoid}
\end{subfigure}
\hfill
\begin{subfigure}[b]{0.2\textwidth}
         \centering
         \includegraphics[width=1\textwidth]{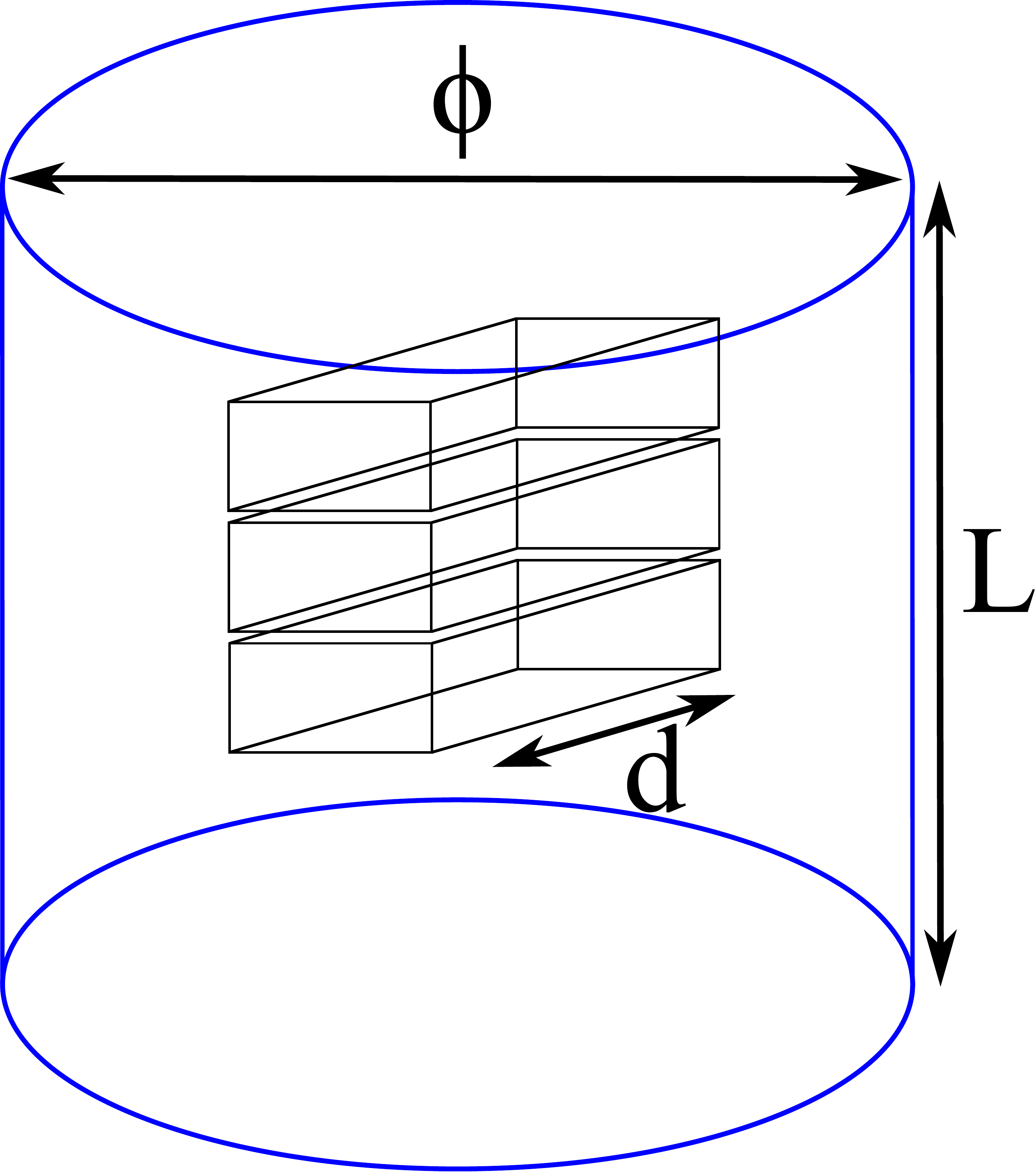}
         \caption{}
         \label{fig:MulticavitiesStackedInHeight_Long_Solenoid}
\end{subfigure}
\hfill
\begin{subfigure}[b]{0.2\textwidth}
         \centering
         \includegraphics[width=1\textwidth]{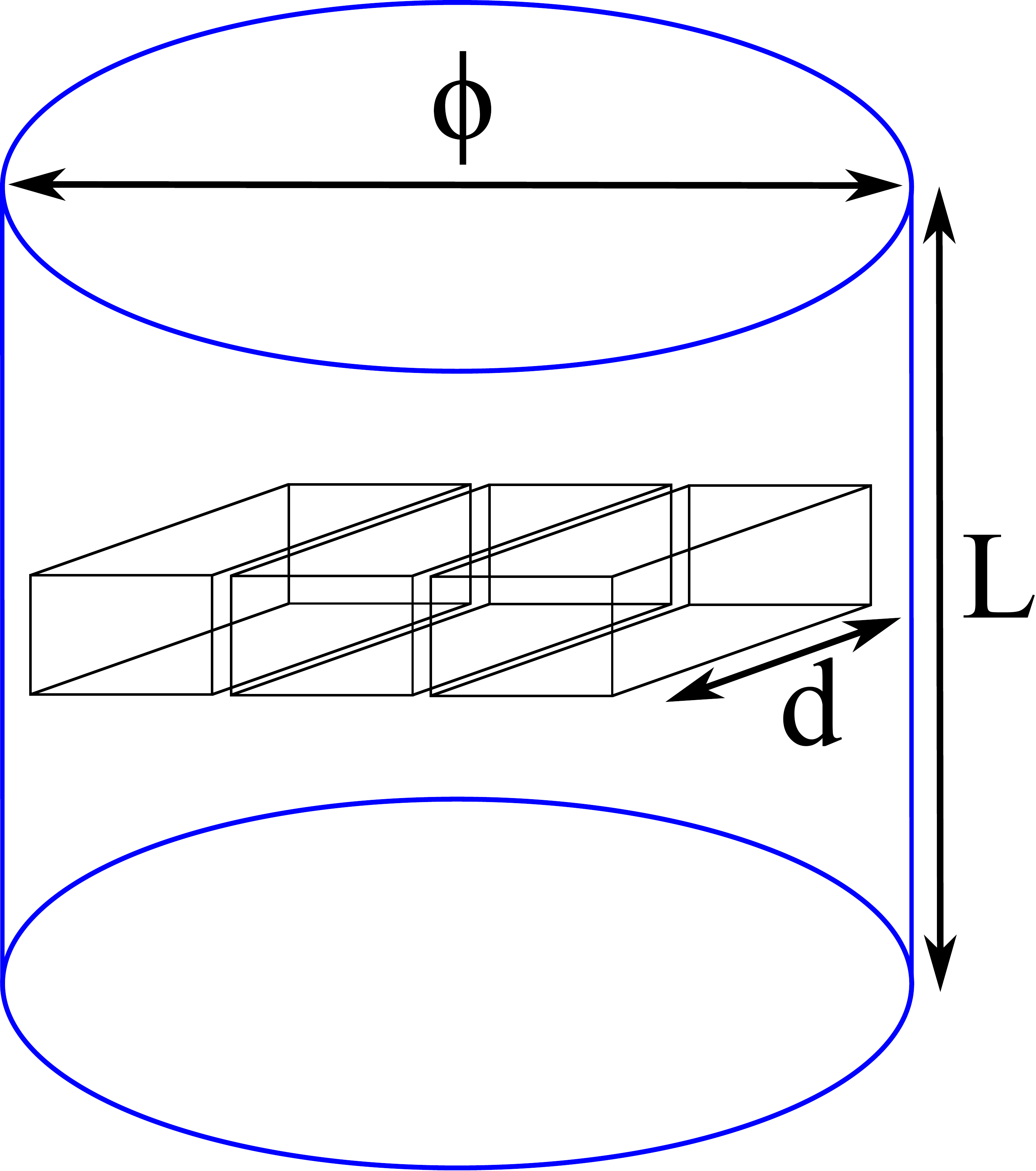}
         \caption{}
         \label{fig:MulticavitiesStackedInWidth_Long_Solenoid}
\end{subfigure}
\caption{(a) Possibilities to stack three long subcavities to create a multicavity. Implementation of these multicavity stackings in dipole and solenoid magnets. For dipole magnets: (b) in length, (c) in height, and (d) in width. For solenoid magnets: (e) in length, (f) in height, and (g) in width.}
\label{fig:StakingLongSubcavitiesInDipolesAndSolenoids}
\end{figure}

There are three possibilities for coupling (or stacking) the subcavities in a 1D multicavity structure: in length, in height or in width. Figure~\ref{fig:MulticavitiesStackedInSeveralDirections_Long} shows these types of stackings in a multicavity based on three long subcavities.
From Figure~\ref{fig:MulticavitiesStackedInLong_Long_Dipole} to Figure~\ref{fig:MulticavitiesStackedInWidth_Long_Solenoid}\footnote{Note here how for solenoid magnets the vertical direction of the multicavity ($y-$axis in Figure~\ref{fig:MulticavitiesStackedInSeveralDirections_Long}) is now oriented towards the longitudinal direction of the bore (this is, $z-$axis in Figure~\ref{fig:Solenoid}) to align the electric field of the haloscope with the static magnetic field of the magnet.} examples for each type of stacking in both dipole and solenoid magnets are shown. The stacking in length direction is the case employed in RADES (with standard subcavity lengths) so far (see manufactured prototypes in Figure~\ref{fig:RADEScavities}) making use of the scenario from Figure~\ref{fig:MulticavitiesStackedInLong_Long_Dipole} for the CAST dipole magnet.\\

In the case of a multicavity employing the length direction ($z-$axis in Figure~\ref{fig:MulticavitiesStackedInSeveralDirections_Long}), longer subcavities lead to lower coupling values $|k|$ (for the same operating frequency). This occurs due to the distance from the electric field maximum to the coupling iris, since, with higher subcavity lengths less energy reaches the irises. There is a limit in length where the irises cannot provide the correct $k$ value independently on their geometry. For this reason, the use of very large subcavities in a multicavity design stacked in length is not possible. When designing, one has to find the maximum length where the physical coupling is still realizable. This effect is independent of the number of subcavities.\\

For this type of multicavities, there is a great room to increase the number of subcavities and their lengths if a dipole or similar magnet is employed due to the great length of the bore ($10$~m in the case of BabyIAXO). However, for solenoid magnets, the 1D multicavity concept using the length for the stacking of the subcavities (Figure~\ref{fig:MulticavitiesStackedInLong_Long_Solenoid}) is not the best proposal since the limiting dimension is the diameter. Thus, for the greatest solenoid bore from Table~\ref{tab:magnets} (the one from MRI (ADMX-EFR) magnet) a maximum haloscope length of $650$~mm (the diameter of the bore) is imposed. This can easily be achieved with one single cavity. In case of necessity of using 1D multicavites with stacking in length, a multicavity based on not very long subcavities ($13$ subcavities of $d=50$~mm, for example) could be designed, although it will be shown below that there are more efficient solutions to increase the volume of a multicavity in solenoid magnets.\\

For the other two stacking possibilities ($x$ and $y-$axis in Figure~\ref{fig:MulticavitiesStackedInSeveralDirections_Long}), the energy that reaches the irises is very high due to the lower distance from the center of the cavity (where the maximum electric field is stored) to the iris. Then, the use of any subcavity length independently of the interresonator coupling value $k$ employed is possible. Nevertheless, there are some limitations again for these new staking directions, due to the bore size in both dipoles and solenoids. In dipoles, for the long multicavity stacked in height (Figure~\ref{fig:MulticavitiesStackedInHeight_Long_Dipole}) and in width (Figure~\ref{fig:MulticavitiesStackedInWidth_Long_Dipole}) the longitudinal axis of the bore should be employed as the limit for the length of the subcavities, and the diameter of the transversal section of the bore limits the stacking direction of all the subcavities. This implies a great freedom in length $d$ ($L_{BabyIAXO}=10$~m), but also a limitation in the number of subcavities $N$ that can be stacked ($\phi_{BabyIAXO}=600$~mm).\\

In solenoids, for long subcavities stacked in height (Figure~\ref{fig:MulticavitiesStackedInHeight_Long_Solenoid}), the longitudinal axis of the subcavities can be oriented in any radial bore axis ($d$ limited to $\phi_{MRI}=650$~mm) and the subcavity stacking in the longitudinal bore axis ($N$ limited to $L_{MRI}=800$~mm). On the other hand, for stacking a long multicavity in width for solenoid magnets (Figure~\ref{fig:MulticavitiesStackedInWidth_Long_Solenoid}), both array stacking and longitudinal axis of the subcavities should consider any radial bore axis. In this case, both $d$ and $N$ are not limited to $\phi_{MRI}=650$~mm, but to a lower value due to the cylindrical shape of the bore. If the length of the subcavities covers the whole radial axis ($d\approx\phi_{MRI}=650$~mm), only one subcavity could be added. However, if the length $d$ is reduced to a more moderated value, $N$ can be increased. For example, if a square area footprint is desired at the MRI solenoid bore, a maximum $d=\frac{\phi_{MRI}}{2}\sqrt{2}=459.62$~mm value (equation for fitting a square into a circle) should be implemented. A more efficient geometry could be implemented in this case by using different lengths for each subcavity taking advantage of almost all the bore circle area for increasing even more the volume of the haloscopes (envisaged work is expected in this subject).\\

\begin{figure}[h]
\centering
\begin{subfigure}[b]{0.49\textwidth}
         \centering
         \includegraphics[width=1\textwidth]{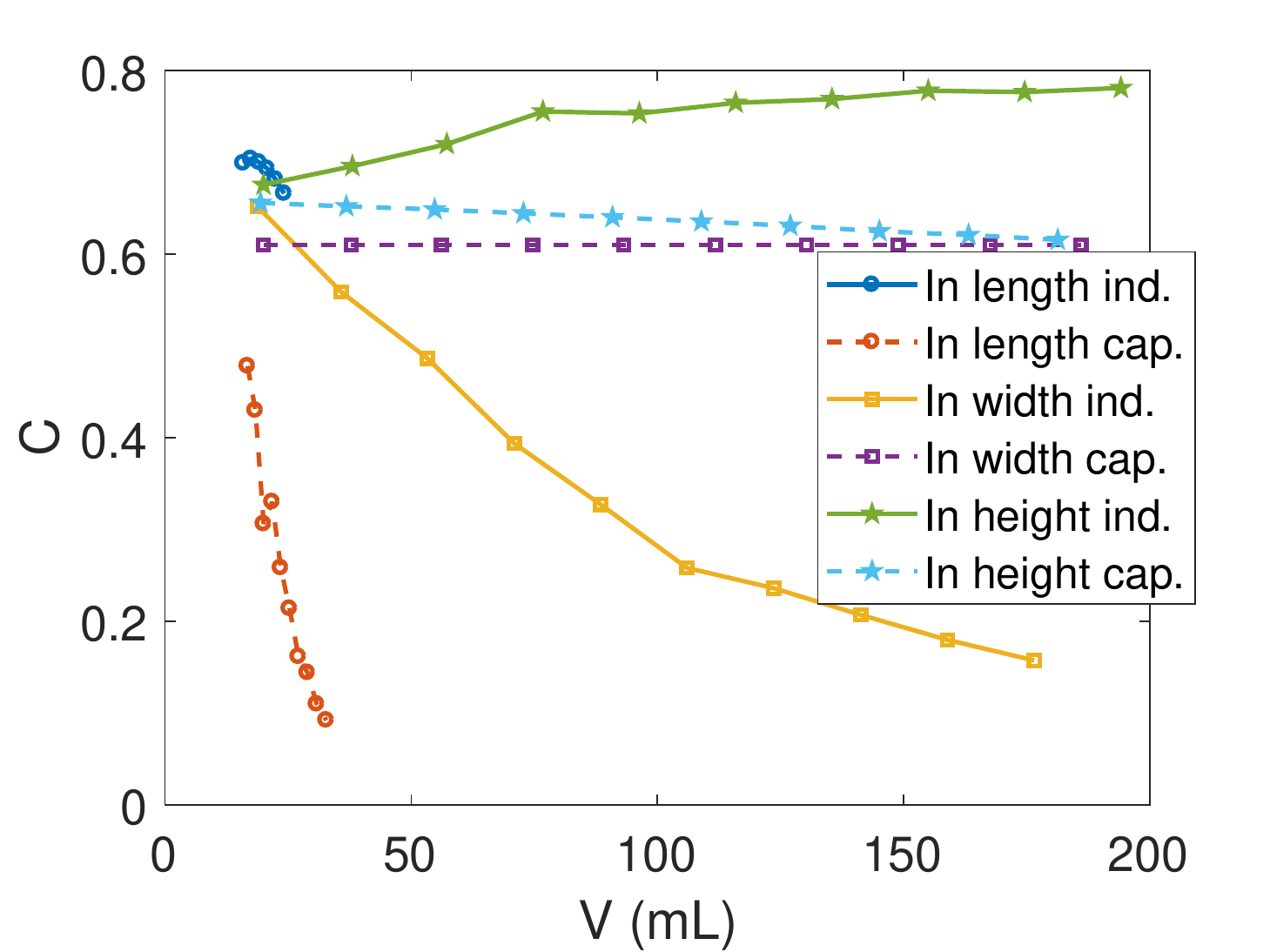}
         \caption{}
         \label{fig:C_3couplingDirections_Long}
\end{subfigure}
\hfill
\begin{subfigure}[b]{0.49\textwidth}
         \centering
         \includegraphics[width=1\textwidth]{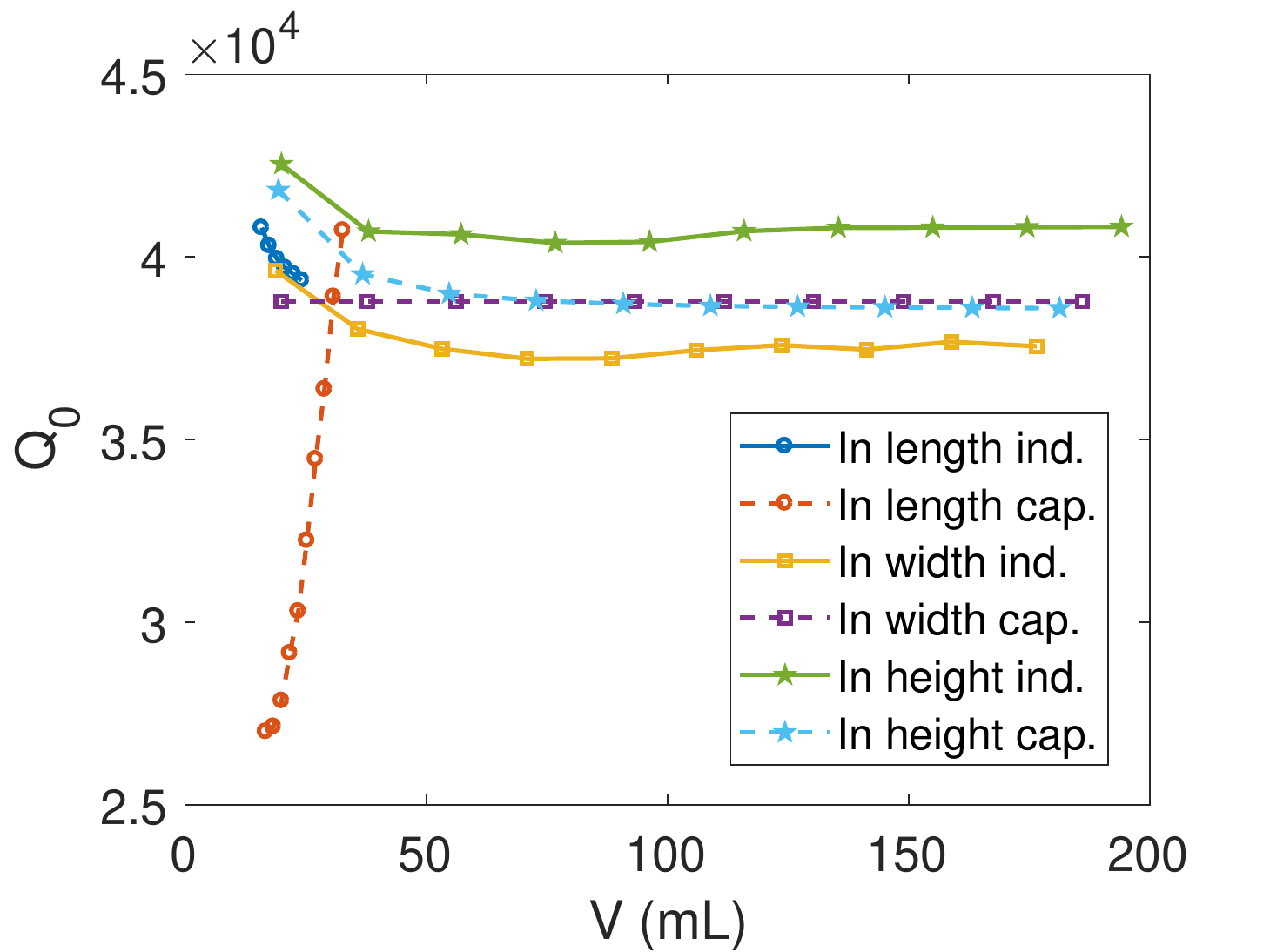}
         \caption{}
         \label{fig:Q0_3couplingDirections_Long}
\end{subfigure}
\hfill
\begin{subfigure}[b]{0.49\textwidth}
         \centering
         \includegraphics[width=1\textwidth]{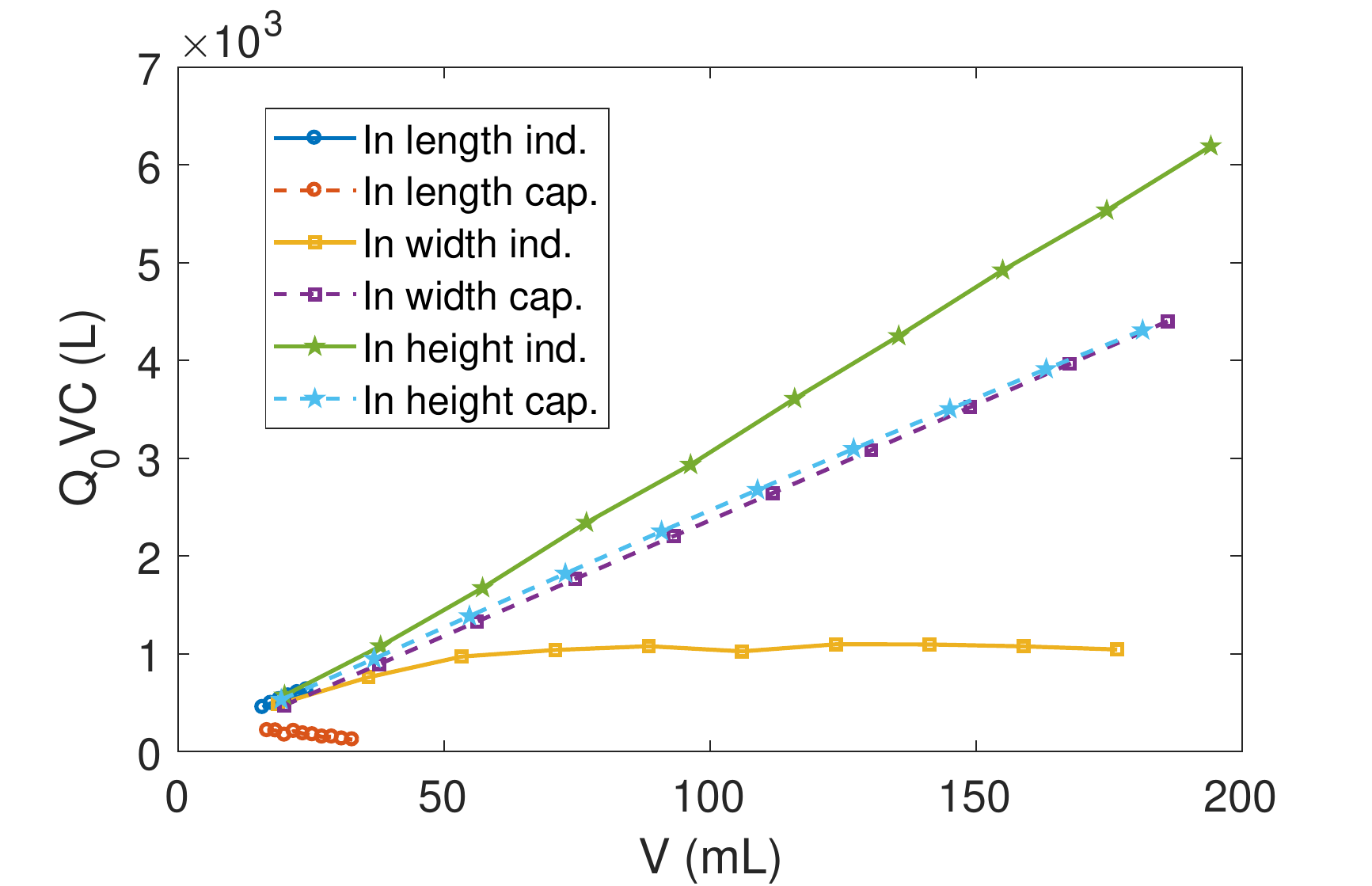}
         \caption{}
         \label{fig:QVC_3couplingDirections_Long}
\end{subfigure}
\caption{Influence on the design parameters of each type of coupling (inductive or capacitive) when the coupling is introduced along each direction (longitudinal (length), horizontal (width) or vertical (height)) in a structure composed of two coupled long subcavities (with $b=10.16$~mm and $a=17.85$~mm): (a) form factor, (b) quality factor, and (c) $Q_0 \times V \times C$ factor. The volume depends only on the length as the number of subcavities is fixed to two.}
\label{fig:Parameters_3couplingDirections}
\end{figure}

In Figures~\ref{fig:C_3couplingDirections_Long}, \ref{fig:Q0_3couplingDirections_Long} and \ref{fig:QVC_3couplingDirections_Long} it is observed a comparison study of these three types of coupling directions for X band in a two subcavities structure employing an inductive or capacitive iris varying the volume, which depends only on the subcavity length since the height ($b=10.16$~mm), the width ($a=22.86$~mm) and the number of subcavities ($N=2$ for simplicity) is fixed.
These results are valid for both dipole and solenoid magnets as long as the approximation of $\vec{B}_e=B_e \, \hat{y}$ for dipoles and $\vec{B}_e=B_e \, \hat{z}$ for solenoids is accomplished. In that case, the form factor $C$ is equal in both situations.\\

In multicavities with long subcavities, the limit in the number of subcavities $N$ and in the subcavity length $d$ is imposed by the mode separation, according to the results provided in Figure~\ref{fig:Allind_vs_Allcap_vs_Alt_vs_Indi_ModeMixing}, similarly to the limit in the length of single cavities (described in section~\ref{SubSec:IndLong}). Nevertheless, the size of the magnet bores is generally rather smaller than these limits for any stacking direction in a multicavity haloscope. In the example of Figures~\ref{fig:C_3couplingDirections_Long}, \ref{fig:Q0_3couplingDirections_Long} and \ref{fig:QVC_3couplingDirections_Long} it is reduced to $N=2$ for simplicity.\\

A coupling value of $|k|=0.025$ is used for this study, which is a typical value employed in the RADES collaboration. Figure~\ref{fig:Parameters_3couplingDirections} shows how for the longitudinal (or in length) coupling option the curves (both inductive and capacitive) are limited to volume values lower than $50$~mL. This is due to the length limitation in the subcavities for this kind of coupling direction as previously explained (the required coupling $|k|=0.025$ cannot be obtained with larger lengths). For the other four curves there is no such limitation so this study can be continued with higher volumes if necessary. Analysing the $Q_0 \times V \times C$ plot in Figure~\ref{fig:QVC_3couplingDirections_Long} it can be seen that the vertical (or in height) direction is the best option for long subcavities. However, depending on the type and dimensions of the magnet the $x-$axis direction option could be more appropriate.\\

\begin{figure}[h]
\centering
\begin{subfigure}[b]{0.49\textwidth}
         \centering
         \includegraphics[width=1\textwidth]{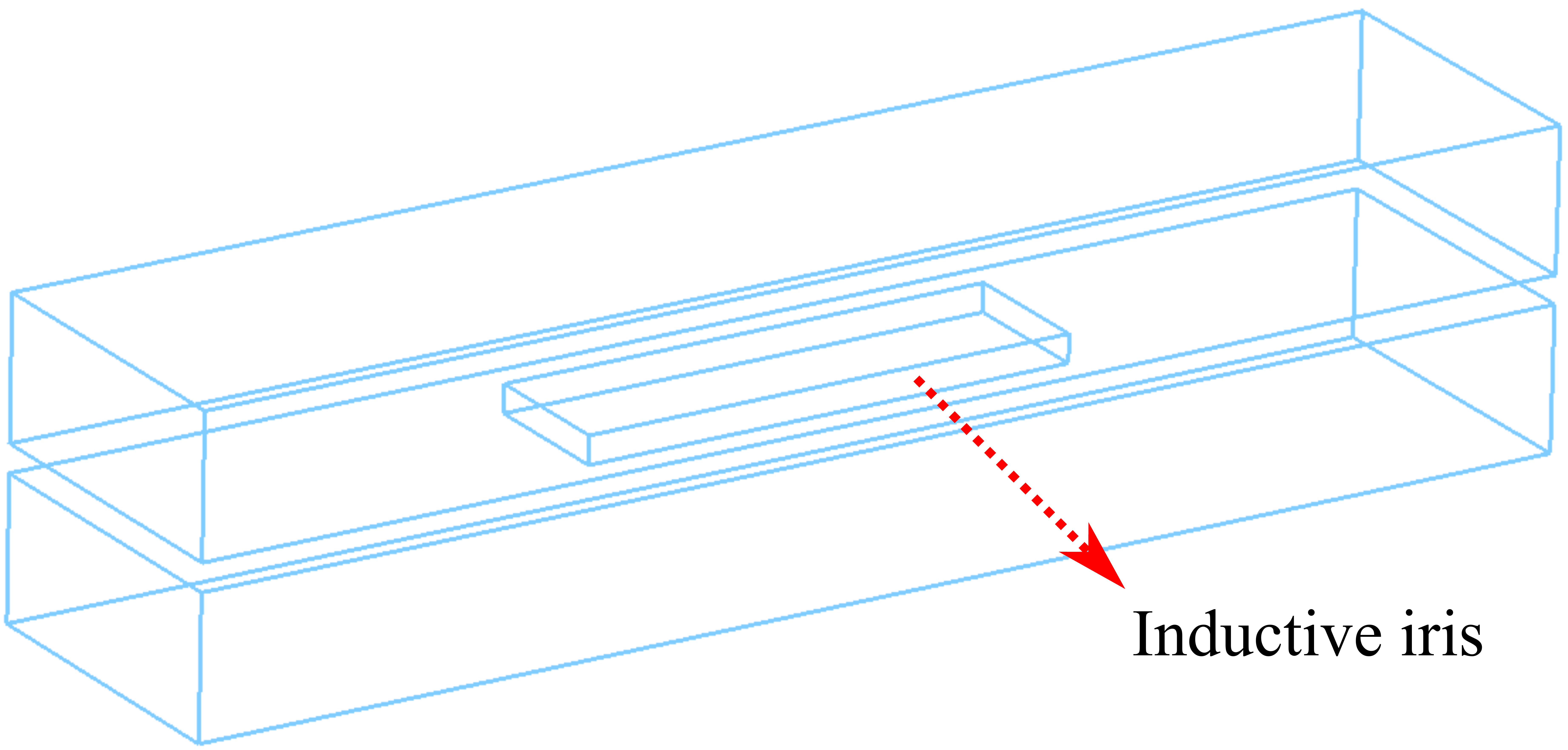}
         \caption{}
         \label{fig:LongWithVerticalCoupling_Ind}
\end{subfigure}
\hfill
\begin{subfigure}[b]{0.45\textwidth}
         \centering
         \includegraphics[width=1\textwidth]{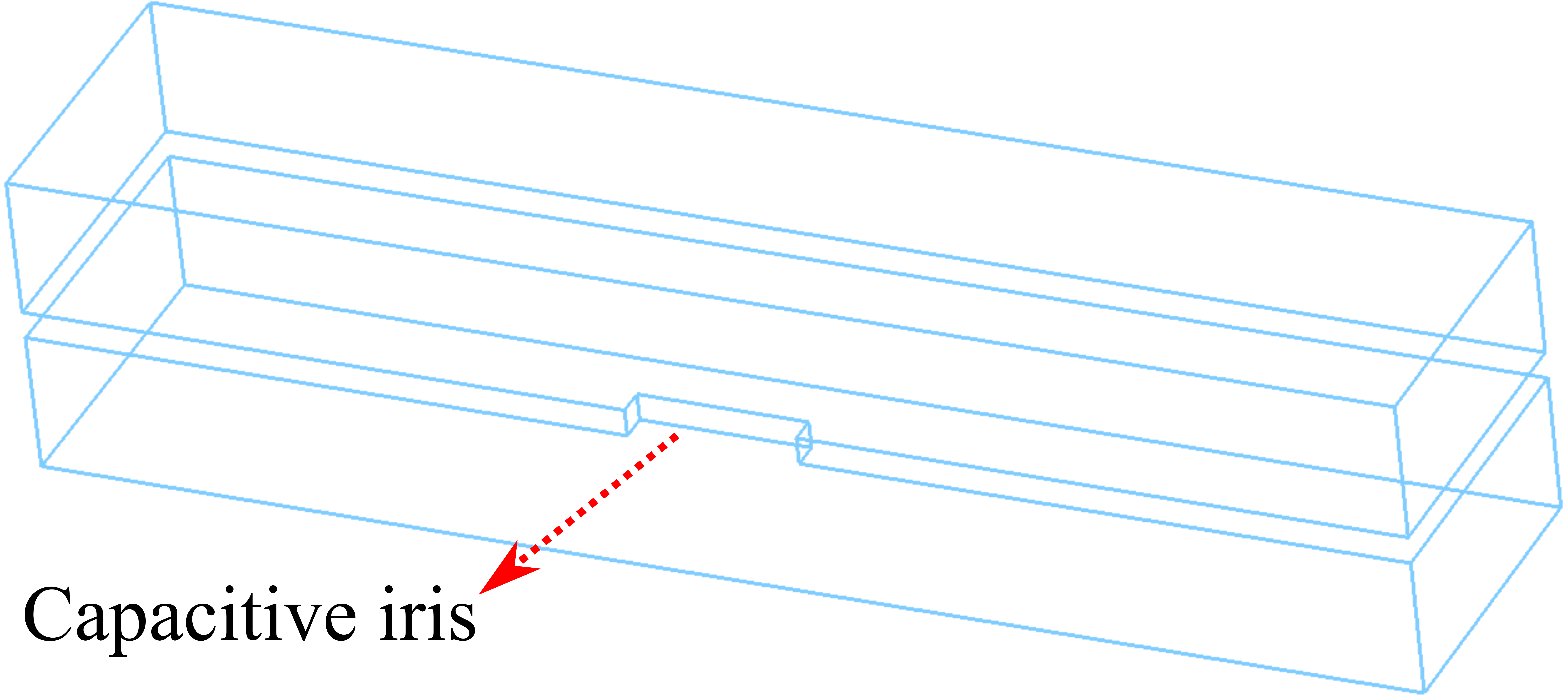}
         \caption{}
         \label{fig:LongWithVerticalCoupling_Cap}
\end{subfigure}
\caption{Sketch of a multicavity based on 2 subcavities stacked in height employing (a) an inductive iris window, and (b) a capacitive iris window.}
\label{fig:LongWithVerticalCoupling}
\end{figure}

For the vertical coupling direction it is not obvious how to design an inductive/capacitive iris. For this reason, a previous study varying the position and dimensions of a rectangular window has been carried out to find the inductive and capacitive behaviour. For an inductive operation the window is positioned at the center of the subcavity with a quasi-square or rectangular shape (see Figure~\ref{fig:LongWithVerticalCoupling_Ind}).
For a capacitive iris it is displaced to one side along the width with a thin rectangular shape (see Figure~\ref{fig:LongWithVerticalCoupling_Cap}).\\
 
All these studies have been carried out for the all-inductive and all-capacitive multicavity cases. However, as seen previously, the alternating case is the one that provides the largest frequency separation between adjacent modes. Therefore, as a proof of concept an alternating multicavity haloscope coupled in the vertical axis and based on $N=4$ long subcavities of $d=100$~mm has been designed.\\

The selected physical $|k|$ value for the interresonator couplings is $0.025$, and the resulting coupling matrix (utilising equation \ref{eq:Mij_kij_fB}) is:
\begin{gather}\label{eq:CouplingMatrix_1D_LongExample}
 \bf{M} =
  \begin{pmatrix}
   -0.5 & 0.5 & 0 & 0 \\
   0.5 & 0 & -0.5 & 0 \\
   0 & -0.5 & 0 & 0.5 \\
   0 & 0 & 0.5 & -0.5
   \end{pmatrix}
\end{gather}
which has been employed for the design of the structure with the methods described in \cite{Cameron,RADES_paper1}. Considering the non-zero off-diagonal elements, an alternating behaviour can be observed (positive sign for capacitive couplings and negative for inductive couplings).\\

\begin{figure}[h]
\centering
\begin{subfigure}[b]{0.49\textwidth}
         \centering
         \includegraphics[width=1\textwidth]{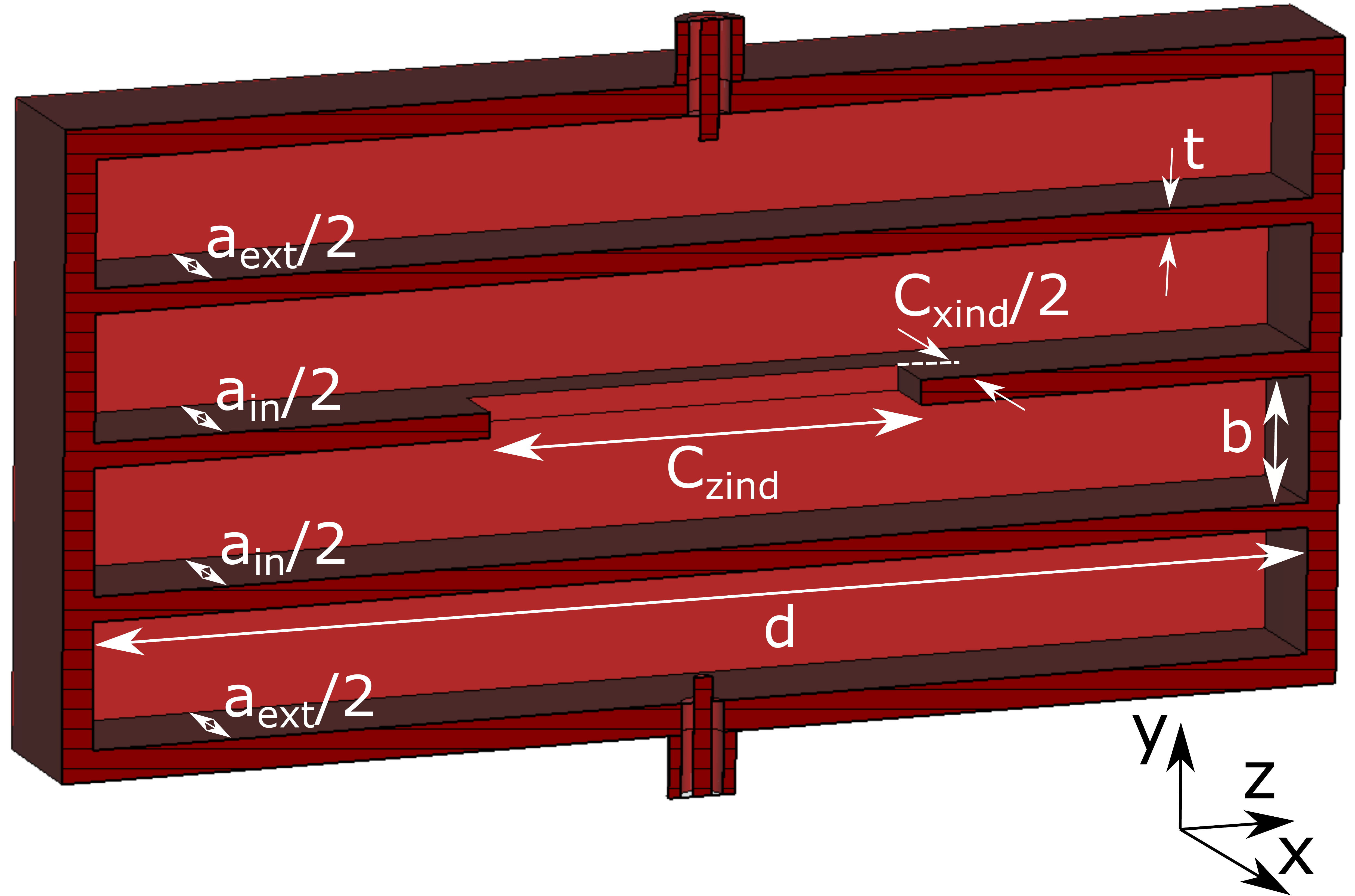}
         \caption{}
         \label{fig:4cav_Longd100mm_VerticalCoupling_Alt_left}
\end{subfigure}
\hfill
\begin{subfigure}[b]{0.49\textwidth}
         \centering
         \includegraphics[width=1\textwidth]{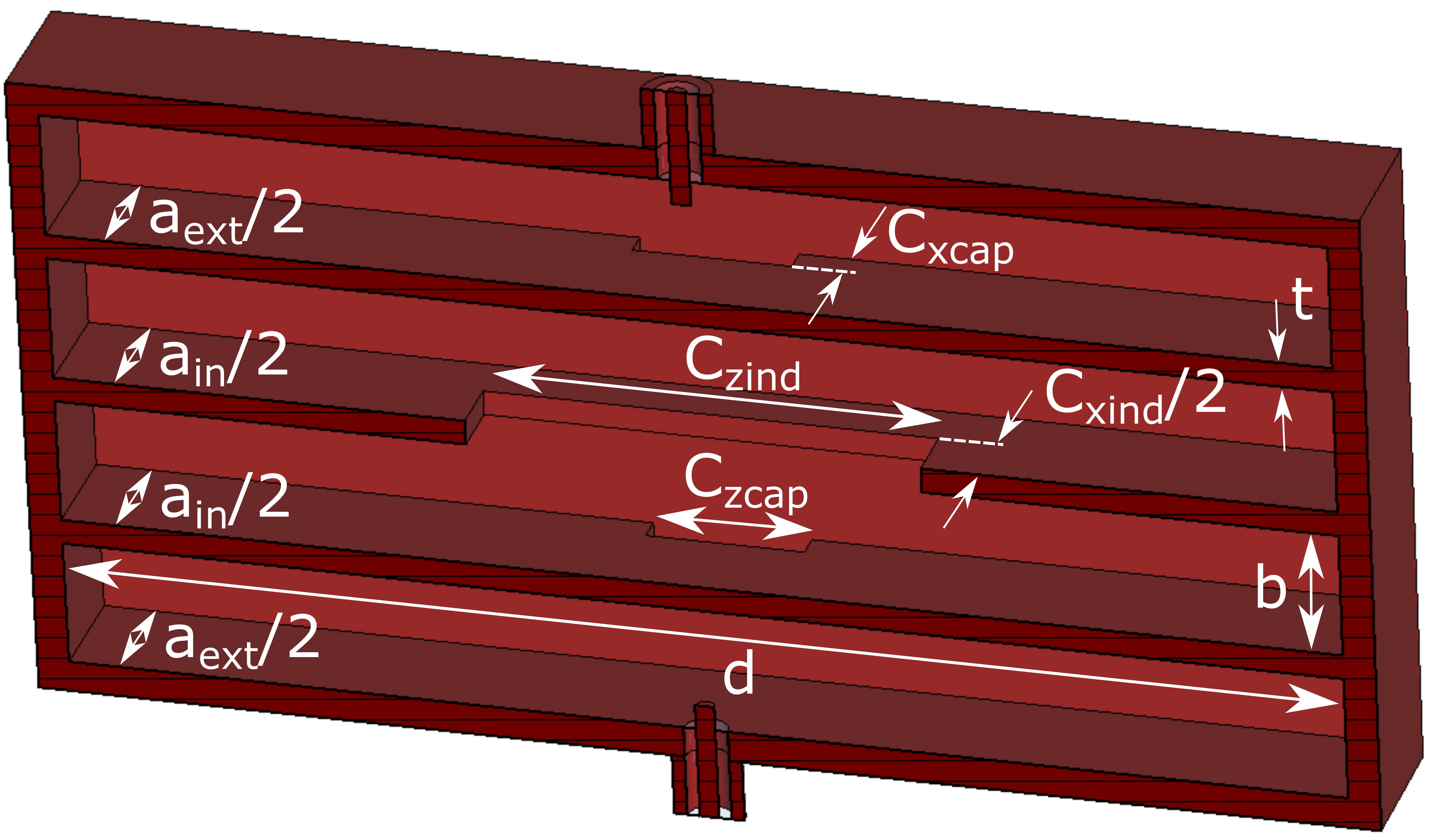}
         \caption{}
         \label{fig:4cav_Longd100mm_VerticalCoupling_Alt_right}
\end{subfigure}
\hfill
\begin{subfigure}[b]{0.6\textwidth}
         \centering
         \includegraphics[width=1\textwidth]{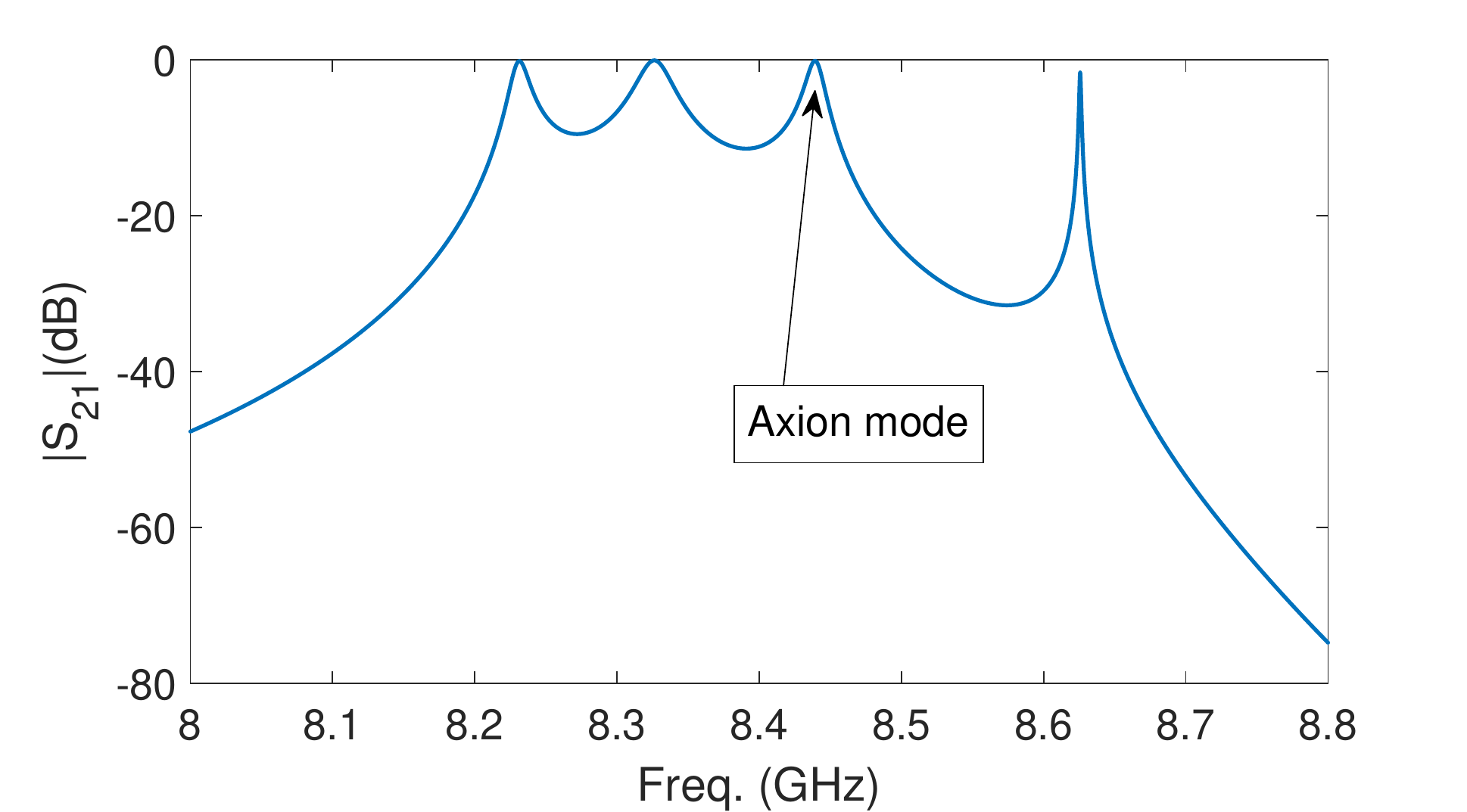}
         \caption{}
         \label{fig:4cav_Longd100mm_VerticalCoupling_Alt_S21}
\end{subfigure}
\caption{Alternating 1D vertically-coupled multicavity haloscope design based on four long subcavities with two capacitive and one inductive irises: (a) left piece of the structure, (b) right piece of the structure, and (c) $S_{21}$ scattering parameter magnitude as a function of the frequency.}
\label{fig:4cav_Longd100mm_VerticalCoupling_Alt}
\end{figure}

Figures~\ref{fig:4cav_Longd100mm_VerticalCoupling_Alt_left} and \ref{fig:4cav_Longd100mm_VerticalCoupling_Alt_right} show the final aspect of the haloscope evidencing the geometry and position of each type of coupling in this kind of multicavity (subcavities stacked in height).
In Figure~\ref{fig:4cav_Longd100mm_VerticalCoupling_Alt_S21} the simulated $S_{21}$ scattering parameter magnitude as a function of the frequency is shown.\\
 
The dimensions of the structure according to Figures~\ref{fig:4cav_Longd100mm_VerticalCoupling_Alt_left} and \ref{fig:4cav_Longd100mm_VerticalCoupling_Alt_right} are: length ($z-$axis) and height ($y-$axis) of all the subcavities $d=100$~mm and $b=10.16$~mm, respectively; internal subcavities width $a_{in}=18.3$~mm ($x-$axis); external subcavities width $a_{ext}=17.9$~mm; capacitive iris length $C_{zcap}=12.4$~mm ($z-$axis); capacitive iris width $C_{xcap}=2$~mm ($x-$axis); inductive iris length $C_{zind}=35.5$~mm; inductive iris width $C_{xind}=10$~mm and thickness of all the irises $t=2$~mm ($y-$axis). The capacitive windows are positioned at one side in width ($x-$axis) and centered in length ($z-$axis), while the inductive window is placed at the center both in width and length of the subcavities.\\

\begin{figure}[h]
\centering
\begin{subfigure}[b]{0.49\textwidth}
         \centering
         \includegraphics[width=1\textwidth]{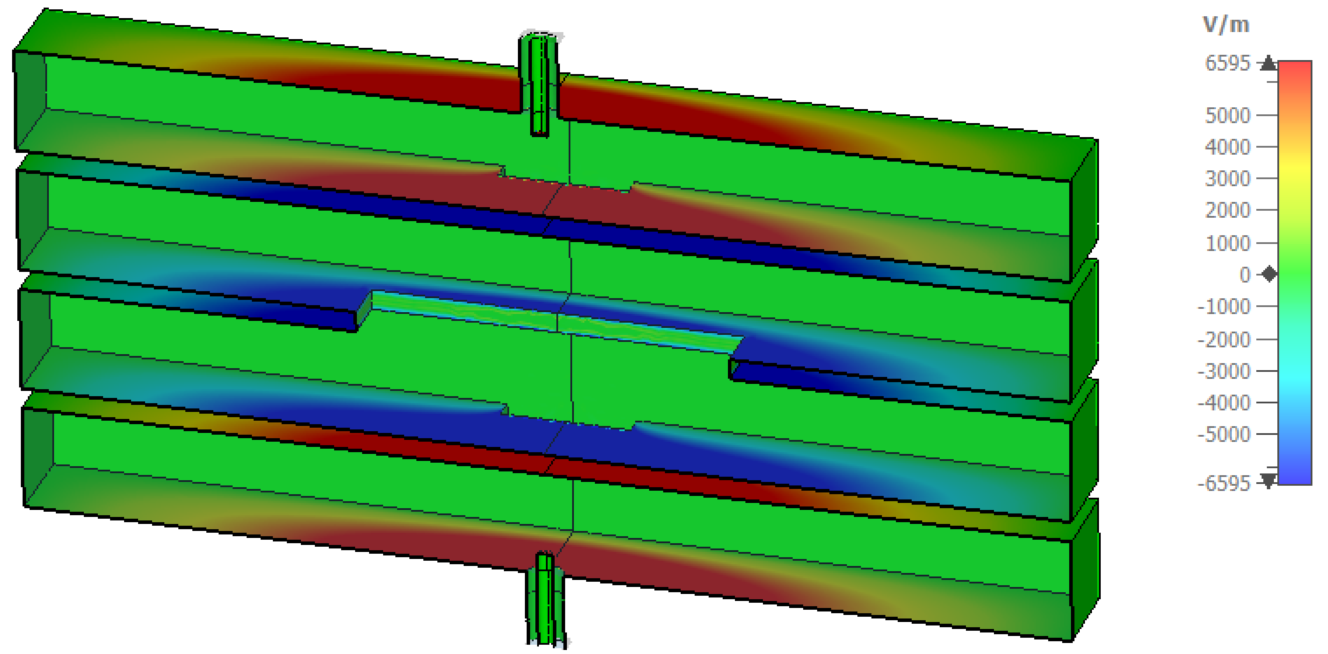}
         \caption{}
         \label{fig:4cav_Longd100mm_VerticalCoupling_Alt_EfieldMode1}
\end{subfigure}
\hfill
\begin{subfigure}[b]{0.49\textwidth}
         \centering
         \includegraphics[width=1\textwidth]{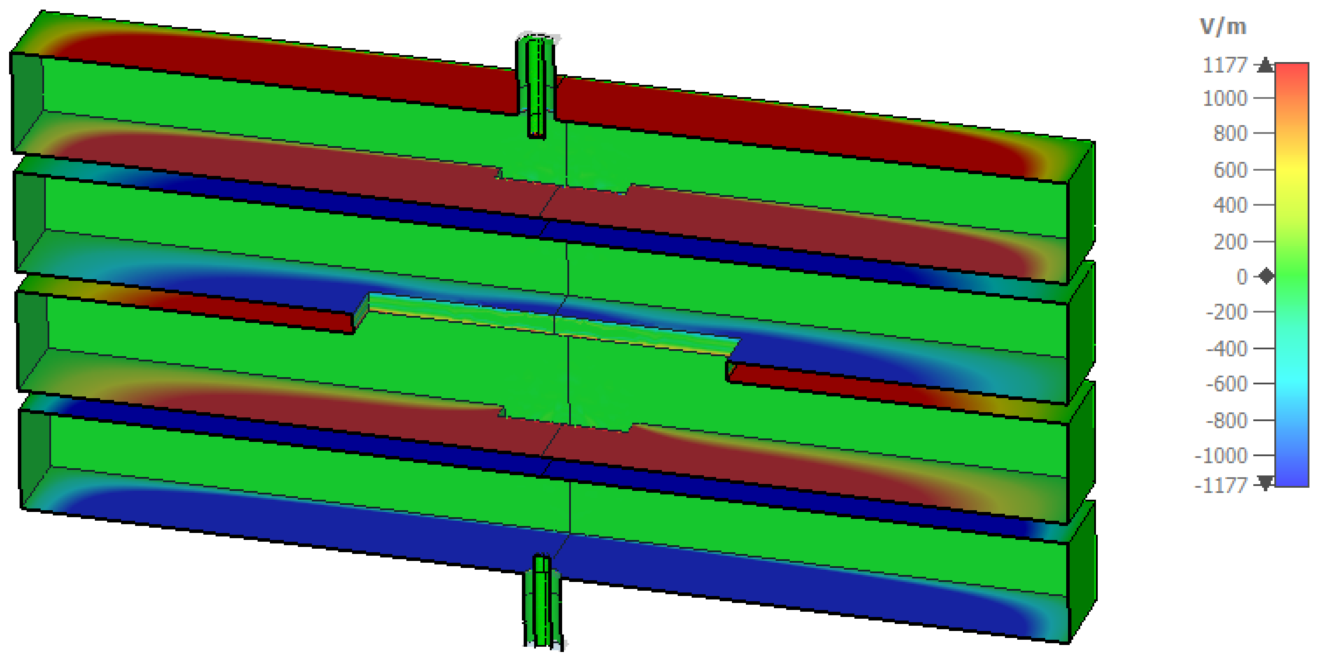}
         \caption{}
         \label{fig:4cav_Longd100mm_VerticalCoupling_Alt_EfieldMode2}
\end{subfigure}
\hfill
\begin{subfigure}[b]{0.49\textwidth}
         \centering
         \includegraphics[width=1\textwidth]{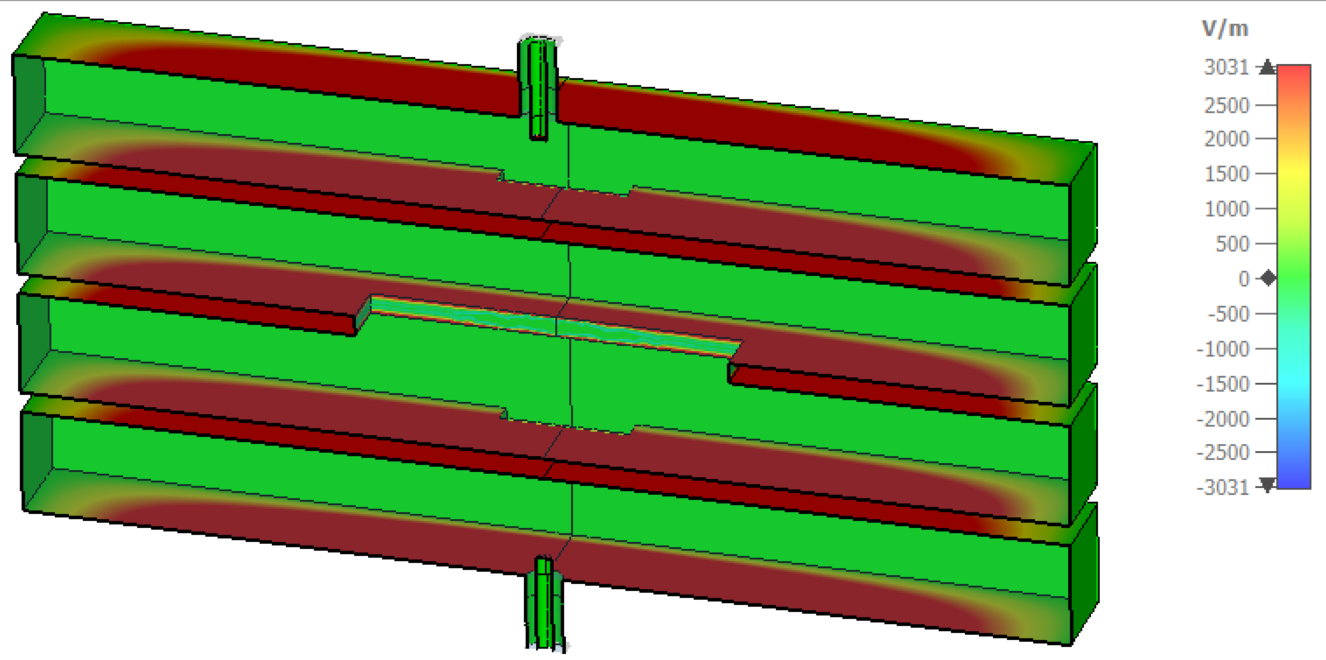}
         \caption{}
         \label{fig:4cav_Longd100mm_VerticalCoupling_Alt_EfieldMode3}
\end{subfigure}
\hfill
\begin{subfigure}[b]{0.49\textwidth}
         \centering
         \includegraphics[width=1\textwidth]{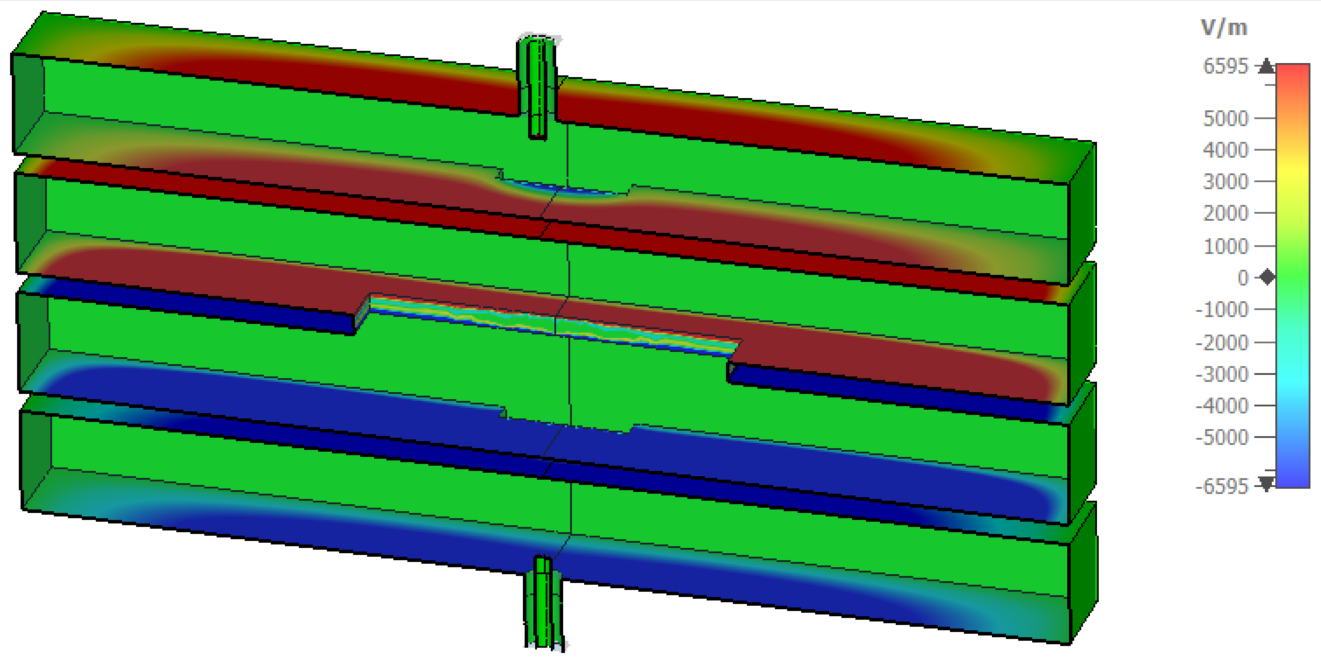}
         \caption{}
         \label{fig:4cav_Longd100mm_VerticalCoupling_Alt_EfieldMode4}
\end{subfigure}
\caption{Magnitude of the electric field of the configuration modes in the structure shown in Figure~\ref{fig:4cav_Longd100mm_VerticalCoupling_Alt}: (a) [+ \text{--} \text{--} +], (b) [+ \text{--} + \text{--}], (c) [+ + + +] and (d) [+ + \text{--} \text{--}], where $"+"$ and $"\text{--}"$ represent a positive and negative E-field level in each subcavity, respectively.}
\label{fig:4cav_Longd100mm_VerticalCoupling_Alt_Efield}
\end{figure}

As it can be seen, the number of resonances inside the working band (from $8$ to $8.8$~GHz) is four, which, as expected, matches with the number of subcavities $N=4$ (four configuration modes of the $TE_{101}$ resonance in the coupled cavity system). If the magnitude of the electric field of these four eigenmodes (see Figure~\ref{fig:4cav_Longd100mm_VerticalCoupling_Alt_Efield}) is observed, the axion mode is identified as the third one (the one with all the subcavities in synchrony \cite{RADES_paper2}), verifying the alternating behaviour (since there is an even number of subcavities).
The configuration eigenmodes associated to the $TE_{101}$ resonant mode that appear in Figure~\ref{fig:4cav_Longd100mm_VerticalCoupling_Alt_Efield} are enumerated in Table~\ref{tab:4cav_Longd100mm_VerticalCoupling_Alt_Modes}, together with the resonant frequencies.\\

\begin{table}[h]
\begin{tabular}{|c|c|c|c|c|c|c|}
\hline
Frequency (GHz) & Configuration\\ \hline\hline
8.231 & [+ \text{--} \text{--} +] \\ \hline
8.326 & [+ \text{--} + \text{--}] \\ \hline
8.439 & [+ + + +] \\ \hline
8.626 & [+ + \text{--} \text{--}] \\ \hline
\end{tabular}
\centering
\caption{\label{tab:4cav_Longd100mm_VerticalCoupling_Alt_Modes} Description of the configuration modes of the $TE_{101}$ mode that appear in Figure~\ref{fig:4cav_Longd100mm_VerticalCoupling_Alt_Efield}.}
\end{table}

Due to the low number of subcavities used ($N=4$) the relative mode separation of this structure ($\Delta f= 1.3$~$\%$ or $113$~MHz) is far from our limits. The proximity of the next resonant mode ($TE_{102}$) is not a relevant issue because it is even further in frequency (and outside the frequency range represented) due to the moderate $d$ value.\\

The measured resonant frequency of the axion mode is $f=8.439$~GHz, which is a good result compared to the goal $8.4$~GHz. The quality and form factors are $Q_0=39579$ and $C=0.654$, respectively, which is in line with the results from Figure~\ref{fig:Parameters_3couplingDirections} for the vertical coupling option in inductive and capacitive irises for lengths of $d=100$~mm ($V\approx37$~mL for two subcavities). This validates the theoretical analysis presented in this section. The resulting total volume of the haloscope is $V=74$~mL and the total $Q_0 \times V \times C$ is $1915.47$~L, which is $9.56$ times that of a single standard WR-90 cavity.

\subsection{Tall subcavities}
\label{SubSec:MulticavTall}
\begin{figure}[h]
\centering
\begin{subfigure}[b]{0.69\textwidth}
         \centering
         \includegraphics[width=0.45\textwidth]{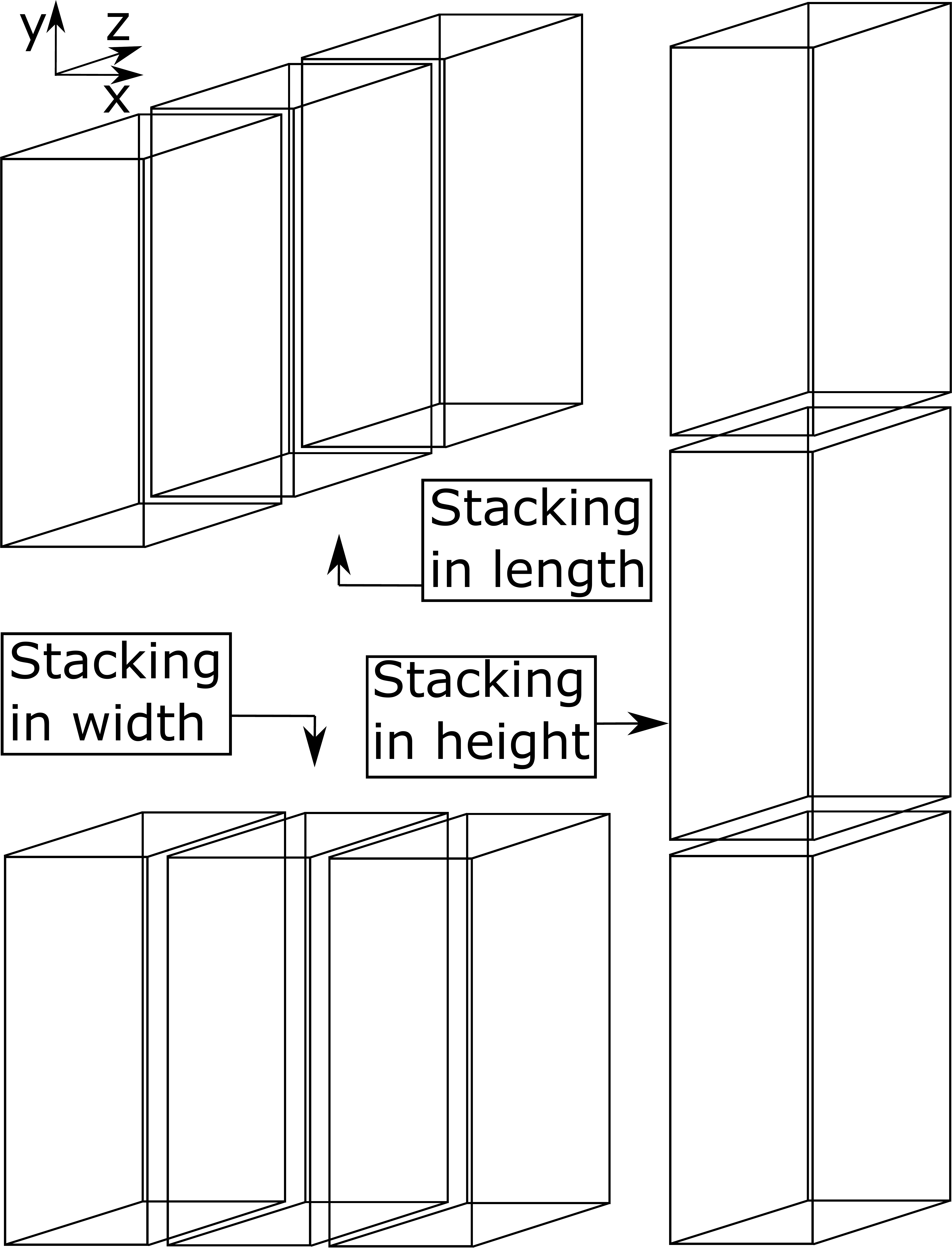}
         \caption{}
         \label{fig:MulticavitiesStackedInSeveralDirections_Tall}
\end{subfigure}
\hfill
\begin{subfigure}[b]{0.32\textwidth}
         \centering
         \includegraphics[width=1\textwidth]{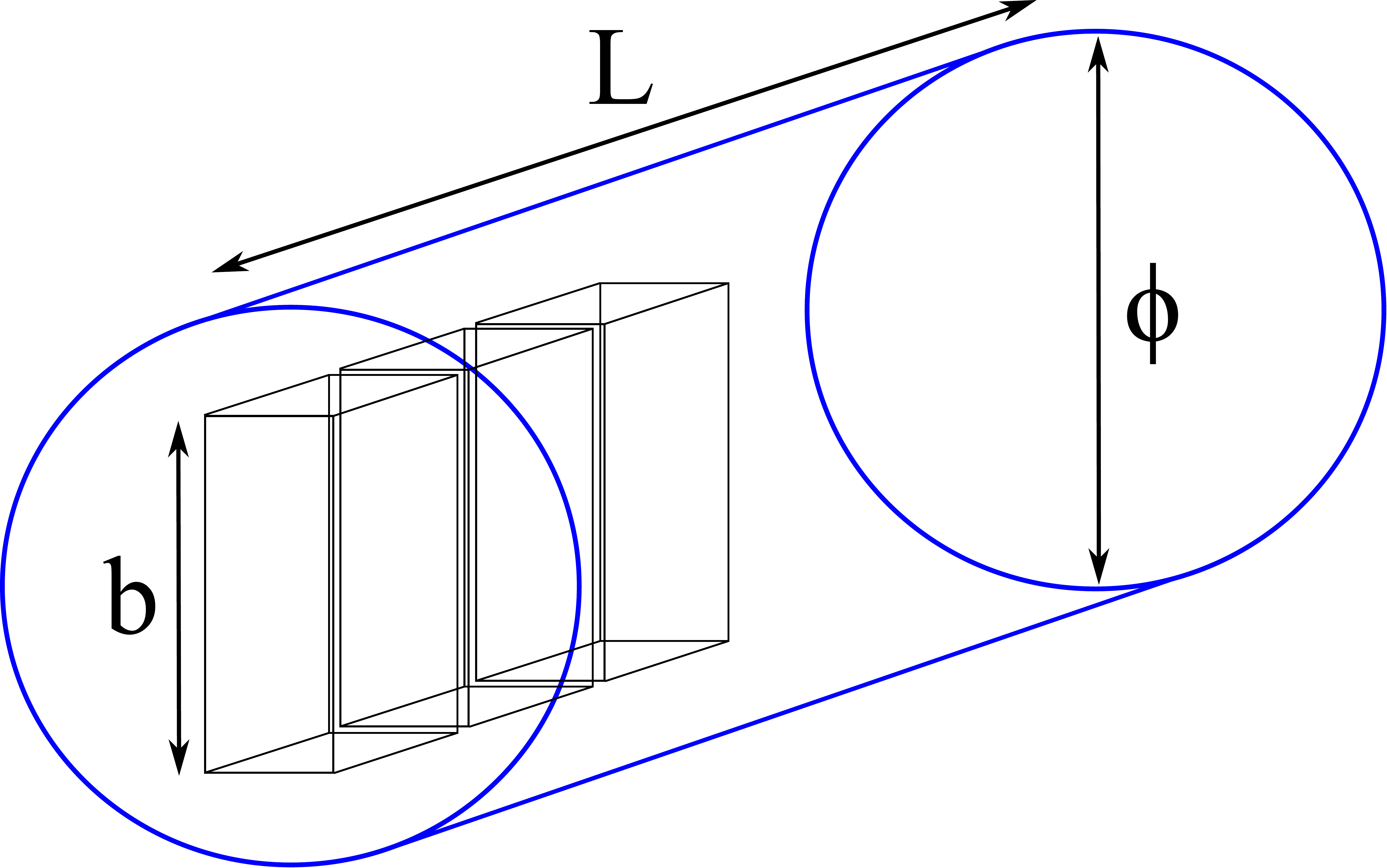}
         \caption{}
         \label{fig:MulticavitiesStackedInLong_Tall_Dipole}
\end{subfigure}
\hfill
\begin{subfigure}[b]{0.32\textwidth}
         \centering
         \includegraphics[width=1\textwidth]{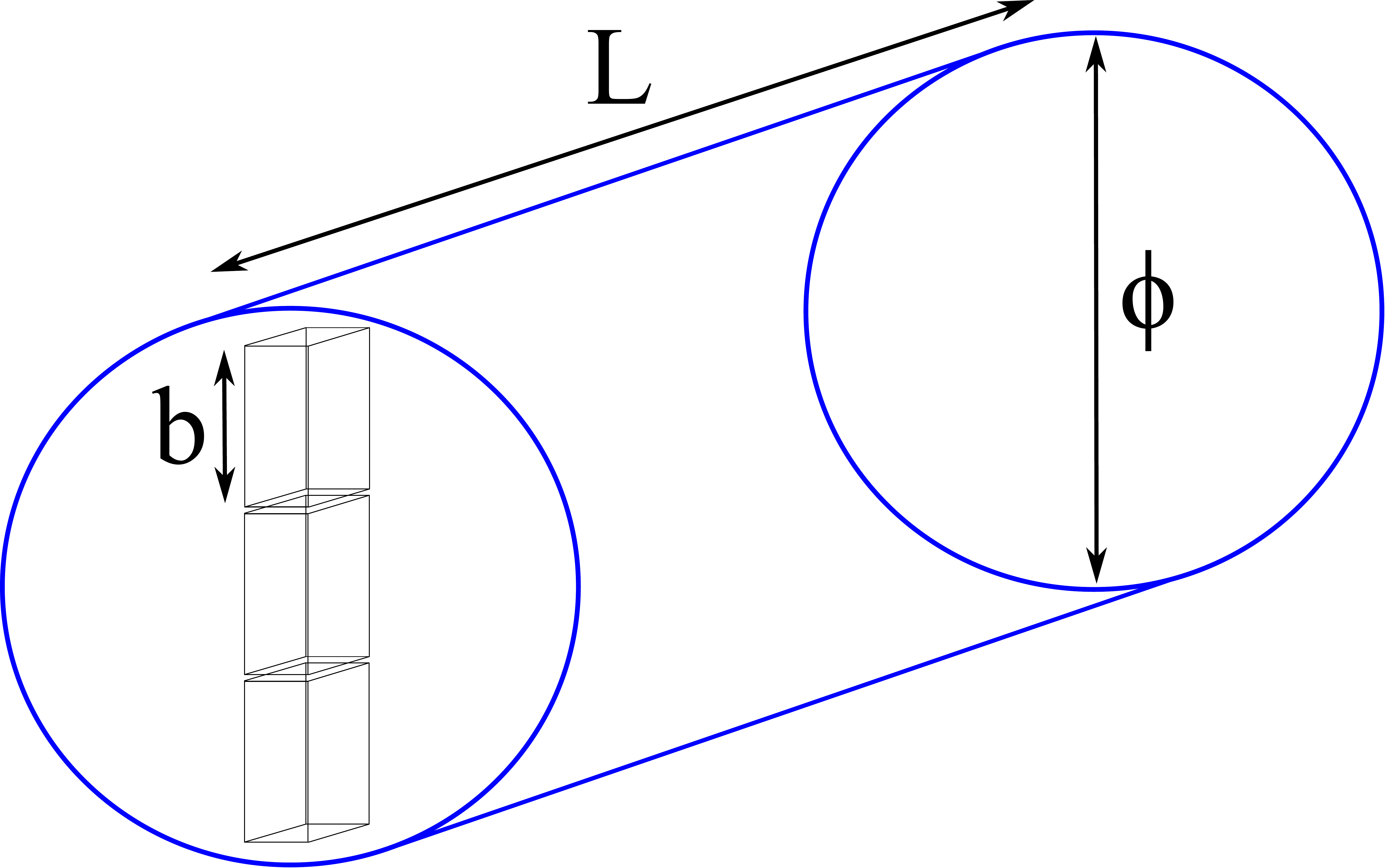}
         \caption{}
         \label{fig:MulticavitiesStackedInHeight_Tall_Dipole}
\end{subfigure}
\hfill
\begin{subfigure}[b]{0.32\textwidth}
         \centering
         \includegraphics[width=1\textwidth]{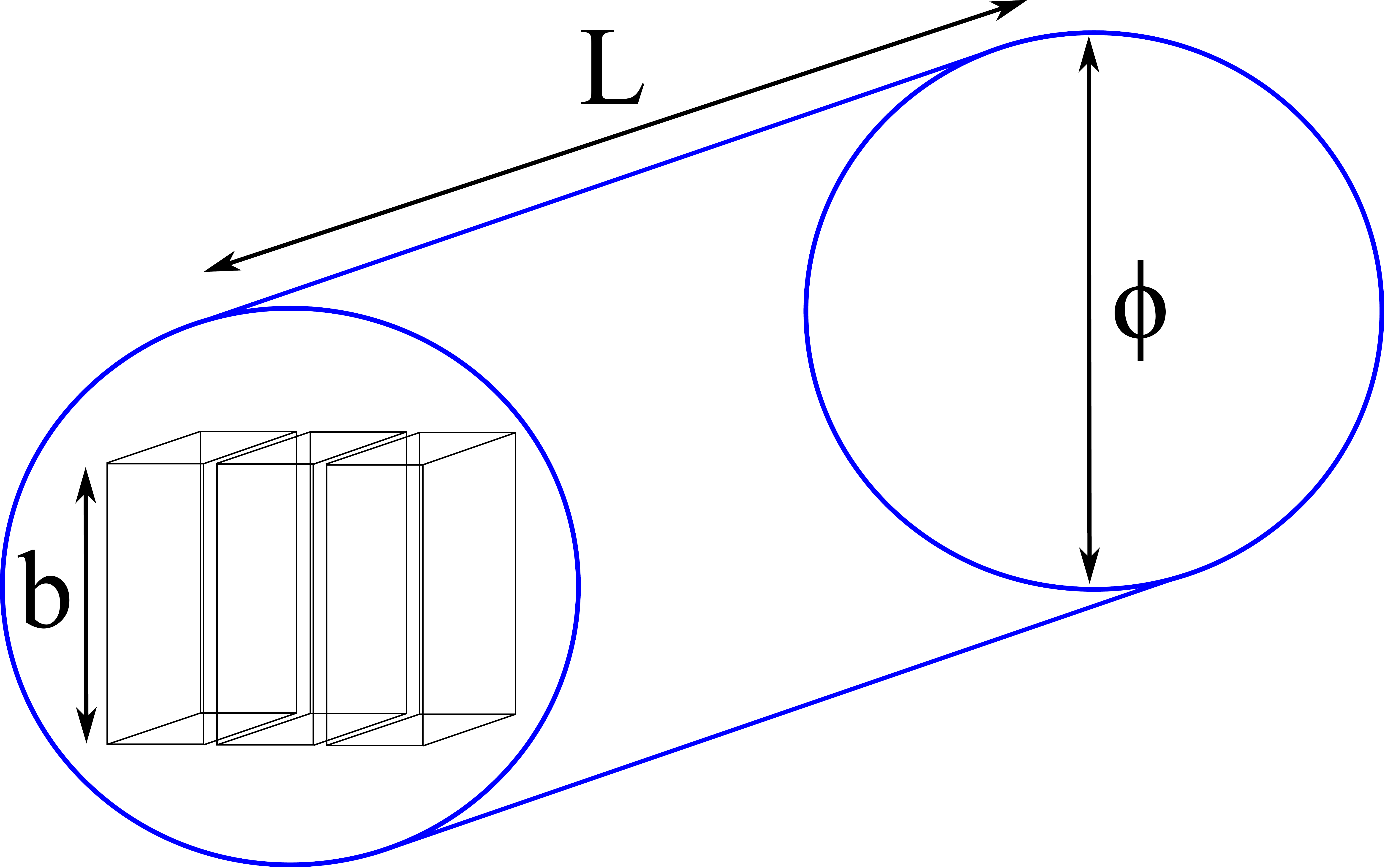}
         \caption{}
         \label{fig:MulticavitiesStackedInWidth_Tall_Dipole}
\end{subfigure}
\hfill
\begin{subfigure}[b]{0.2\textwidth}
         \centering
         \includegraphics[width=1\textwidth]{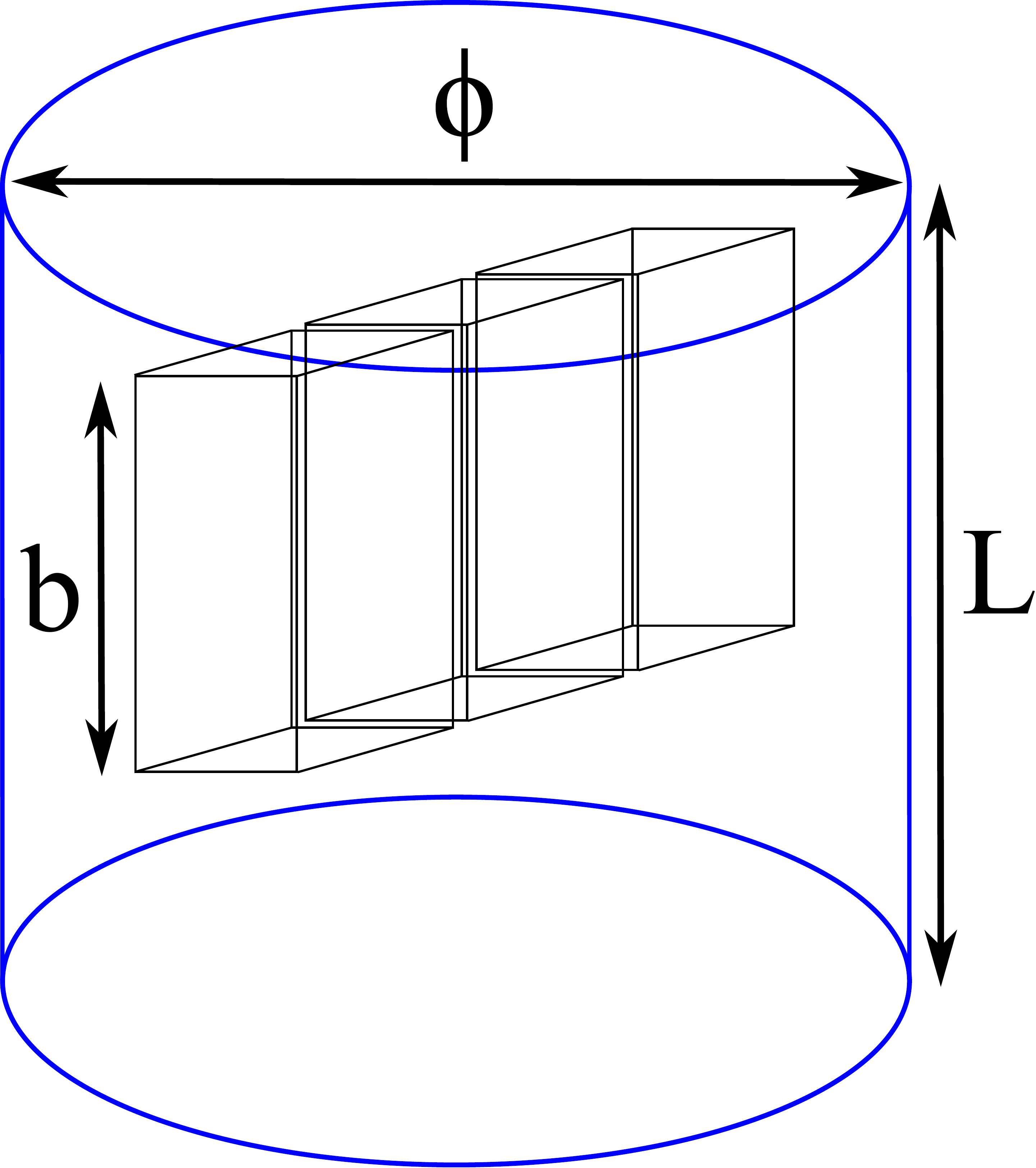}
         \caption{}
         \label{fig:MulticavitiesStackedInLong_Tall_Solenoid}
\end{subfigure}
\hfill
\begin{subfigure}[b]{0.2\textwidth}
         \centering
         \includegraphics[width=1\textwidth]{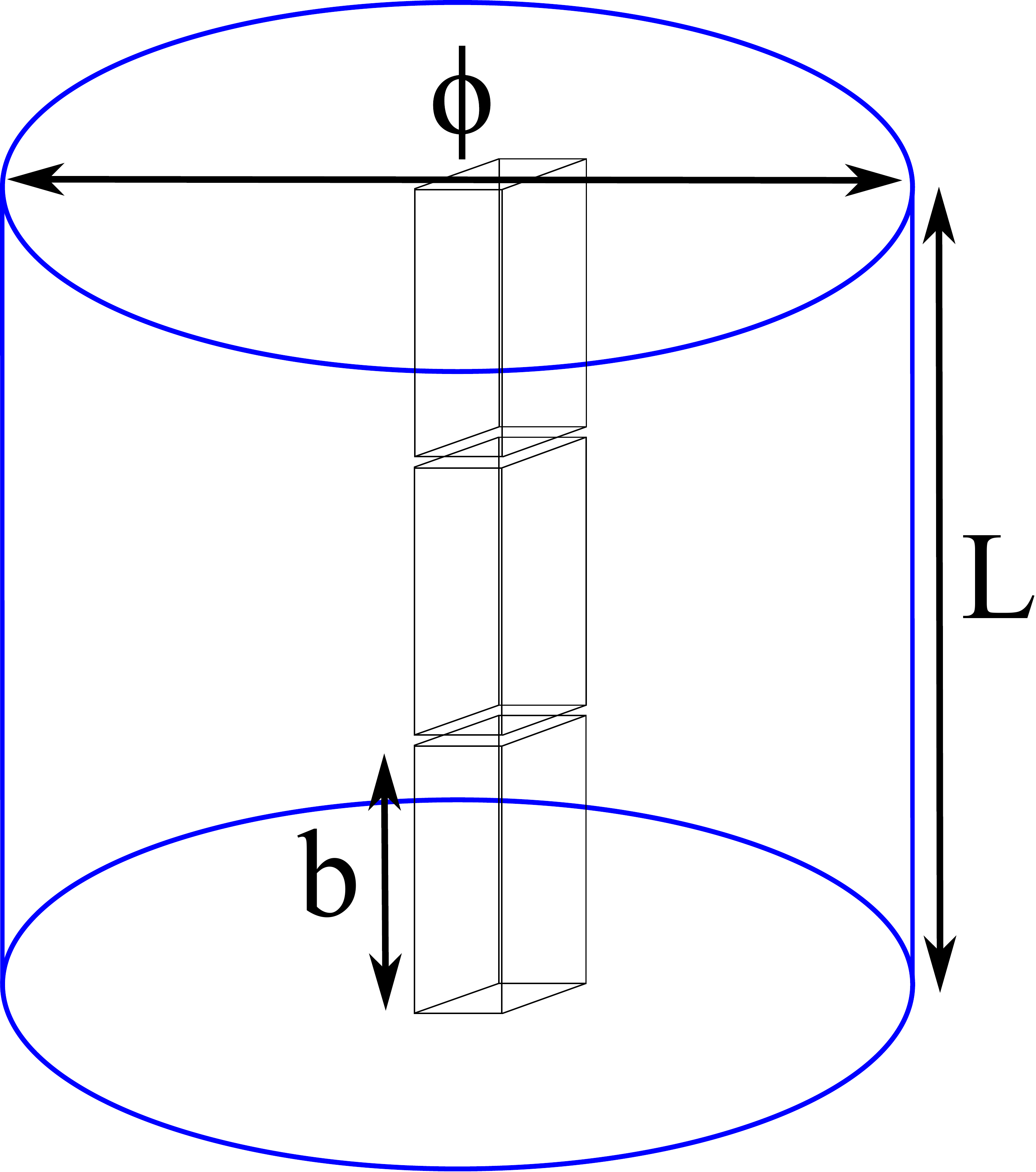}
         \caption{}
         \label{fig:MulticavitiesStackedInHeight_Tall_Solenoid}
\end{subfigure}
\hfill
\begin{subfigure}[b]{0.2\textwidth}
         \centering
         \includegraphics[width=1\textwidth]{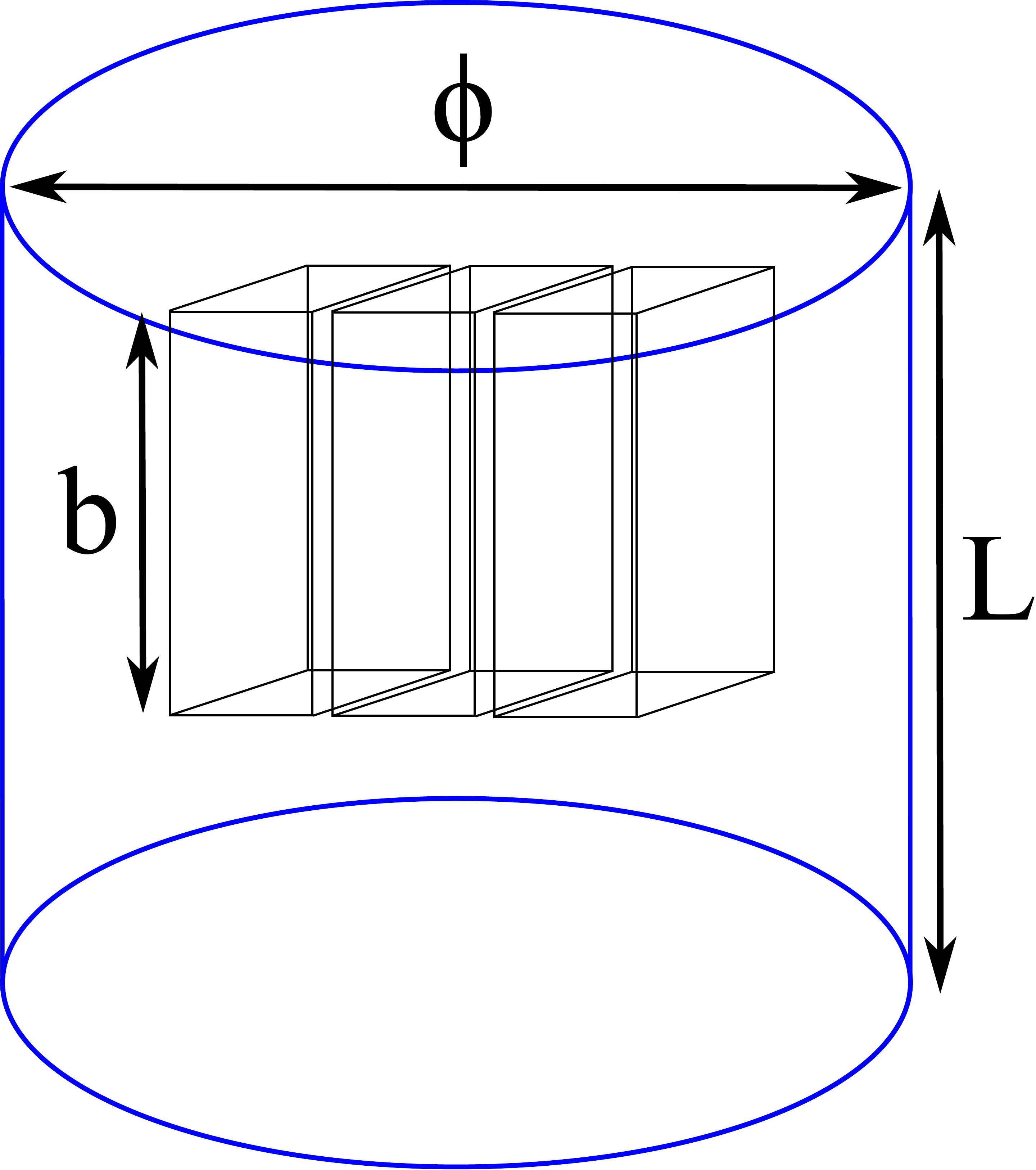}
         \caption{}
         \label{fig:MulticavitiesStackedInWidth_Tall_Solenoid}
\end{subfigure}
\caption{(a) Possibilities to stack three tall subcavities in different directions to create a multicavity. Implementation of these multicavity stackings in dipole and solenoid magnets. For dipole magnets: (b) in length, (c) in height, and (d) in width. For solenoid magnets: (e) in length, (f) in height, and (g) in width.}
\label{fig:StakingTallSubcavitiesInDipolesAndSolenoids}
\end{figure}

As discussed in section~\ref{SubSec:IndTall} the tall structure is a good alternative for increasing the volume of haloscopes. This concept can be applied also in 1D multicavity structures increasing the vertical dimension of the subcavities. As in the case of the long subcavities in the previous section, there are three possibilities for coupling (or stacking) the subcavities in a 1D multicavity structure: in length, in height or in width. Figure~\ref{fig:MulticavitiesStackedInSeveralDirections_Tall} depicts these types of stackings in a multicavity based on three tall subcavities.
From Figure~\ref{fig:MulticavitiesStackedInLong_Tall_Dipole} to Figure~\ref{fig:MulticavitiesStackedInWidth_Tall_Solenoid} examples for the three stacking directions in both dipole and solenoid magnets are displayed.\\

For dipole magnets, the tall subcavities stacked in length (Figure~\ref{fig:MulticavitiesStackedInLong_Tall_Dipole}) have to orient their heights towards the $y-$axis which implies a limitation in the $b$ value ($\phi_{BabyIAXO}=600$~mm), while the array stacking direction has to be oriented towards the longitudinal bore axis yielding to a great freedom to increase the number of subcavities $N$ ($L_{BabyIAXO}=10$~m). In the case of tall subcavities stacked in height for dipole magnets (Figure~\ref{fig:MulticavitiesStackedInHeight_Tall_Dipole}) both the subcavity heights and array should be oriented towards the $y-$axis, so a serious restriction is presented in both $b$ and $N$ ($\phi_{BabyIAXO}=600$~mm limiting the total haloscope height). In case of necessity of using 1D multicavites with stacking in height, a multicavity based on not very tall subcavities (15 subcavities of $b=40$~mm, for example) could be designed although it will be shown below that there are more efficient solutions to increase the volume of a multicavity in dipole magnets. Finally, for tall multicavities stacked in width for dipoles (Figure~\ref{fig:MulticavitiesStackedInWidth_Tall_Dipole}) both the subcavity heights and array should be oriented on the transverse plane of the bore, but in different directions (observing Figure~\ref{fig:Dipole}, $x-$axis for the stacking direction and $y-$axis for the subcavity heights). This situation is very similar to the long subcavities stacked in width for solenoids (see Figure~\ref{fig:MulticavitiesStackedInWidth_Long_Solenoid}) so analogous considerations are applied here for the limitation in the $b$ and $N$ values taking into account the cylindrical shape of the bore.\\

For solenoid magnets, the tall subcavities stacked in length (Figure~\ref{fig:MulticavitiesStackedInLong_Tall_Solenoid}) have to orient their heights towards the longitudinal bore axis (maximum cavity height value of $b=L_{MRI}=800$~mm) and the array stacking direction towards any radial bore axis ($N$ limited to $\phi_{MRI}=650$~mm). In the case of tall subcavities stacked in height for solenoid magnets (Figure~\ref{fig:MulticavitiesStackedInHeight_Tall_Solenoid}), both the subcavity heights and array should be oriented towards the longitudinal bore axis, so a serious restriction is presented in both $b$ and $N$ ($L_{MRI}=800$~mm for the total haloscope height). This situation is not efficient, similarly to the long multicavities using the length for the stacking of the subcavities for solenoids (Figure~\ref{fig:MulticavitiesStackedInLong_Long_Solenoid}), since the greatest solenoid height from Table~\ref{tab:magnets} is $L=800$~mm and this can easily be achieved with one single cavity. The rest of the stacking directions are more suitable for this type of multicavities as both the length and a radial axis are utilised. Finally, for tall multicavities stacked in width in solenoids (Figure~\ref{fig:MulticavitiesStackedInWidth_Tall_Solenoid}), the situation is the same that the tall subcavities stacked in length.\\

\begin{figure}[h]
\centering
\begin{subfigure}[b]{0.49\textwidth}
         \centering
         \includegraphics[width=1\textwidth]{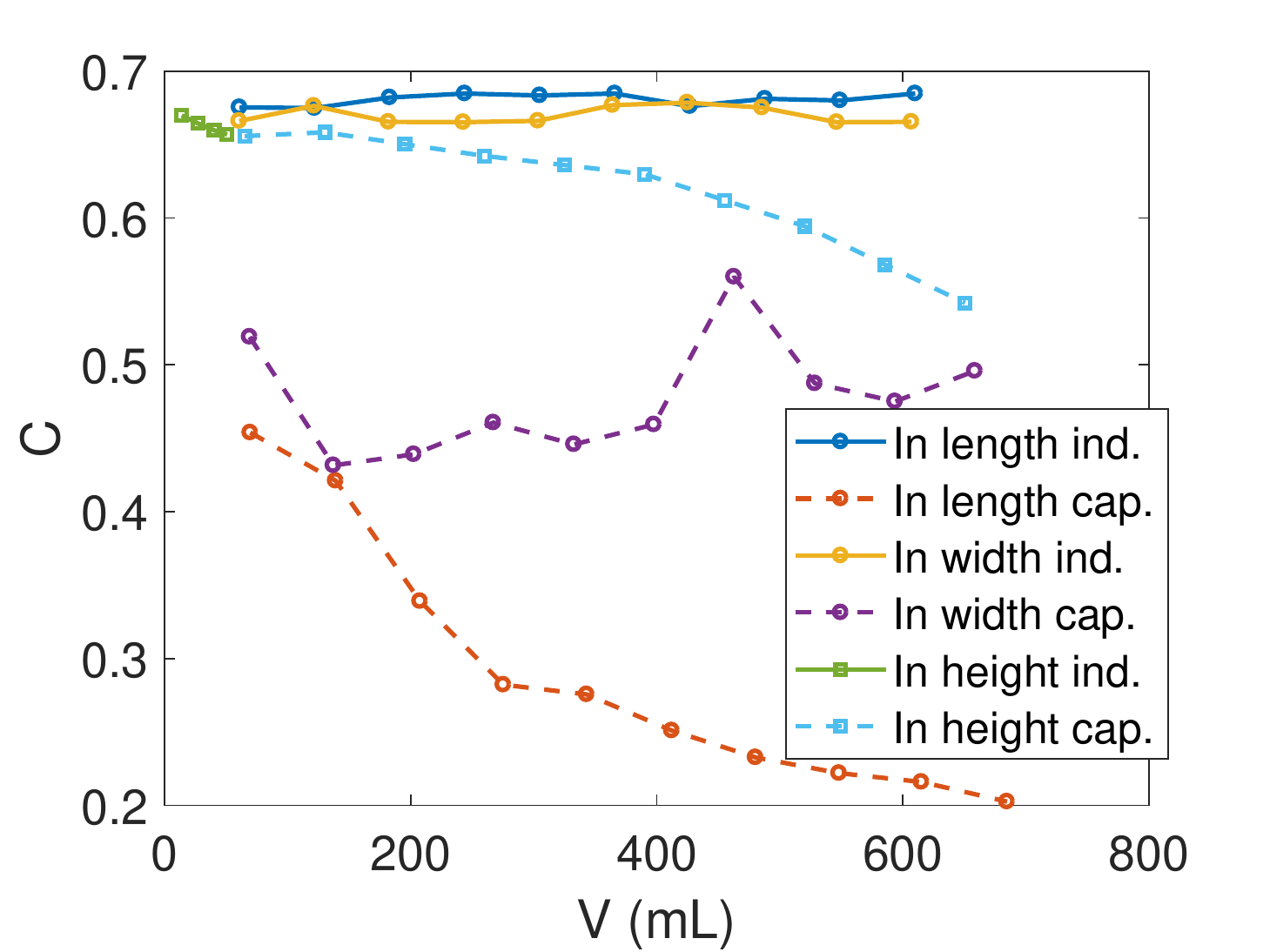}
         \caption{}
         \label{fig:C_3couplingDirections_Tall}
\end{subfigure}
\hfill
\begin{subfigure}[b]{0.49\textwidth}
         \centering
         \includegraphics[width=1\textwidth]{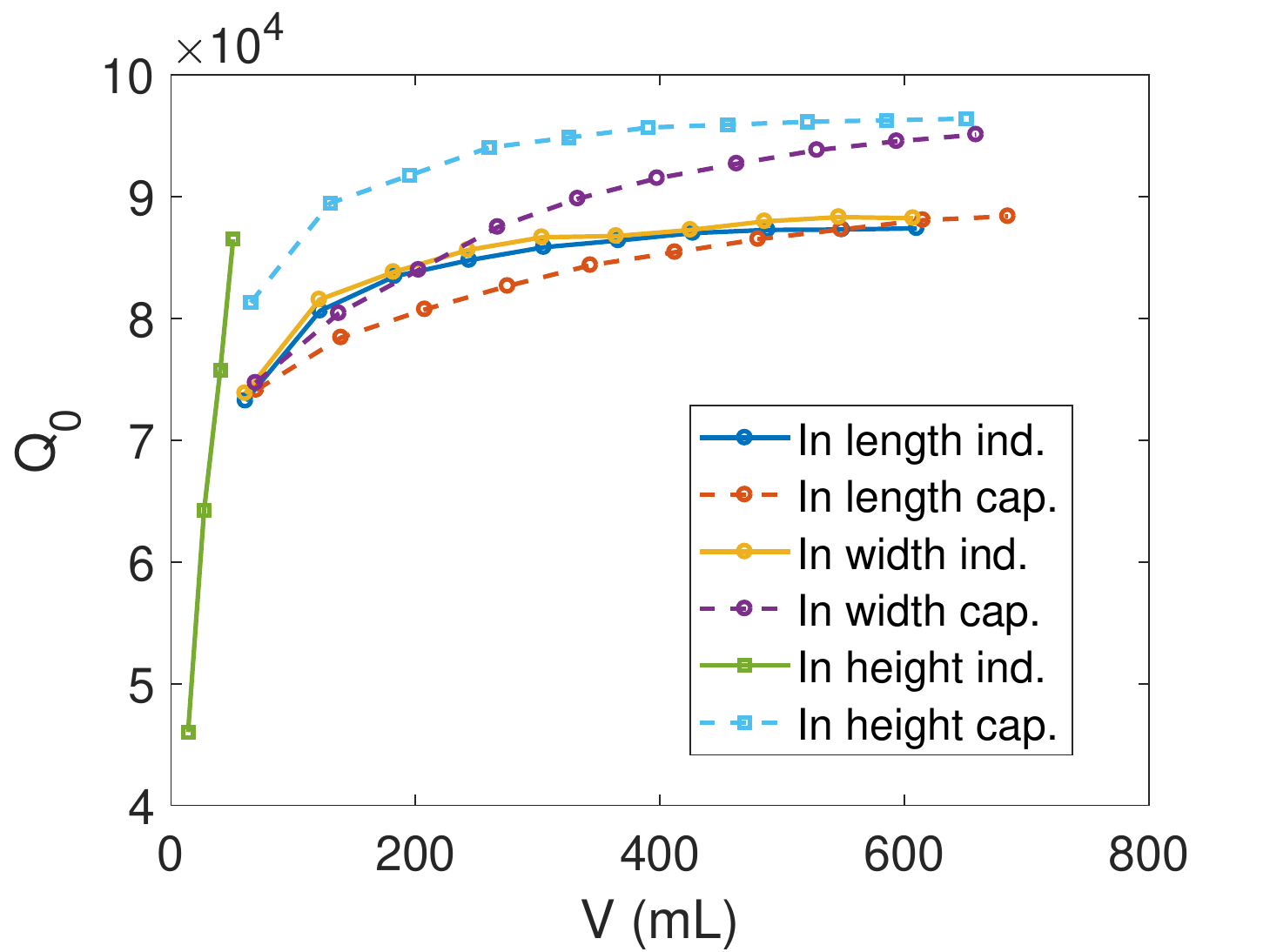}
         \caption{}
         \label{fig:Q0_3couplingDirections_Tall}
\end{subfigure}
\hfill
\begin{subfigure}[b]{0.49\textwidth}
         \centering
         \includegraphics[width=1\textwidth]{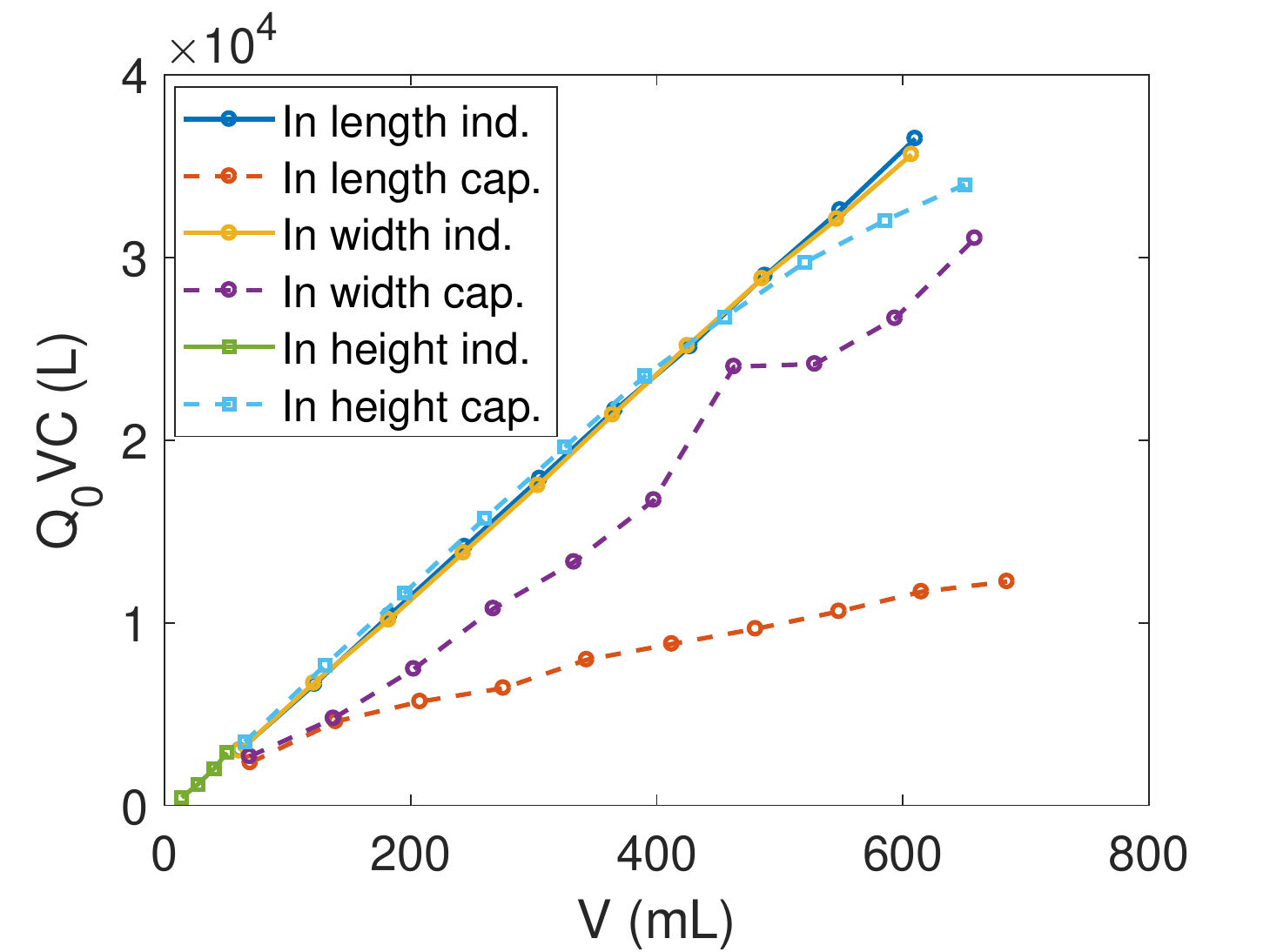}
         \caption{}
         \label{fig:QVC_3couplingDirections_Tall}
\end{subfigure}
\caption{Influence on the design parameters of each type of coupling (inductive or capacitive) when the coupling is introduced along each direction (longitudinal (or in length), in width and vertical (or in height)) in a structure composed of two coupled tall subcavities: (a) form factor, (b) quality factor, and (c) $Q_0 \times V \times C$ factor. The volume depends only on the height as the number of subcavities is fixed to two.}
\label{fig:Parameters_3couplingDirections_Tall}
\end{figure}

A similar study as that of the previous section has been carried out for the direction of coupling in the three axis (in length, in height and in width) employing a physical coupling value of $|k|=0.025$, and increasing the height of the subcavities in a prototype based on $N=2$ subcavities. In Figure~\ref{fig:Parameters_3couplingDirections_Tall} the behaviour of these types of couplings versus the multicavity volume (which depends only on the height of the subcavities $b$) is shown.\\

Figure~\ref{fig:Parameters_3couplingDirections_Tall} shows that the coupling along the vertical direction with inductive iris has a height limitation with volume values around $50$~mL to provide for the correct coupling value. The situation is similar to what happens for long subcavities with the in-length coupling direction. For the other options, there is no such restriction and the limit is imposed by the mode clustering issue. Figure~\ref{fig:QVC_3couplingDirections_Tall} shows that the in-lenght and in-width coupling with inductive irises and the in-height coupling direction with capacitive irises are the best options for the tall subcavities due to its high $Q_0 \times V \times C$ values when $b$ (or the volume $V$) increases.\\

Regardless of the magnet dimensions, the limit in the number of subcavities $N$ and in the subcavity height $b$ is imposed by the same criteria as the single cavities (as was the case for long multicavities): the mode separation described in section~\ref{SubSec:IndLong}, but according to the results provided in Figure~\ref{fig:Allind_vs_Allcap_vs_Alt_vs_Indi_ModeMixing}. Nevertheless, as stated previously, the size of the magnet bores is generally much smaller than the haloscope limits for any stacking direction in a X band multicavity design.\\

\begin{figure}[h]
\centering
\begin{subfigure}[b]{0.59\textwidth}
         \centering
         \includegraphics[width=1\textwidth]{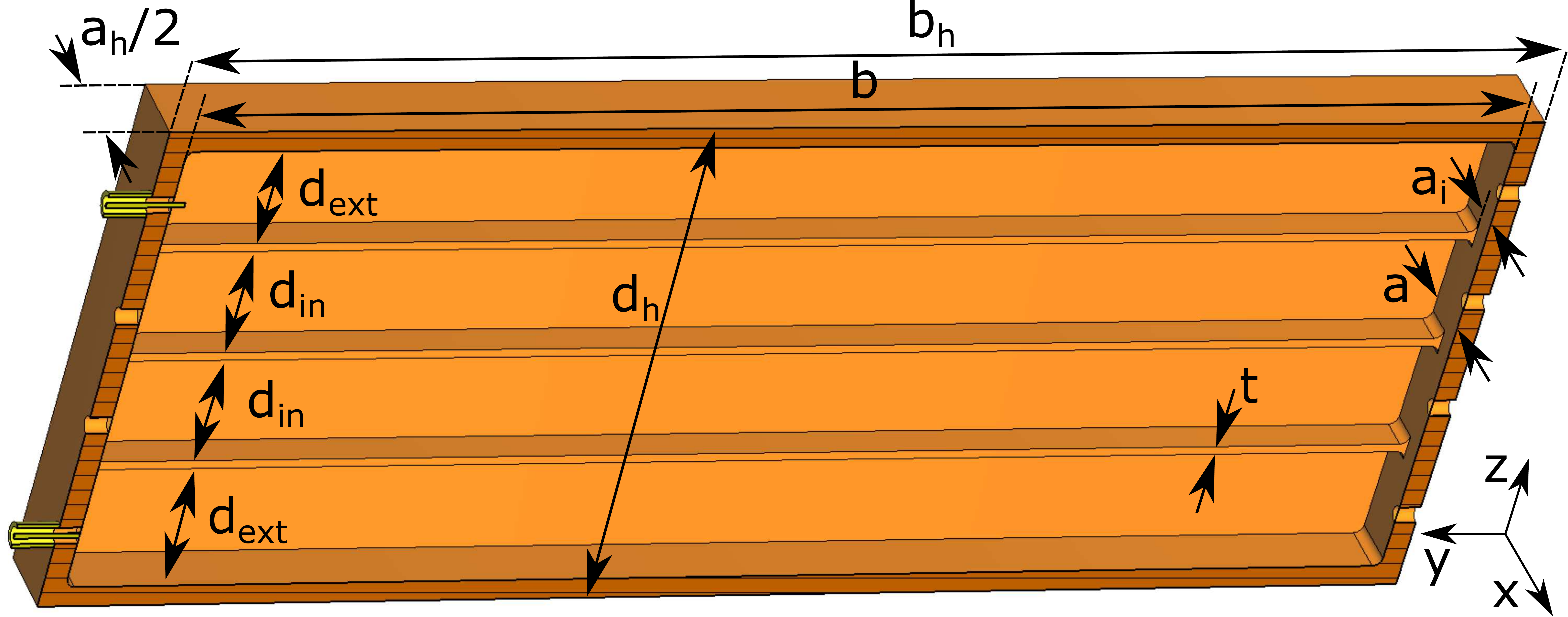}
         \caption{}
         \label{fig:4cav_Allind_TallSubcavities_Model}
\end{subfigure}
\hfill
\begin{subfigure}[b]{0.39\textwidth}
         \centering
         \includegraphics[width=1\textwidth]{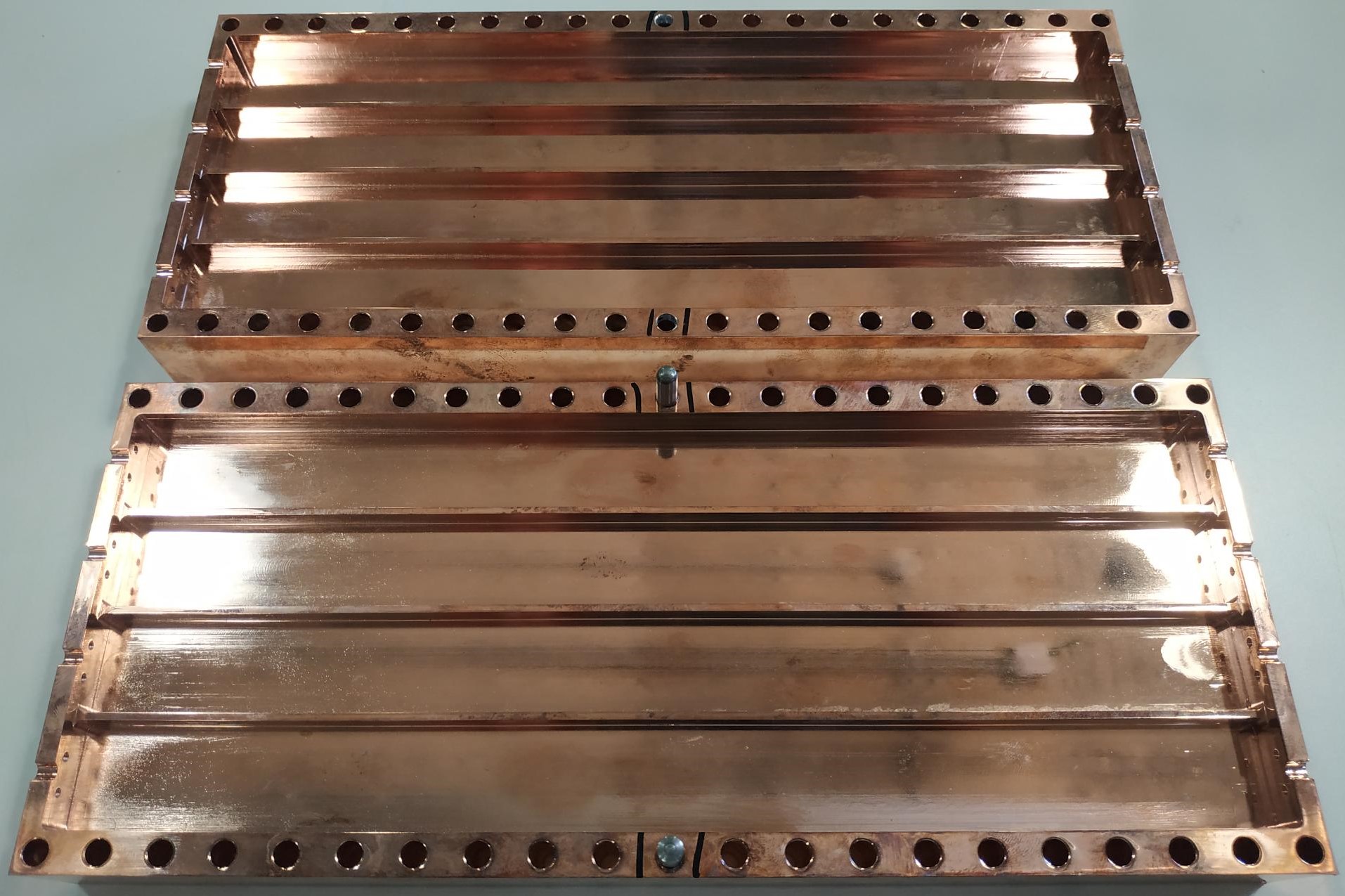}
         \caption{}
         \label{fig:4cav_Allind_TallSubcavities_Manufactured}
\end{subfigure}
\hfill
\begin{subfigure}[b]{0.65\textwidth}
         \centering
         \includegraphics[width=1\textwidth]{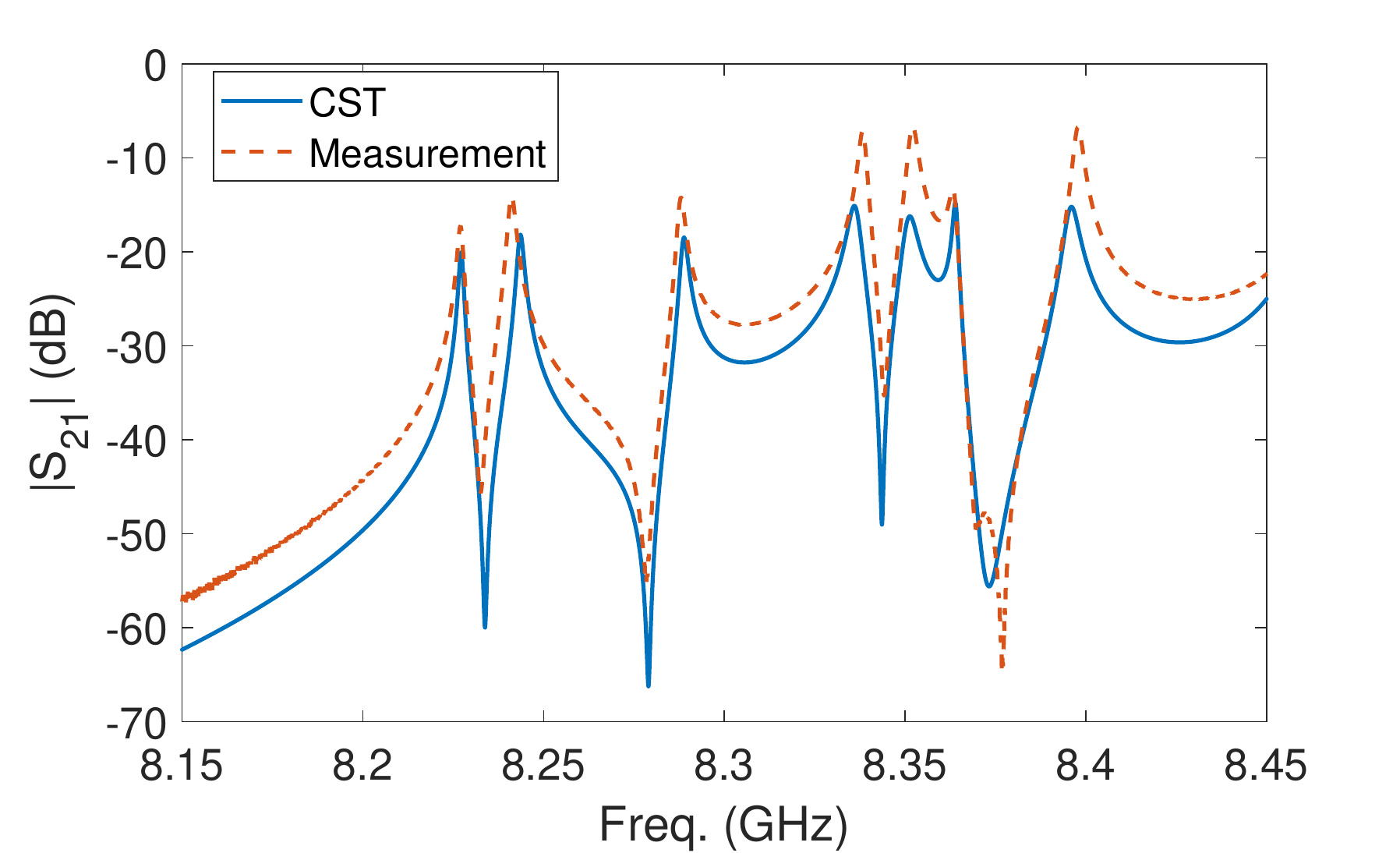}
         \caption{}
         \label{fig:4cav_Allind_TallSubcavities_pS_CST_vs_Meas}
\end{subfigure}
\caption{Manufactured haloscope prototype based on four tall subcavities with three inductive irises employing the longitudinal coupling direction: (a) model of one of the two symmetrical halves, (b) picture of the manufactured structure divided in two symmetrical halves, and (c) $S_{21}$ scattering parameter magnitude as a function of the frequency (simulation versus real measurements at room temperature).}
\label{fig:4cav_Allind_TallSubcavities}
\end{figure}

The design and manufacturing of an all-inductive 1D multicavity of $N=4$ tall subcavities with $b=300$~mm employing the in length stacking direction (Figure~\ref{fig:4cav_Allind_TallSubcavities_Model}) has been carried out for validation.
This structure could fit perfectly in both BabyIAXO dipole and MRI solenoid magnets (with the orientations depicted in Figures~\ref{fig:MulticavitiesStackedInLong_Tall_Dipole} and \ref{fig:MulticavitiesStackedInLong_Tall_Solenoid}, respectively).\\

Again, the selected $|k|$ value for the physical couplings is $0.025$ and the extracted coupling matrix is the following:
\begin{gather}\label{eq:CouplingMatrix_1D_TallExample}
 \bf{M} =
  \begin{pmatrix}
   0.5 & -0.5 & 0 & 0 \\
   -0.5 & 1 & -0.5 & 0 \\
   0 & -0.5 & 1 & -0.5 \\
   0 & 0 & -0.5 & 0.5
   \end{pmatrix}
\end{gather}
The signs of the $M_{12}$, $M_{23}$ and $M_{34}$ elements (and their symmetrical pairs) are negative because the structure is based on all inductive irises.\\

In Figure~\ref{fig:4cav_Allind_TallSubcavities_Manufactured} a picture of the manufactured prototype is shown. Figure~\ref{fig:4cav_Allind_TallSubcavities_pS_CST_vs_Meas} plots the $S_{21}$ scattering parameter magnitude as a function of the frequency for both simulation and measurements, which are in good agreement.\\

The dimensions of the prototype are: height and width of all the subcavities $b=300$~mm and $a=22.86$~mm, respectively, internal subcavities lengths $d_{in}=26$~mm, external subcavities lengths $d_{ext}=27$~mm, inductive width $a_i=9$~mm and thickness of the irises $t=2$~mm. The total dimension of the haloscope taking into account the external copper thickness of $t_{Cu}=5$~mm are width $a_h=32.86$~mm, height $b_h=310$~mm and length $d_h=122$~mm.\\

From simulations, considering copper walls, a quality factor value of $Q_0^{2K}=76000$ is obtained for the axion mode (at $8.227$~GHz) at cryogenic temperatures, and $Q_0^{300K}=13200$ at room temperature. The measurements from the manufactured structure provide a value of $Q_0^{300K}=7300$ ($55.3\%$ of the simulation result), which corresponds with a typical reduction of manufactured $Q_0$ compared with other RADES structures \cite{RADES_paper1}. The degradation in $Q_0$ can be explained from different reasons, mainly due to manufacturing roughness and RF surface current discontinuities as a result of the fabrication cuts.\\

Regarding the form factor, a value of $C=0.625$ has been obtained (from simulations), which can be further increased with an optimisation process. The configuration modes associated to the modes (resonances) that appear in the plot are enumerated in Table~\ref{tab:4cav_Allind_TallSubcavities_Modes}.\\

\begin{table}[h]
\begin{tabular}{|c|c|c|c|c|c|c|}
\hline
Freq. (GHz) & Resonant mode & Configuration\\ \hline\hline
8.227 & $TE_{101}$ & [+ + + +] \\ \hline
8.241 & $TE_{111}$ & [+ + + +] \\ \hline
8.288 & $TE_{121}$ & [+ + + +] \\ \hline
8.338 & $TE_{101}$ & [+ + \text{--} \text{--}] \\ \hline
8.352 & $TE_{111}$ & [+ + \text{--} \text{--}] \\ \hline
8.363 & $TE_{131}$ & [+ + + +] \\ \hline
8.398 & $TE_{121}$ & [+ + \text{--} \text{--}] \\ \hline
\end{tabular}
\centering
\caption{\label{tab:4cav_Allind_TallSubcavities_Modes} Description of the configuration and resonant modes that appear in Figure~\ref{fig:4cav_Allind_TallSubcavities_pS_CST_vs_Meas} from real measurements.}
\end{table}

The relative mode separation of this structure is $\Delta f= 0.17$~$\%$ ($14.3$~MHz) which matches with Figure~\ref{fig:ModeMixing_vs_b-a} for $b/a=300/22.86=13.12$. The proximity of the next configuration mode ([+ + \text{--} \text{--}]) is not a problem because it is relatively far in frequency ($1.35$~$\%$ or $111.4$~MHz), and given the low number of subcavities ($N=4$), this is expected. The resulting total volume of the haloscope is $V=743$~mL, and the total $Q_0 \times V \times C$ is $3.53\times 10^4$~L, which is $176.14$ times that of a single standard WR-90 cavity.\\

Tall structures provide an additional benefit to alleviate the mode clustering issue: the presence of transmission zeros. They are created due to the interaction between cavity higher order modes when they are close in frequency. For example, a transmission zero appears at the right side of the axion mode ($TE_{101}$) due to the interaction between this mode and the $TE_{111}$ resonance (phase cancellation between signals coupled to both modes \cite{Guglielmi:1995}).

\subsection{Large subcavities}
\label{SubSec:MulticavLongAndTall}
A powerful strategy to increase the volume of a haloscope is to combine all the previous ideas: 1D multicavity concept with large (that is, long and tall) subcavities. Regarding the best stacking direction in large multicavities, it is expected to be very similar to that in long multicavities and tall multicavities. Considering only the mode separation limits, and observing Figures~\ref{fig:Parameters_3couplingDirections} and \ref{fig:Parameters_3couplingDirections_Tall}, it can be concluded that the coupling direction option with the best $Q_0 \times V \times C$ factor value is the vertical (or in height) one with capacitive iris for both dipole and solenoid magnets. However, the best stacking direction option also depends on the magnet employed due to its dimensions taking into account the illustrations shown in Figures~\ref{fig:StakingLongSubcavitiesInDipolesAndSolenoids} and \ref{fig:StakingTallSubcavitiesInDipolesAndSolenoids}.\\

For example, in the BabyIAXO dipole magnet a long and tall multicavity structure stacked in width (similar to the cases from Figures~\ref{fig:MulticavitiesStackedInWidth_Long_Dipole} and \ref{fig:MulticavitiesStackedInWidth_Tall_Dipole} but with long and tall subcavities) with $a=17.85$~mm, $b=\frac{\phi_{BabyIAXO}}{2}\sqrt{2}-2t_{Cu}=414.26$~mm, and $d=1600$~mm (limited to avoid mode clustering issues, as depicts Table~\ref{tab:IndLongAndTallQVC}) with $N=\lfloor \frac{b}{a+t} \rfloor=20$ subcavities (where $t=2$~mm is the thickness of the irises) could be implemented, which implies a volume of $V=236.75$~L. In the MRI (AMDX-EFR) solenoid magnet a multicavity stacked in width (similar to the cases from Figures~\ref{fig:MulticavitiesStackedInWidth_Long_Solenoid} and \ref{fig:MulticavitiesStackedInWidth_Tall_Solenoid} but with long and tall subcavities) with $a=17.85$~mm, $b=L_{MRI}=800$~mm, and $d=\frac{\phi_{MRI}}{2}\sqrt{2}-2t_{Cu}=449.62$~mm, with $N=\lfloor \frac{d}{a+t} \rfloor=22$ subcavities could be implemented, providing a volume of $V=141.52$~L. Comparing the volume in both examples the BabyIAXO case provides a greater value. However, despite the lower volume value ($141.52$~L versus $236.75$~L) and observing equation~\ref{eq:ga}, the lower system temperature ($0.1$~K versus $4.2$~K) and the higher magnetic field ($9$~T versus $2.5$~T) of the MRI (AMDX-EFR) magnet (see Table~\ref{tab:magnets}) make this bore much more recommended between these two examples.\\

\begin{figure}[h]
\centering
\begin{subfigure}[b]{0.3\textwidth}
         \centering
         \includegraphics[width=1\textwidth]{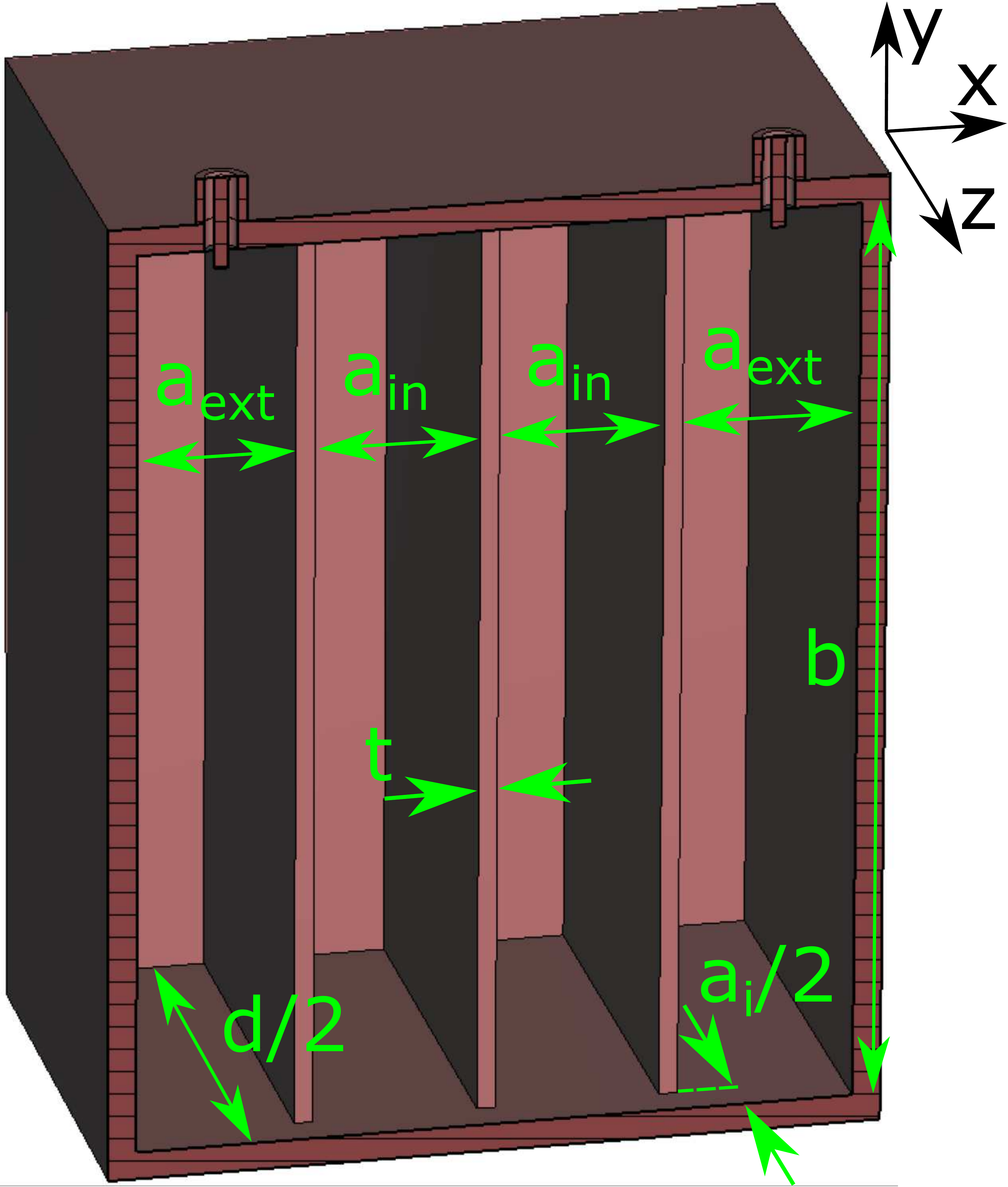}
         \caption{}
         \label{fig:4cav_Allind_LongTallSubcavities_InWidthCoup_Model}
\end{subfigure}
\hfill
\begin{subfigure}[b]{0.59\textwidth}
         \centering
         \includegraphics[width=1\textwidth]{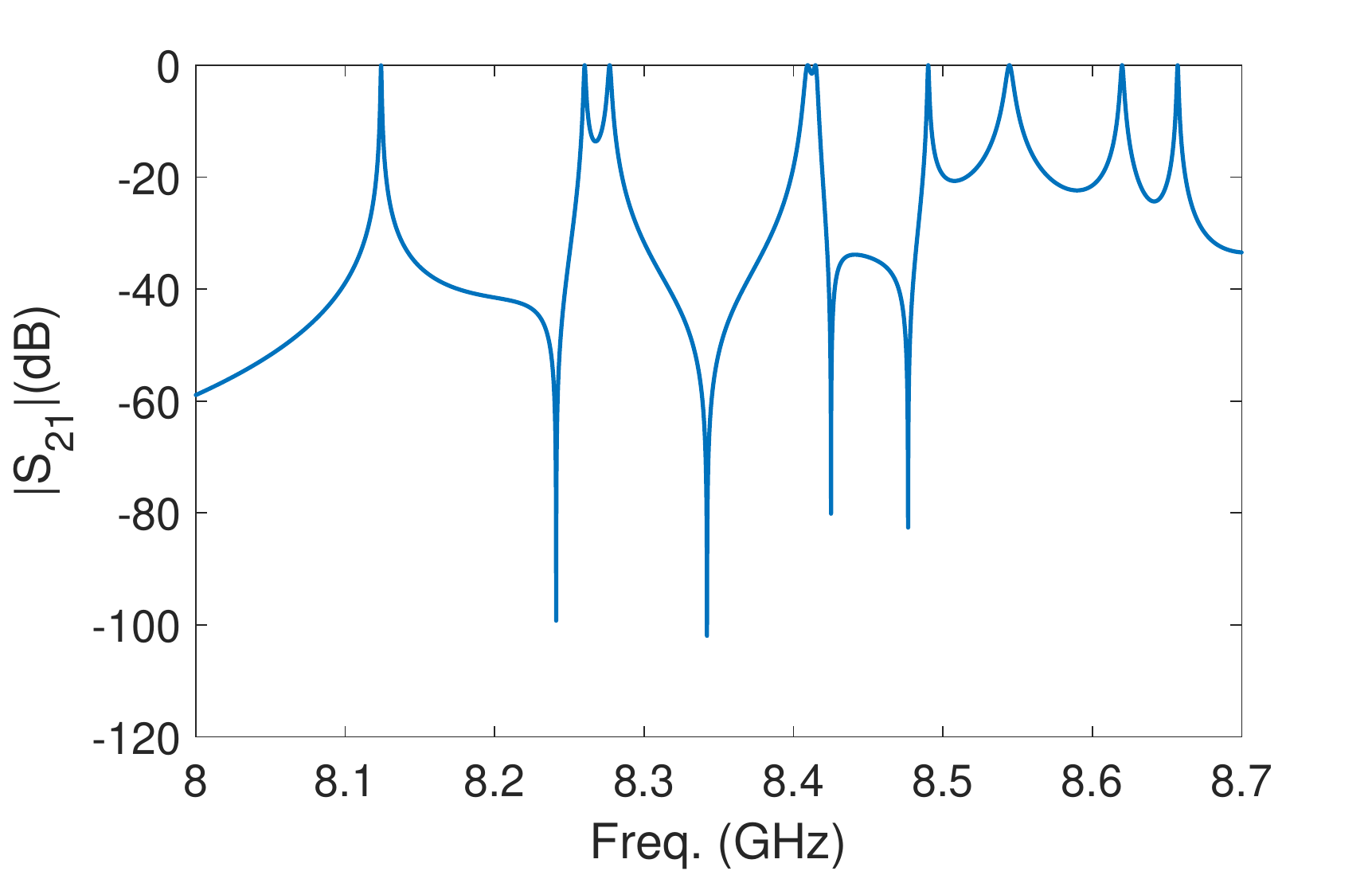}
         \caption{}
         \label{fig:4cav_Allind_LongTallSubcavities_InWidthCoup_S21}
\end{subfigure}
\caption{All-inductive multicavity haloscope design based on four subcavities with three inductive irises combining the long and tall cavity concepts: (a) picture of a symmetrical half of the structure, and (b) simulated magnitude of $S_{21}$ scattering parameter as a function of the frequency for $T=2$~K.}
\label{fig:4cav_Allind_LongTallSubcavities_InWidthCoup}
\end{figure}

An all-inductive multicavity haloscope based on $N=4$ large subcavities of $d=100$~mm and $b=100$~mm employing the in-width direction for the couplings has been designed in this paper as a preliminary proof of concept. Figure~\ref{fig:4cav_Allind_LongTallSubcavities_InWidthCoup_Model} shows the physical model of the haloscope.
In this case the selected physical coupling $|k|$ value and the resulting coupling matrix is that to the four tall subcavities design (see equation~\ref{eq:CouplingMatrix_1D_TallExample}) from the previous section. Figure~\ref{fig:4cav_Allind_LongTallSubcavities_InWidthCoup_S21} shows the CST simulation for the $S_{21}$ scattering parameter magnitude as a function of the frequency. The dimensions of the structure are: length and height of all the subcavities $d=100$~mm and $b=100$~mm, respectively; internal subcavities width $a_{in}=17.6$~mm; external subcavities width $a_{ext}=17.85$~mm; and width and thickness of all the inductive irises $a_i=8.9$~mm and $t=2$~mm, respectively.\\

A quality and form factor values of $Q_0=58327$ (at $T=2$~K) and $C=0.464$, respectively are obtained for the $TE_{101}$ mode. The form factor for this structure was not optimized (envisaged work is expected in this topic). The eigenmodes that appear in the plot are listed in Table~\ref{tab:4cav_Allind_LongTallSubcavities_InWidthCoup_Modes}.\\

\begin{table}[h]
\begin{tabular}{|c|c|c|c|c|c|c|}
\hline
Freq. (GHz) & Resonant mode & Configuration\\ \hline\hline
8.124 & $TE_{101}$ & [+ + + +] \\ \hline
8.26 & $TE_{111}$ & [+ + + +] \\ \hline
8.277 & $TE_{101}$ & [+ + \text{--} \text{--}] \\ \hline
8.409 & $TE_{101}$ & [+ \text{--} \text{--} +] \\ \hline
8.415 & $TE_{111}$ & [+ + \text{--} \text{--}] \\ \hline
8.49 & $TE_{101}$ & [+ \text{--} + \text{--}] \\ \hline
8.544 & $TE_{111}$ & [+ \text{--} \text{--} +] \\ \hline
8.62 & $TE_{111}$ & [+ \text{--} + \text{--}] \\ \hline
8.657 & $TE_{121}$ & [+ + + +] \\ \hline
\end{tabular}
\centering
\caption{\label{tab:4cav_Allind_LongTallSubcavities_InWidthCoup_Modes} List of the configuration and resonant modes shown in Figure~\ref{fig:4cav_Allind_LongTallSubcavities_InWidthCoup_S21}.}
\end{table}

The relative mode separation of this prototype is $\Delta f= 1.67$~$\%$ ($136$~MHz) which is not far from the value provided in Figure~\ref{fig:ModeMixing_vs_b-a} for $b/a=100/17.7=5.65$ ($1$~$\%$). The position of the ports (at the center of the subcavities) avoids the excitation of the $TE_{102}$ resonance since this mode has a zero electric field at that position. However, if this mode appeared it would satisfy the mode separation shown in Figure~\ref{fig:ModeMixing_vs_d-a} for $d/a=100/17.7=5.65$ ($4.45$~$\%$). Similarly to the previous structure (tall multicavity with $N=4$), the distance of the next configuration mode ([+ + \text{--} \text{--}]) is not a problem because it is far in frequency ($1.88$~$\%$ or $153$~MHz) due to the low number of subcavities employed in this multicavity structure. Similar to the example shown in the previous section (tall structures), the response of this multicavity also shows transmission zeros produced between resonant modes, aiding their separation when they are close in frequency.\\

The volume of this haloscope is $V=714$~mL, resulting to a $Q_0 \times V \times C$ value of $1.9\times 10^4$~L, which is $96.44$ times that of a single standard WR-90 cavity.

\section{2D and 3D multicavities}
\label{Sec:2D3D}
\begin{figure}[h]
\centering
\begin{subfigure}[b]{0.55\textwidth}
         \centering
         \includegraphics[width=1\textwidth]{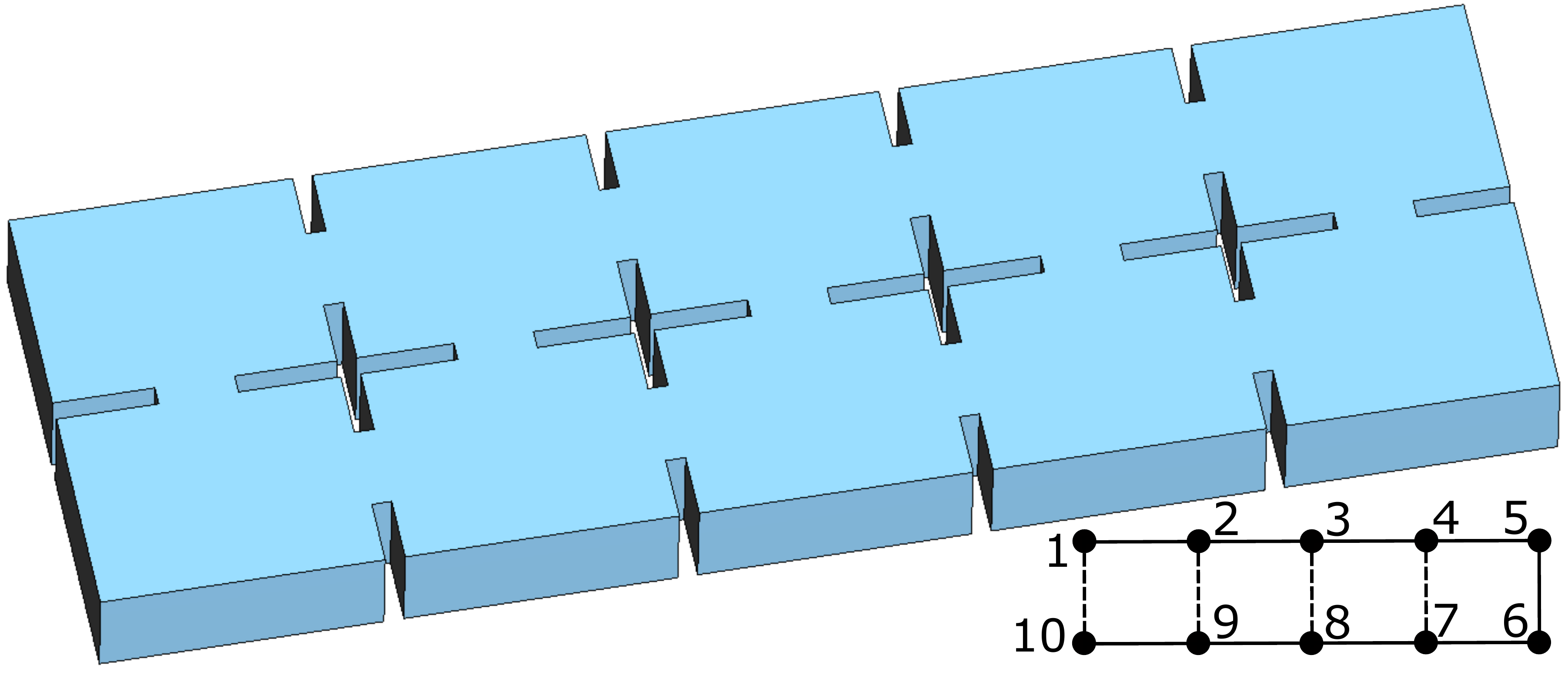}
         \caption{}
         \label{fig:2Dhorizontal}
\end{subfigure}
\hfill
\begin{subfigure}[b]{0.49\textwidth}
         \centering
         \includegraphics[width=1\textwidth]{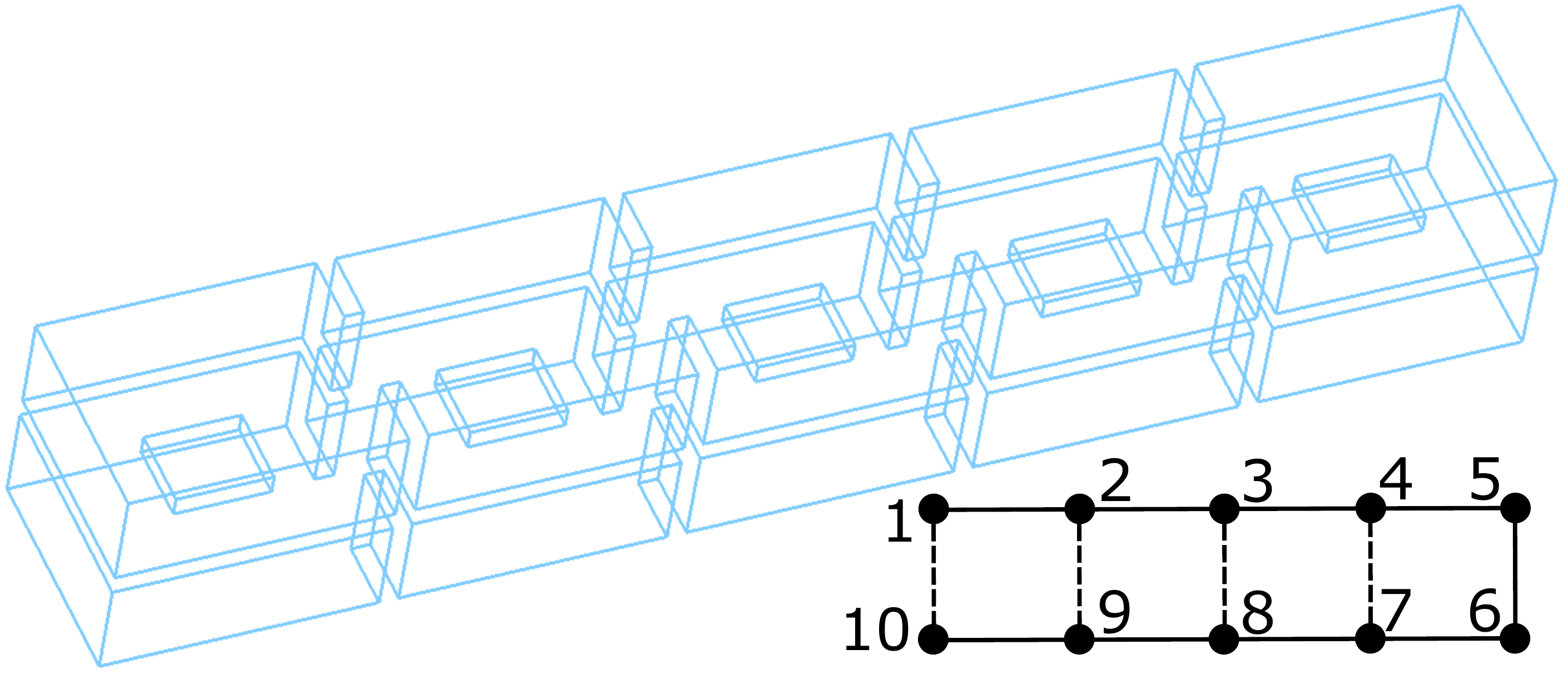}
         \caption{}
         \label{fig:2Dvertical}
\end{subfigure}
\hfill
\begin{subfigure}[b]{0.49\textwidth}
         \centering
         \includegraphics[width=1\textwidth]{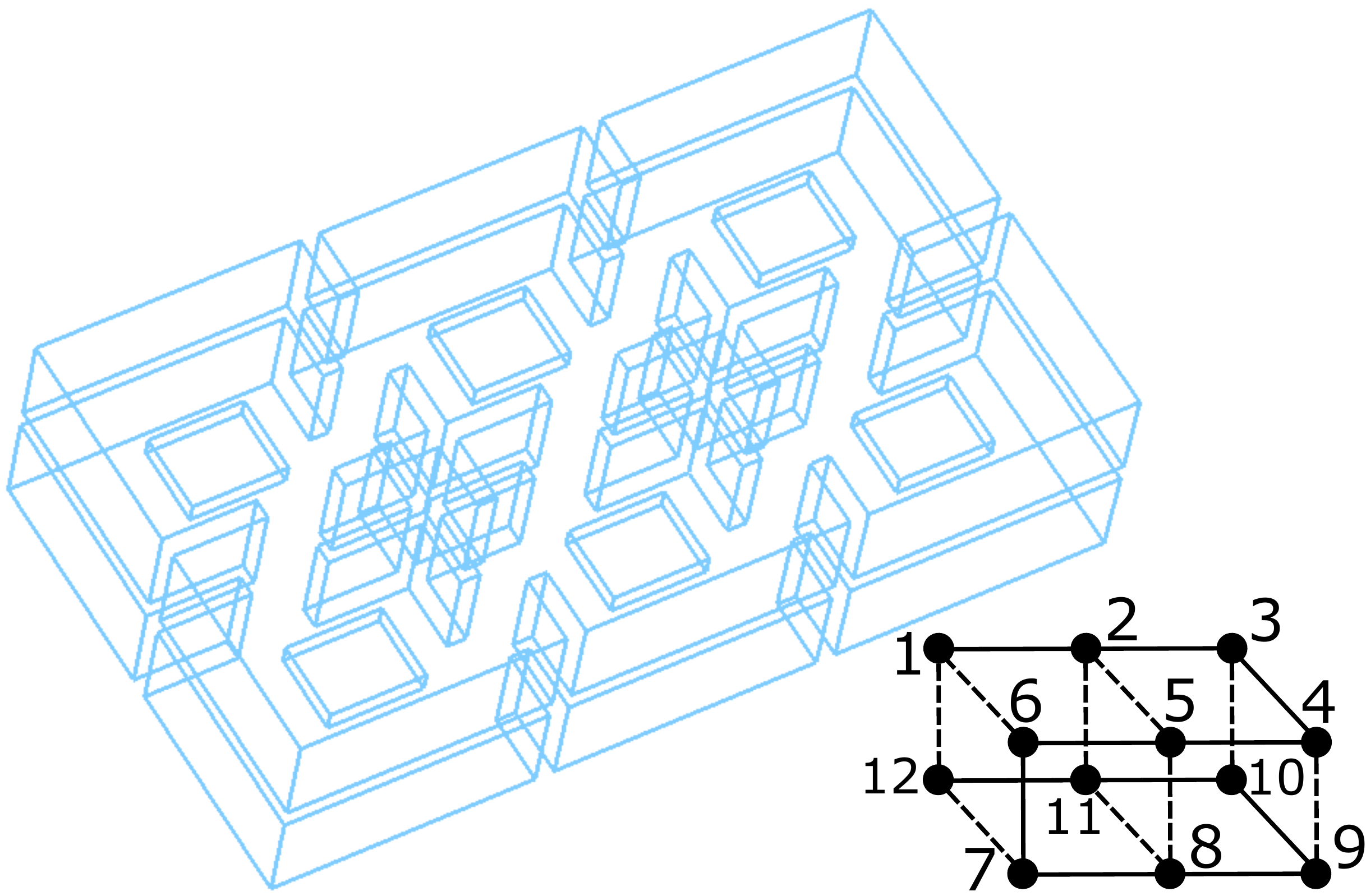}
         \caption{}
         \label{fig:3D2vertical2horizontal}
\end{subfigure}
\caption{Examples of 2D and 3D structures with cross-couplings: (a) horizontally folded structure based on ten subcavities with four cross-couplings (a topology diagram is shown at the right-bottom corner), (b) vertically folded structure with the same topology properties, and (c) 3D structure based on 12 subcavities with nine cross-couplings. In the topology diagrams, a solid line denotes the main coupling path while a dashed line indicates a cross-coupling.}
\label{fig:2Dand3Dtopologies}
\end{figure}

A straightforward generalization of 1D multicavities leads to the definition of 2D and 3D multicavities, which can be interesting geometries in order to fit the available room in some magnets. In addition, they can provide transmission zeros which can be used for rejecting nearby modes to the axion one. These topologies employ a type of interresonator coupling known as {\em cross-coupling} and they are created by irises that connect non-adjacent subcavities \cite{Cameron}. To achieve this goal, the simplest way is to fold the array of subcavities either vertically or horizontally thus making possible to introduce iris windows between non-adjacent cavities, therefore obtaining a $2D$ array of subcavities. If the original in-line topology is folded along two different axis the resulting structure would be a $3D$ array of subcavities. In Figure~\ref{fig:2Dand3Dtopologies} several examples of 2D (vertically or horizontally folded) and 3D multicavity structures can be observed.
Note that diagonal cross-couplings could also be incorporated if needed (for instance, between the first and the second last subcavities), but they are not included in these examples.\\

The coupling matrices associated to the 2D and 3D examples shown in the topology diagrams from Figure~\ref{fig:2Dand3Dtopologies} are the following:
\begin{equation}\label{eq:CouplingMatrix_2D_Example}
\bf{M_{2D}}=\left(
\begin{smallmatrix}
   \Omega_1 & M_{1,2} & 0 & 0 & 0 & 0 & 0 & 0 & 0 & M_{1,10} \\
   M_{1,2} & \Omega_2 & M_{2,3} & 0 & 0 & 0 & 0 & 0 & M_{2,9} & 0 \\
   0 & M_{2,3} & \Omega_3 & M_{3,4} & 0 & 0 & 0 & M_{3,8} & 0 & 0 \\
   0 & 0 & M_{3,4} & \Omega_4 & M_{4,5} & 0 & M_{4,7} & 0 & 0 & 0 \\
   0 & 0 & 0 & M_{4,5} & \Omega_5 & M_{5,6} & 0 & 0 & 0 & 0 \\
   0 & 0 & 0 & 0 & M_{5,6} & \Omega_6 & M_{6,7} & 0 & 0 & 0 \\
   0 & 0 & 0 & M_{4,7} & 0 & M_{6,7} & \Omega_7 & M_{7,8} & 0 & 0 \\
   0 & 0 & M_{3,8} & 0 & 0 & 0 & M_{7,8} & \Omega_8 & M_{8,9} & 0 \\
   0 & M_{2,9} & 0 & 0 & 0 & 0 & 0 & M_{8,9} & \Omega_9 & M_{9,10} \\
   M_{1,10} & 0 & 0 & 0 & 0 & 0 & 0 & 0 & M_{9,10} & \Omega_{10}
\end{smallmatrix}
\right)
\end{equation}
\begin{equation}\label{eq:CouplingMatrix_3D_Example}
\bf{M_{3D}}=\left(
\begin{smallmatrix}
   \Omega_1 & M_{1,2} & 0 & 0 & 0 & M_{1,6} & 0 & 0 & 0 & 0 & 0 & M_{1,12} \\
   M_{1,2} & \Omega_2 & M_{2,3} & 0 & M_{2,5} & 0 & 0 & 0 & 0 & 0 & M_{2,11} & 0 \\
   0 & M_{2,3} & \Omega_3 & M_{3,4} & 0 & 0 & 0 & 0 & 0 & M_{3,10} & 0 & 0 \\
   0 & 0 & M_{3,4} & \Omega_4 & M_{4,5} & 0 & 0 & 0 & M_{4,9} & 0 & 0 & 0 \\
   0 & M_{2,5} & 0 & M_{4,5} & \Omega_5 & M_{5,6} & 0 & M_{5,8} & 0 & 0 & 0 & 0 \\
   M_{1,6} & 0 & 0 & 0 & M_{5,6} & \Omega_6 & M_{6,7} & 0 & 0 & 0 & 0 & 0 \\
   0 & 0 & 0 & 0 & 0 & M_{6,7} & \Omega_7 & M_{7,8} & 0 & 0 & 0 & M_{7,12} \\
   0 & 0 & 0 & 0 & M_{5,8} & 0 & M_{7,8} & \Omega_8 & M_{8,9} & 0 & M_{8,11} & 0 \\
   0 & 0 & 0 & M_{4,9} & 0 & 0 & 0 & M_{8,9} & \Omega_9 & M_{9,10} & 0 & 0 \\
   0 & 0 & M_{3,10} & 0 & 0 & 0 & 0 & 0 & M_{9,10} & \Omega_{10} & M_{10,11} & 0 \\
   0 & M_{2,11} & 0 & 0 & 0 & 0 & 0 & M_{8,11} & 0 & M_{10,11} & \Omega_{11} & M_{11,12} \\
   M_{1,12} & 0 & 0 & 0 & 0 & 0 & M_{7,12} & 0 & 0 & 0 & M_{11,12} & \Omega_{12}
\end{smallmatrix}
\right)
\end{equation}\\

The three main diagonals of both matrices have the same behaviour as in equation~\ref{eq:CouplingMatrix_1D}. However, for 2D and 3D topologies an anti-diagonal with non-zero value appears due to the new cross-couplings. In addition, for 3D structures other cross-couplings can appear due to the folding introduced in the two axes (horizontal and vertical) as depicts the model shown in Figure~\ref{fig:3D2vertical2horizontal}. This is the case of the elements $M_{1,6}$, $M_{2,5}$, $M_{7,12}$ and $M_{8,11}$ (and its symmetrical pairs).\\

As a first proof of concept, a rigorous study has been carried out in which different types of topologies have been tested on an all-inductive 2D multicavity structure. The study was conducted on $N = 6$ subcavities folded horizontally (three subcavities per row), each subcavity with standard dimensions. The main objective of this study is to find a topology that rejects the next eigenmode to the axion one in order to improve the mode clustering issue. This has been achieved with only one cross-coupling just placing a window iris between the first and last ($6^{th}$) subcavity. This prototype has been designed, optimized and manufactured.\\

In the design of this structure the following coupling matrix has been employed for the development of the geometry parameters:
\begin{gather}\label{eq:CrossCoupling_Matrix}
 \bf{M} =
  \begin{pmatrix}
   1 & -0.5 & 0 & 0 & 0 & -0.5 \\
   -0.5 & 1 & -0.5 & 0 & 0 & 0 \\
   0 & -0.5 & 1 & -0.5 & 0 & 0 \\
   0 & 0 & -0.5 & 1 & -0.5 & 0 \\
   0 & 0 & 0 & -0.5 & 1 & -0.5 \\
   -0.5 & 0 & 0 & 0 & -0.5 & 1
   \end{pmatrix}
\end{gather}

\begin{figure}[h]
\centering
\includegraphics[width=0.6\textwidth]{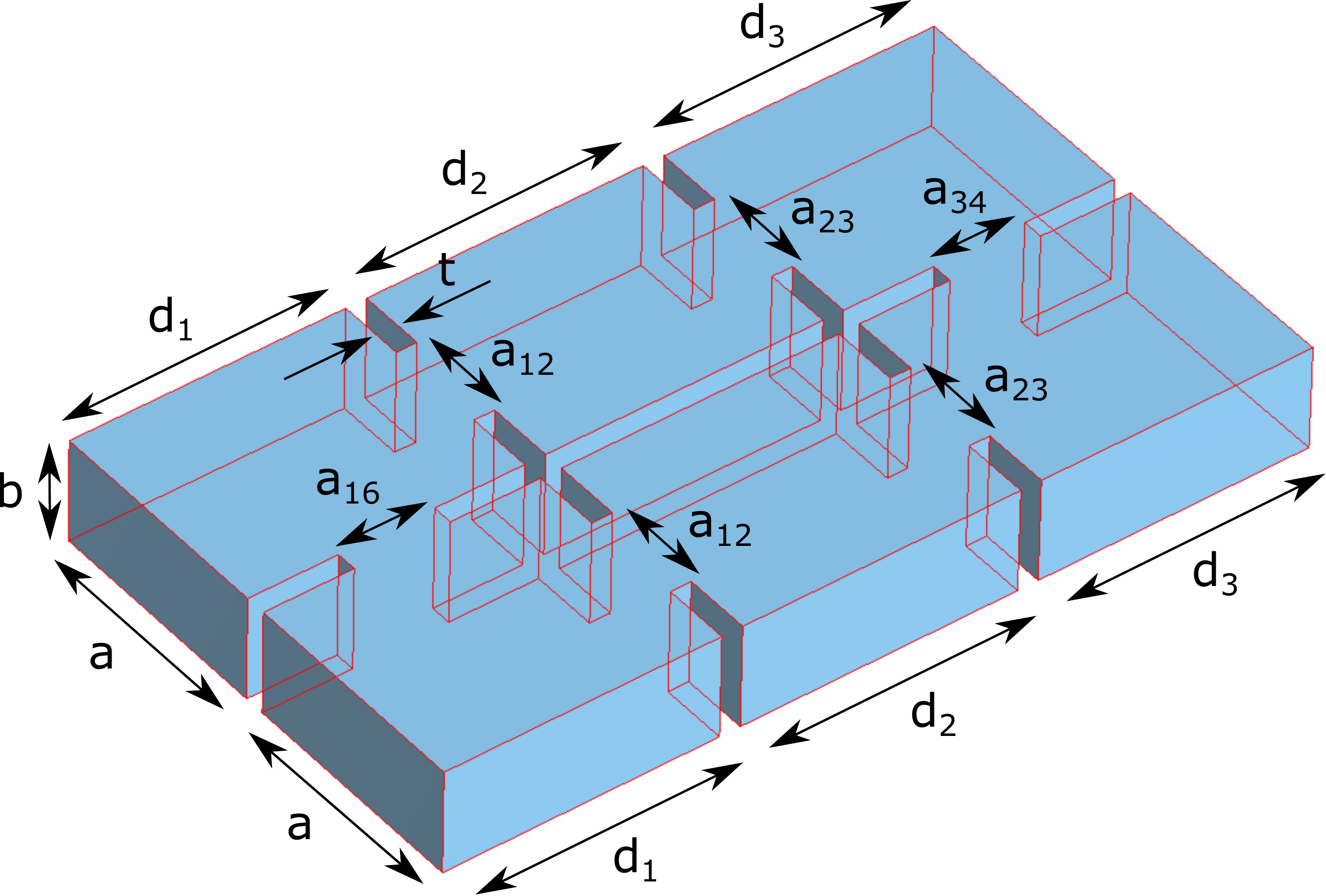}
\caption{Model of a 6 subcavities 2D multicavity structure folded horizontally (two symmetrical rows with three subcavities per row). It is based on an all-inductive coupled multicavity structure. An inductive iris between the $1^{st}$ and $6^{th}$ subcavity is introduced thanks to the physical folding applied in the horizontal plane.}
\label{fig:6cav_M16_Model}
\end{figure}

As it can be seen in equation~\ref{eq:CrossCoupling_Matrix}, a non zero value is selected for the elements $M_{16}$ and $M_{61}$ due to the use of a cross-coupling iris. From this coupling matrix it can be observed that the sign for all the interresonator couplings is negative ($k<0$). This indicates that the structure can be implemented with all irises of inductive type (even for the cross-coupling one). The model of this structure is shown in Figure~\ref{fig:6cav_M16_Model}.\\

\begin{figure}[h]
\centering
\begin{subfigure}[b]{1\textwidth}
         \centering
         \includegraphics[width=1\textwidth]{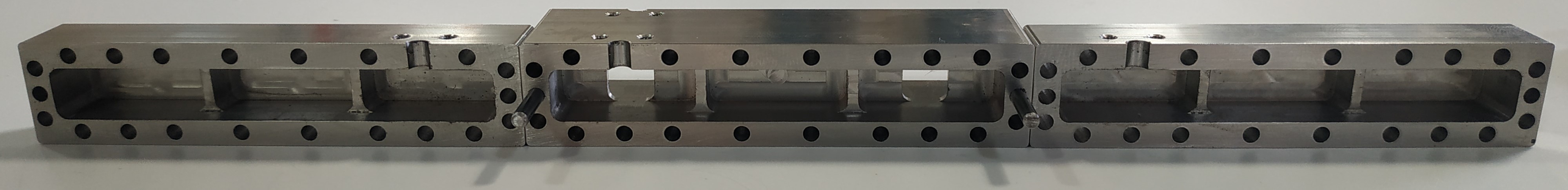}
         \caption{}
         \label{fig:6cav_M16_Picture_316L}
\end{subfigure}
\hfill
\begin{subfigure}[b]{0.39\textwidth}
         \centering
         \includegraphics[width=1\textwidth]{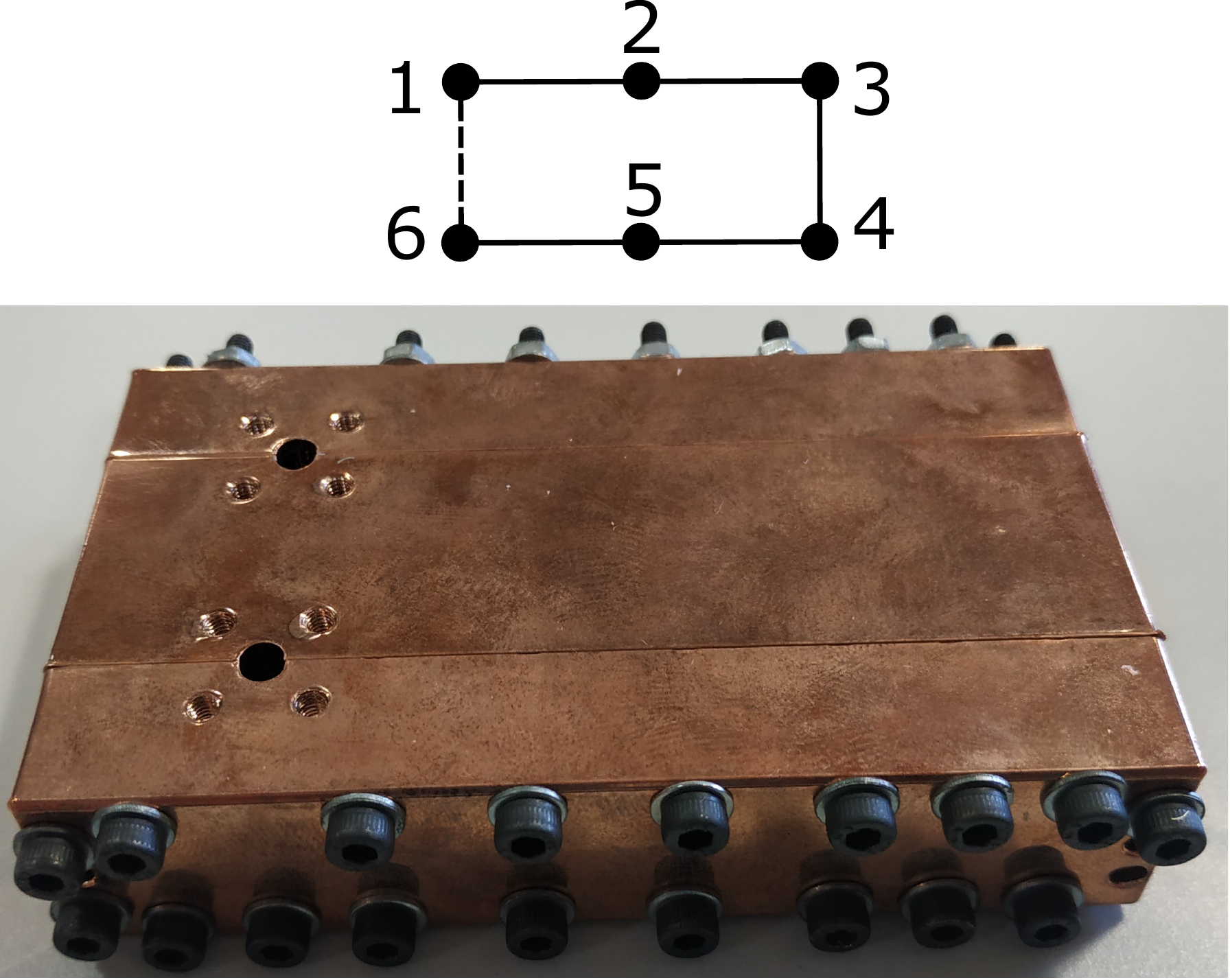}
         \caption{}
         \label{fig:6cav_M16_Picture_Copper}
\end{subfigure}
\hfill
\begin{subfigure}[b]{0.59\textwidth}
         \centering
         \includegraphics[width=1\textwidth]{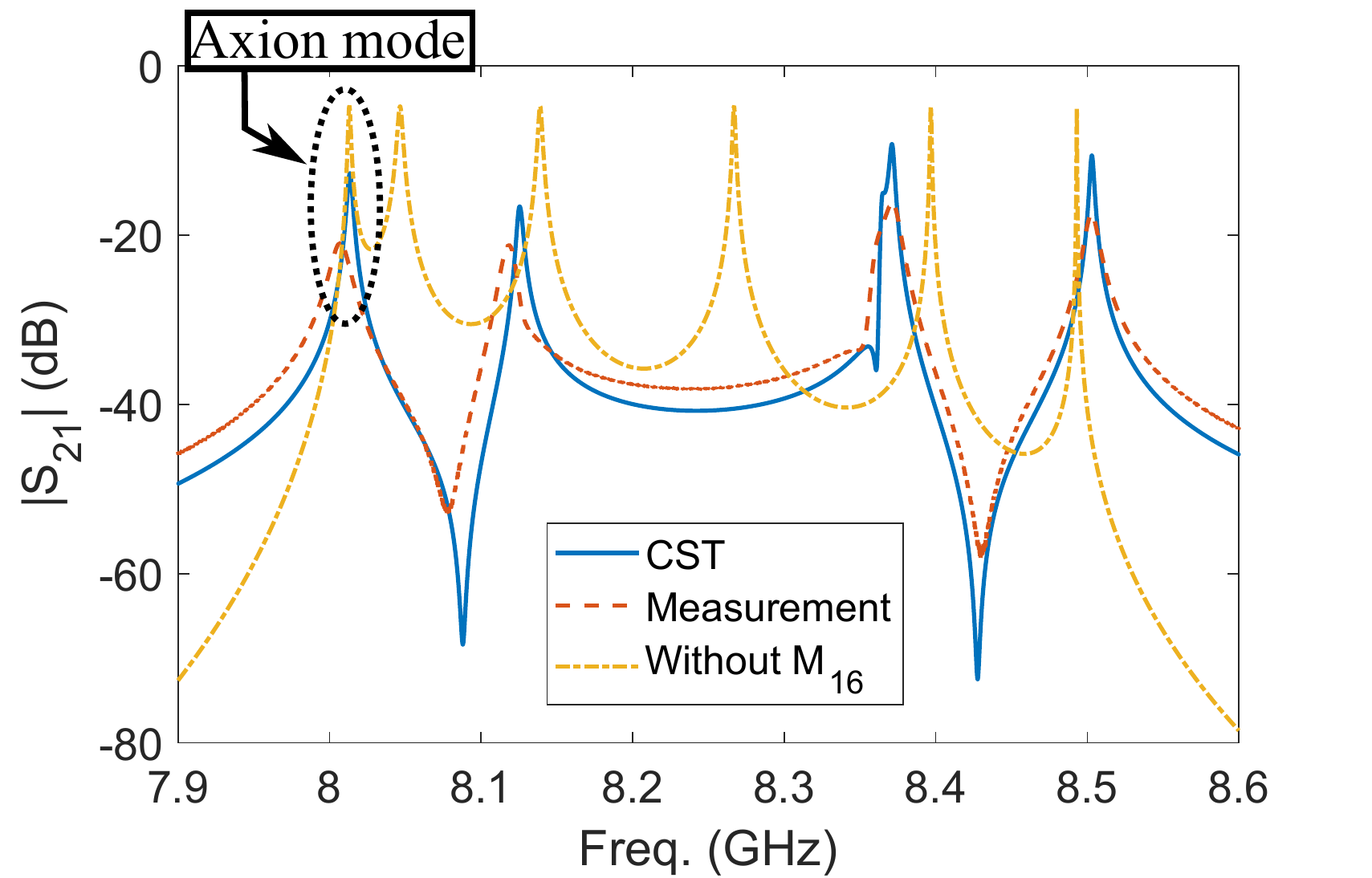}
         \caption{}
         \label{fig:6cav_M16_Sp}
\end{subfigure}
\caption{Structure based on six subcavities with a cross-coupling between the first and the last subcavities: (a) picture of the manufactured pieces (three parts), (b) topology diagram and copper coated final structure already mounted, and (c) $S_{21}$ scattering parameter magnitude as a function of the frequency (simulation versus measurements at room temperature). The analytical response of a 6-subcavities structure without the cross-coupling $M_{16}$ has been added to show the cancellation of the closest mode using one transmission zero.}
\label{fig:6cav_M16}
\end{figure}

Two pictures of the fabricated prototype are shown in Figures~\ref{fig:6cav_M16_Picture_316L} and \ref{fig:6cav_M16_Picture_Copper}.
The optimisation process has been based on adjusting the frequency position of one of the transmission zeros to cancel the closest mode. This cancellation of the closest mode can be observed in Figure~\ref{fig:6cav_M16_Sp} which plots the $S_{21}$ scattering parameter magnitude as a function of the frequency for the optimized design in comparison with experimental results and with the response of a multicavity based on 6 subcavities without cross-couplings.\\

The final dimensions of this structure (see Figure~\ref{fig:6cav_M16_Model}) are: $a=22.86$~mm, $b=10.16$~mm, $d_1=26.516$~mm, $d_2=26.845$~mm, $d_3=26.503$~mm, $a_{12}=a_{23}=9.921$~mm, $a_{34}=8.894$~mm, $a_{16}=9.203$~mm, and thickness of all the inductive irises $t=2$~mm.\\

As it is shown in Figure~\ref{fig:6cav_M16_Sp}, the mode separation from the axion mode (obtained at $8.013$~GHz) to the next eigenmode is $\Delta f=127$~MHz in simulation and $111$~MHz in measurements (without cross-coupling it is $34$~MHz). A good agreement is observed between simulation results and measurements. From simulations employing copper material a $Q_0^{2K}=40000$ is predicted for the axion mode at cryogenic temperatures and $Q_0^{300K}=6800$ at room temperature. The measurements from the manufactured structure at room temperature provide a value of $Q_0=4000$ with the copper coated structure, a $60\%$ of the simulation result. The obtained form factor for this mode is $C=0.702$. Note how for this multicavity structure a form factor higher than the theoretical one for a single cavity is obtained. This occurs due to the use of several subcavities connected by irises, in which one of the configuration modes is cancelled by one transmission zero. Therefore, these transmission zeros provide a good avenue not only for improving the mode clustering, but also to slightly increase the form factor. Even higher performances could be achieved when 2D and 3D geometries are combined with long, tall or large subcavities (and with the alternating coupling concept) regarding the main topics covered for multicavities in this work: the $Q_0 \times V \times C$ factor, the mode clustering ($\Delta f$), the realizable interresonator physical coupling ($k$), the use of transmission zeros by cross-coupling and the bore sizes in dipole and solenoid magnets (see examples in Figures~\ref{fig:StakingLongSubcavitiesInDipolesAndSolenoids} and \ref{fig:StakingTallSubcavitiesInDipolesAndSolenoids}).\\

In the case of this prototype, the total volume obtained is $V=38$~mL, and the $Q_0 \times V \times C$ factor is $1067$~L, which is $5.33$ times that of a single standard WR-90 cavity.\\

\begin{figure}[h]
\centering
\begin{subfigure}[b]{0.55\textwidth}
         \centering
         \includegraphics[width=1\textwidth]{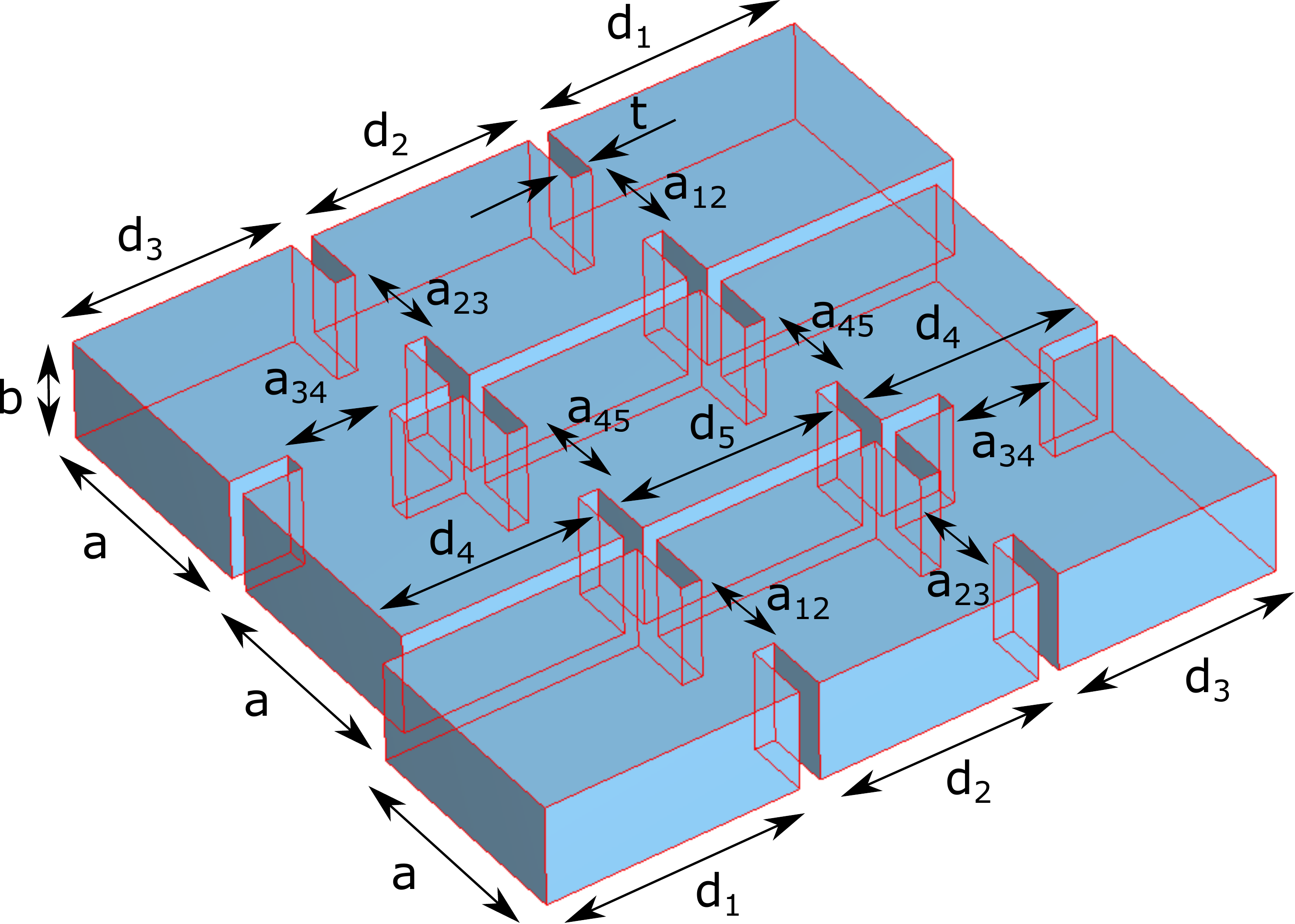}
         \caption{}
         \label{fig:Meandro_3x3_Model}
\end{subfigure}
\hfill
\begin{subfigure}[b]{0.44\textwidth}
         \centering
         \includegraphics[width=1\textwidth]{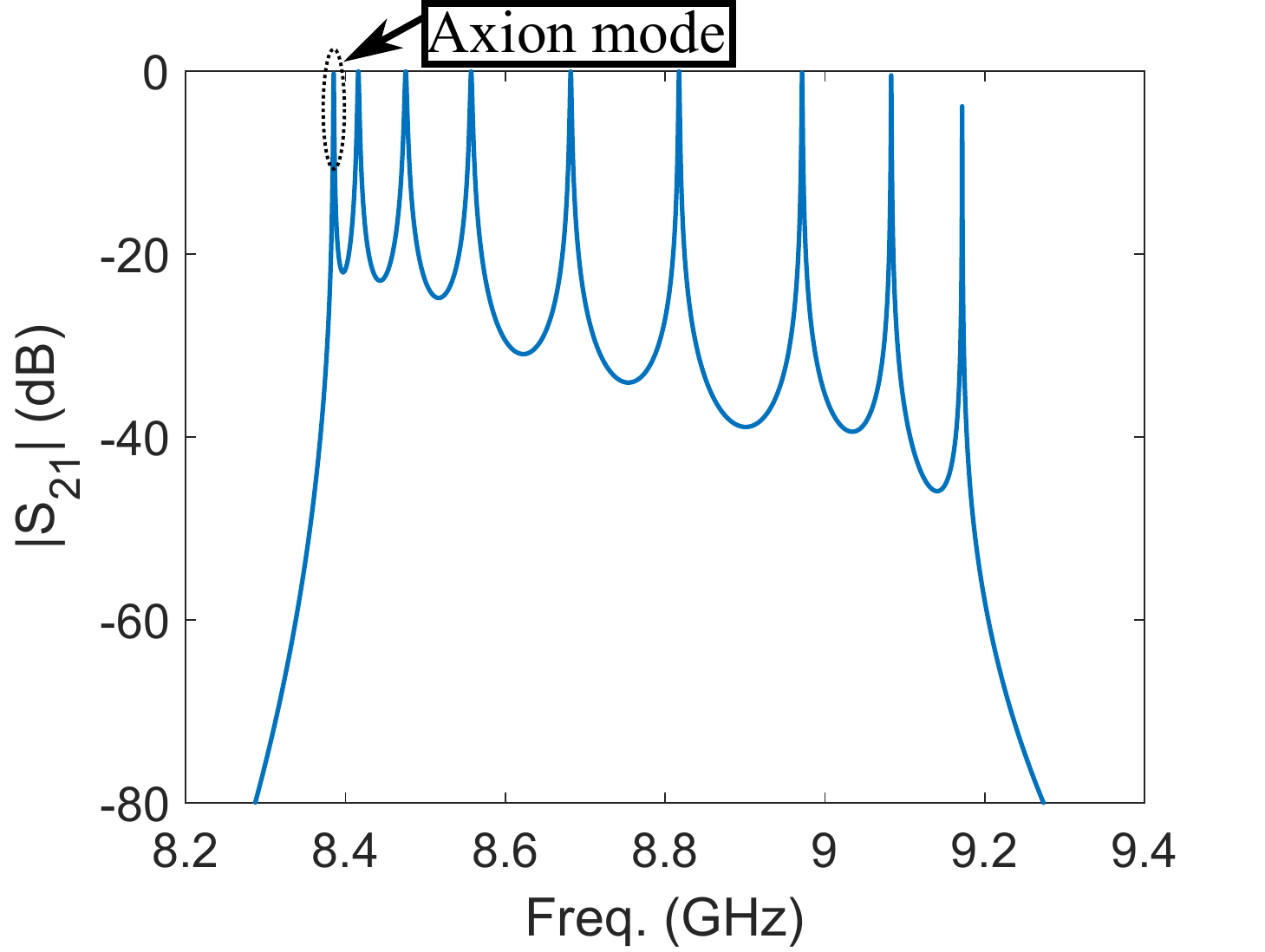}
         \caption{}
         \label{fig:Meandro_3x3_Sp}
\end{subfigure}
\caption{Structure based on nine subcavities with a meander shape (2D geometry): (a) picture of the model, (b) simulated $S_{21}$ scattering parameter magnitude as a function of the frequency.}
\label{fig:Meander}
\end{figure}

As a final study of this work, an interesting 2D geometry is now proposed to use more efficiently the available magnet bore footprint. The design is simpler than the previous one, since it has no cross-couplings (and therefore there are not transmission zeros). The idea is based on introducing a meander multicavity geometry, as it is shown in Figure~\ref{fig:Meandro_3x3_Model}.
Also, in Figure~\ref{fig:Meandro_3x3_Sp} the $S_{21}$ scattering parameter magnitude as a function of the frequency is shown. The axion mode is the first resonance in the response since the structure is based on all inductive irises.\\

Although the geometry of this structure is 2D, topologically it is 1D, since it implements only couplings from adjacent resonators. This can be observed in its coupling matrix:
\begin{equation}
\label{eq:CouplingMatrix_2D_Meander}
\bf{M}=\left(
\begin{smallmatrix}
   0.5 & -0.5 & 0 & 0 & 0 & 0 & 0 & 0 & 0 \\
   -0.5 & 1 & -0.5 & 0 & 0 & 0 & 0 & 0 & 0 \\
   0 & -0.5 & 1 & -0.5 & 0 & 0 & 0 & 0 & 0 \\
   0 & 0 & -0.5 & 1 & -0.5 & 0 & 0 & 0 & 0 \\
   0 & 0 & 0 & -0.5 & 1 & -0.5 & 0 & 0 & 0 \\
   0 & 0 & 0 & 0 & -0.5 & 1 & -0.5 & 0 & 0 \\
   0 & 0 & 0 & 0 & 0 & -0.5 & 1 & -0.5 & 0 \\
   0 & 0 & 0 & 0 & 0 & 0 & -0.5 & 1 & -0.5 \\
   0 & 0 & 0 & 0 & 0 & 0 & 0 & -0.5 & 0.5
\end{smallmatrix}
\right)
\end{equation}

The final dimensions of this structure (see Figure~\ref{fig:Meandro_3x3_Model}) are: $a=22.86$~mm, $b=10.16$~mm, $d_1=25.2$~mm, $d_2=d_3=d_4=d_5=22$~mm, and width and thickness of all the inductive irises $a_{12}=a_{23}=a_{34}=a_{45}=10.25$~mm and $t=2$~mm, respectively. This design provides an axion mode frequency of $f_a=8.385$~GHz, and a form and quality factors of $C=0.684$ and $Q_0^{2K}=41475$, respectively which corresponds with very good results compared to previous designs from RADES and the ones shown in this work. The mode clustering value $\Delta f$ is $30.8$~MHz ($0.37\%$), in accordance with the results from Figure~\ref{fig:Allind_vs_Allcap_vs_Alt_vs_Indi_ModeMixing}. The resulting total volume of the haloscope is $V=49.14$~mL, and the total $Q_0 \times V \times C$ is $1394.05$~L, which is $6.96$ times that of a single standard WR-90 cavity.\\

The benefits of this haloscope geometry are based on its quasi-square shape. In this case, the detector provides a footprint of $72.58\times 76.4$~mm$^2$ (along width and length), while in a 1D geometry these nine subcavities give an elongated shape of $22.86\times 286.4$~mm$^2$. This quasi-square area may be more appropriate in some cases where the magnet bore is limited in all dimensions (as it occurs with some solenoid magnets, see for instance Figure~\ref{fig:MulticavitiesStackedInWidth_Long_Solenoid}). In practise some of the ideas proposed in this paper can be combined to use more efficiently the available space in magnet bores extensively used by the axion search community.

\section{Conclusions and prospects}
\label{Conclusions}
In this work, the volume limits of rectangular haloscopes have been explored. The increase of this parameter improves the axion detection sensitivity, which has been a major motivation in recent years. Different strategies for increasing the volume, taking into account certain constraints such as the frequency separation between adjacent modes (mode clustering) and the variation of the form and quality factors, are presented. Also, exhaustive studies with single cavities and 1D multicavities and, in a more introductory way, 2D and 3D multicavities achieving large $Q_0 \times V \times C$ factors, are shown in this work. The compatibility of these haloscopes with the largest dipole and solenoid magnets in the axion community has been demonstrated. Several practical designs have been manufactured and measured, providing good results in quality factor and mode clustering, illustrating the capabilities of some of these studies while serving as validation.\\

It has been found that among the single cavities, large cavities provide the best $Q_0 \times V \times C$ performance. In addition, it has also been shown that, despite their greater complexity in the design process, the use of multicavities can lead to an improvement in this factor. Nevertheless, when searching the axion in a range of masses, the increase in volume is limited by the number of mode crossings that can be tolerated. On the other hand, novel results have been obtained in this paper where the appearance of transmission zeros in some multicavity designs allows to shift or suppress modes close to the axion one and thus it reduces both the mode clustering at one frequency and the possible mode crossings in a range of frequencies. These techniques are intended to serve as a manual for any experimental axion group wishing to search for volume limits in the design of a haloscope based on rectangular cavities to be placed inside both dipole and solenoid magnets. Nevertheless, these strategies and analysis are also useful for any application where increasing the volume of the device, for a given frequency, is a goal.\\

A wide range of promising possibilities opens up from this analysis, depending on the type and configuration of the data taking magnet. The strategies described in this work allow to make the best use of the bore space with the aim of maximizing the sensitivity of axion search experiments. In this sense, the study of the geometry limits employing several ideas proposed in this work, as the alternating coupling in 1D multicavites or the long, tall and large subcavities in 2D/3D multicavites is a recommendable task for the design of a high competitive haloscope in the axion community. Also, the extrapolation of all these studies and tests to cylindrical cavities is being investigated by the authors for a future work.

\acknowledgments
This work was performed within the RADES group. We thank our colleagues for their support. In addition, this work has been funded by the grant PID2019-108122GB-C33, funded by MCIN/AEI/10.13039/501100011033/ and by "ERDF A way of making Europe". JMGB thanks the grant FPI BES-2017-079787, funded by MCIN/AEI/10.13039/501100011033 and by "ESF Investing in your future". Also, this project has received partial funding through the European Research Council under grant ERC-2018-StG-802836 (AxScale).

\bibliographystyle{JHEP.bst}
\bibliography{mybibfile.bib}

\providecommand{\href}[2]{#2}\begingroup\raggedright\begin{thebibliography}{10}

\bibitem{Peccei:1977Jun}
R.~Peccei and H.~Quinn, \emph{{CP conservation in the presence of
  pseudoparticles}},
  \href{https://doi.org/https://doi.org/10.1103/PhysRevLett.38.1440}{\emph{Phys.
  Rev. Lett.} {\bfseries 38} (1977) 1440}.

\bibitem{Peccei:1977Sep}
R.~Peccei and H.~Quinn, \emph{{Constraints imposed by CP conservation in the
  presence of pseudoparticles}},
  \href{https://doi.org/https://doi.org/10.1103/PhysRevD.16.1791}{\emph{Phys.
  Rev. D} {\bfseries 16} (1977) 1791}.

\bibitem{Weinberg:1978}
S.~Weinberg, \emph{{A new light boson?}},
  \href{https://doi.org/https://doi.org/10.1103/PhysRevLett.40.223}{\emph{Phys.
  Rev. Lett.} {\bfseries 40} (1978) 223}.

\bibitem{Wilczek:1978}
F.~Wilczek, \emph{{Problem of strong P and T invariance in the presence of
  instantons}},
  \href{https://doi.org/https://doi.org/10.1103/PhysRevLett.40.279}{\emph{Phys.
  Rev. Lett.} {\bfseries 40} (1978) 279}.

\bibitem{Preskill:1983}
J.~Preskill, M.B.~Wise and F.~Wilczek, \emph{Cosmology of the invisible axion},
  \href{https://doi.org/https://doi.org/10.1016/0370-2693(83)90637-8}{\emph{Physics
  Letters B} {\bfseries 120} (1983) 127}.

\bibitem{Abbott:1983}
L.~Abbott and P.~Sikivie, \emph{A cosmological bound on the invisible axion},
  \href{https://doi.org/https://doi.org/10.1016/0370-2693(83)90638-X}{\emph{Physics
  Letters B} {\bfseries 120} (1983) 133}.

\bibitem{Dine:1983}
M.~Dine and W.~Fischler, \emph{The not-so-harmless axion},
  \href{https://doi.org/https://doi.org/10.1016/0370-2693(83)90639-1}{\emph{Physics
  Letters B} {\bfseries 120} (1983) 137}.

\bibitem{Irastorza:2018dyq}
I.G.~Irastorza and J.~Redondo, \emph{{New experimental approaches in the search
  for axion-like particles}},
  \href{https://doi.org/10.1016/j.ppnp.2018.05.003}{\emph{Prog. Part. Nucl.
  Phys.} {\bfseries 102} (2018) 89}
  [\href{https://arxiv.org/abs/1801.08127}{{\ttfamily 1801.08127}}].

\bibitem{Primakoff:1951}
H.~Primakoff, \emph{{Photoproduction of neutral mesons in nuclear electric
  fields and the mean life of the neutral meson}},
  \href{https://doi.org/https://doi.org/10.1103/PhysRev.81.899}{\emph{Phys.
  Rev.} {\bfseries 81} (1951) 899}.

\bibitem{Sikivie:1983ip}
P.~Sikivie, \emph{{Experimental Tests of the Invisible Axion}},
  \href{https://doi.org/10.1103/PhysRevLett.51.1415}{\emph{Phys. Rev. Lett.}
  {\bfseries 51} (1983) 1415}.

\bibitem{RADES_paper3}
A.~Álvarez Melcón, S.~Arguedas-Cuendis, J.~Baier, K.~Barth, H.~Bräuninger,
  S.~Calatroni et~al., \emph{{First results of the CAST-RADES haloscope search
  for axions at 34.67 microeV}},
  \href{https://doi.org/10.1007/JHEP10(2021)075}{\emph{J. High Energ. Phys.}
  {\bfseries 2021} (2021) 75}
  [\href{https://arxiv.org/abs/2104.13798}{{\ttfamily 2104.13798}}].

\bibitem{RADESreviewUniverse}
A.~Díaz-Morcillo, J.M.~García~Barceló, A.J.~Lozano~Guerrero, P.~Navarro,
  B.~Gimeno, S.~Arguedas~Cuendis et~al., \emph{Design of new resonant
  haloscopes in the search for the dark matter axion: A review of the first
  steps in the rades collaboration},
  \href{https://doi.org/10.3390/universe8010005}{\emph{Universe} {\bfseries 8}
  (2022) }.

\bibitem{kim_CAPP_2020}
D.~Kim, J.~Jeong, S.~Youn, Y.~Kima and Y.K.~Semertzidis, \emph{Revisiting the
  detection rate for axion haloscopes},
  \href{https://doi.org/10.1088/1475-7516/2020/03/066}{\emph{Journal of
  Cosmology and Astroparticle Physics} {\bfseries 2020} (2020) 1 }.

\bibitem{RADES_paper2}
A.~Álvarez Melcón, S.~Arguedas-Cuendis, C.~Cogollos, A.~Díaz-Morcillo,
  B.~Döbrich, J.D.~Gallego et~al., \emph{{Scalable haloscopes for axion dark
  matter detection in the 30 $\mu$eV range with RADES}},
  \href{https://doi.org/10.1007/JHEP07(2020)084}{\emph{Journal of High Energy
  Physics} {\bfseries 084} (2020) 1 }.

\bibitem{Braine:2019fqb}
{\scshape ADMX} collaboration, \emph{{Extended Search for the Invisible Axion
  with the Axion Dark Matter Experiment}},
  \href{https://doi.org/10.1103/PhysRevLett.124.101303}{\emph{Phys. Rev. Lett.}
  {\bfseries 124} (2020) 101303}
  [\href{https://arxiv.org/abs/1910.08638}{{\ttfamily 1910.08638}}].

\bibitem{PhDThesis-Brubaker}
B.M.~Brubaker, \emph{{First results from the HAYSTAC axion search}},  2018.
\newblock https://doi.org/10.48550/arXiv.1801.00835.

\bibitem{BabyIAXO}
A.~Abeln et~al., \emph{{Conceptual design of BabyIAXO, the intermediate stage
  towards the International Axion Observatory}},
  \href{https://doi.org/10.1007/JHEP05(2021)137}{\emph{Journal of High Energy
  Physics} {\bfseries 05} (2021) 137}.

\bibitem{CAST:1999}
K.~Zioutas, C.~Aalseth, D.~Abriola, F.~III, R.~Brodzinski, J.~Collar et~al.,
  \emph{{A decommissioned LHC model magnet as an axion telescope}},
  \href{https://doi.org/https://doi.org/10.1016/S0168-9002(98)01442-9}{\emph{Nuclear
  Instruments and Methods in Physics Research Section A: Accelerators,
  Spectrometers, Detectors and Associated Equipment} {\bfseries 425} (1999)
  480}.

\bibitem{RADES-HTS}
J.~Golm, S.A.~Cuendis, S.~Calatroni, C.~Cogollos, B.~Döbrich, J.D.~Gallego
  et~al., \emph{Thin film (high temperature) superconducting radiofrequency
  cavities for the search of axion dark matter},
  \href{https://doi.org/10.1109/TASC.2022.3147741}{\emph{IEEE Transactions on
  Applied Superconductivity} {\bfseries 32} (2022) 1}.

\bibitem{Aja_2022}
B.~Aja, S.A.~Cuendis, I.~Arregui, E.~Artal, R.B.~Barreiro, F.J.~Casas et~al.,
  \emph{{The Canfranc Axion Detection Experiment (CADEx): search for axions at
  90 GHz with Kinetic Inductance Detectors}},
  \href{https://doi.org/10.1088/1475-7516/2022/11/044}{\emph{Journal of
  Cosmology and Astroparticle Physics} {\bfseries 2022} (2022) 044}.

\bibitem{ADMX-EFR}
C.~Bartram, \emph{{Venturing a glimpse of the dark matter halo with ADMX (Talk
  in Cambridge Workshop on Axion Physics)}},  2021.

\bibitem{Choi:2021}
J.~Choi, S.~Ahn, B.~Ko, S.~Lee and Y.~Semertzidis, \emph{{CAPP-8TB: Axion dark
  matter search experiment around 6.7 $\mu$ev}},
  \href{https://doi.org/10.1016/j.nima.2021.165667}{\emph{Nucl. Instrum.
  Methods. Phys. Res. B} {\bfseries 1013} (2021) 165667}.

\bibitem{RADES_paper1}
A.~Álvarez Melcón, S.~Arguedas-Cuendis, C.~Cogollos, A.~Díaz-Morcillo,
  B.~Döbrich, J.D.~Gallego et~al., \emph{{Axion searches with microwave
  filters: the RADES project}},
  \href{https://doi.org/10.1088/1475-7516/2018/05/040}{\emph{Journal of
  Cosmology and Astroparticle Physics} {\bfseries 040} (2018) 1}.

\bibitem{Balanis:1989}
C.A.~Balanis, \emph{Advanced Engineering Electromagnetics}, John Wiley \& Sons
  (1989).

\bibitem{Q0_Characterization}
C.~Ramella, M.~Pirola and S.~Corbellini, \emph{{Accurate Characterization of
  High-$Q$ Microwave Resonances for Metrology Applications}},
  \href{https://doi.org/10.1109/JMW.2021.3063247}{\emph{IEEE Journal of
  Microwaves} {\bfseries 1} (2021) 610}.

\bibitem{3CavityRADES:2019}
S.~Arguedas~Cuendis et~al., \emph{{The 3 Cavity Prototypes of RADES: An Axion
  Detector Using Microwave Filters at CAST}},
  \href{https://doi.org/10.1007/978-3-030-43761-9_6}{\emph{Springer Proc.
  Phys.} {\bfseries 245} (2020) 45}
  [\href{https://arxiv.org/abs/1903.04323}{{\ttfamily 1903.04323}}].

\bibitem{Cameron}
R.J.~Cameron, C.M.~Kudsia and R.R.~Mansour, \emph{Microwave filters for
  communication systems: fundamentals, design, and applications.}, Wiley,
  second~ed. (2018).

\bibitem{CST}
``\url{https://www.3ds.com/products-services/simulia/products/cst-studio-suite/}.''

\bibitem{Guglielmi:1995}
M.~Guglielmi, F.~Montauti, L.~Pellegrini and P.~Arcioni, \emph{Implementing
  transmission zeros in inductive-window bandpass filters},
  \href{https://doi.org/10.1109/22.402281}{\emph{IEEE Transactions on Microwave
  Theory and Techniques} {\bfseries 43} (1995) 1911}.

\end{thebibliography}\endgroup
\end{document}